\documentclass[pra,12pt,amsmath,amssymb,aps,showpacs,superscriptaddress]{revtex4-1}

\usepackage{amsmath}
\usepackage{graphicx}
\usepackage{amssymb, latexsym}
\usepackage{setspace}
\usepackage[cm]{fullpage}
\usepackage[colorlinks, breaklinks=true, bookmarks=true]{hyperref}
\usepackage{color}

\DeclareMathOperator{\Real}{Re}
\DeclareMathOperator{\Imag}{Im}
 
\DeclareMathOperator{\diag}{diag}
\DeclareMathOperator{\sgn}{sgn}  

\newcommand{\deriv}[2]{\frac{d#1}{d#2}}
\newcommand{\nderiv}[3]{\frac{d^#3#1}{{d#2}^#3}}

\newcommand{\pderivdd}[3]{\frac{\partial^2#1}{\partial#2\partial#3}}
\newcommand{\fnlderiv}[2]{\frac{\delta#1}{\delta#2}}

\newcommand{\derat}[3]{\left.\frac{d#1}{d#2}\right\vert_{#3}}
\newcommand{\fnlderat}[3]{\left.\frac{\delta#1}{\delta#2}\right\vert_{#3}}

\newcommand{\bvec}[1]{\vec{\mathbf{#1}}}
\newcommand{\bvecs}[1]{\vec{\boldsymbol{#1}}}

\newcommand{\abs}[1]{\left\vert #1 \right\vert}
\newcommand{\norm}[1]{\left\Vert #1 \right\Vert}

\newcommand{\au}{\,\text{a.u.}}

\newcommand{\ket}[1]{\left|#1\right>}
\newcommand{\bra}[1]{\left<#1\right|}

\newcommand{\bracketsO}[3]{\left<#1 \left|#2 \right|#3 \right>}

\newcommand{\brackets}[2]{\left<#1 | #2\right>}

\newcommand{\bracketsb}[2]{\left<\left.#1 \right| #2\right>}
\newcommand{\bracketsk}[2]{\left<#1 \left| #2\right.\right>}
\newcommand{\exval}[1]{\left<#1 \right>}
\newcommand{\operator}[1]{\mathbf{\hat{#1}}}

\newcommand{\ie}{i.\,e.\ }
\newcommand{\eg}{e.\,g.\ }

\newcommand{\epsw}{\bar{\epsilon}(\omega)}
\newcommand{\epswu}{\bar{\epsilon}_{unc}(\omega)}
\newcommand{\epswj}{\bar{\epsilon}(\omega_j)}
\newcommand{\epswELk}{\bar{\epsilon}_{EL}^{(k)}(\omega)}
\newcommand{\epsunc}{\epsilon_{unc}}

\newcommand{\muwx}{\overline{\exval{\operator{\mu}_x}}(\omega)}
\newcommand{\Ow}{\overline{\exval{\operator{O}}}(\omega)}
\newcommand{\Cw}{\overline{\exval{\operator{C}}}(\omega)}

\newcommand{\feps}{f_{\epsilon}(\omega)}
\newcommand{\fepsj}{f_{\epsilon}(\omega_j)}
\newcommand{\tfeps}{\tilde{f}_{\epsilon}(\omega)}
\newcommand{\tfepsj}{\tilde{f}_{\epsilon}(\omega_j)}

\newcommand{\fO}{f_{O}(\omega)}

\newcommand{\fC}{f_C(\omega)}

\newcommand{\evC}{\exval{\operator{C}}\!(t)}
\newcommand{\evO}{\exval{\operator{O}}\!(t)}

\newcommand{\opC}{\operator{C}}

\newcommand{\Jepsz}{J_{\epsilon(0)}}
\newcommand{\JepsT}{J_{\epsilon(T)}}

\newcommand{\epswv}{\vec{\bar{\boldsymbol{\epsilon}}}}
\newcommand{\epswr}{\vec{\bar{\boldsymbol{\epsilon}}}_r}
\newcommand{\epswrEL}{\vec{\bar{\boldsymbol{\epsilon}}}_{r,EL}}
\newcommand{\veps}{\bvecs{\epsilon}}
\newcommand{\vg}{\bvec{g}}
\newcommand{\vomega}{\bvecs{\omega}}
\newcommand{\vomr}{\bvecs{\omega}_r}
\newcommand{\vgamma}{\bvecs{\gamma}}
\newcommand{\vdelta}{\bvecs{\delta}}
\newcommand{\vp}{\bvec{p}}
\newcommand{\vq}{\bvec{q}}
\newcommand{\vw}{\bvec{w}}

\newcommand{\ddX}{\ddot X}
\newcommand{\ddXw}{\overline{\exval{\operator{\ddX}}}(\omega)}

\begin{document}

\title{Optimization of high-order harmonic generation by optimal control theory: Ascending a functional landscape in extreme conditions}
\author{Ido Schaefer and Ronnie Kosloff}
\email{ido.schaefer@mail.huji.ac.il}
\email{kosloff1948@gmail.com}
\affiliation{The Institute of Chemistry, The Hebrew University of Jerusalem, Jerusalem 9190401, Israel}
\date{\today}

\begin{abstract}
A theoretical optimization method of high-harmonic-generation (HHG) is developed in the framework of optimal-control-theory (OCT). 
The target of optimization is the emission radiation of a particular frequency.
The OCT formulation includes restrictions on the frequency band of the driving pulse, the permanent ionization probability and the total energy of the driving pulse. The optimization task requires a highly accurate simulation of the dynamics. 
Absorbing boundary conditions are employed, where a complex-absorbing-potential is constructed by an optimization scheme for maximization of the absorption.
A new highly accurate propagation scheme is employed, which can address explicit time dependence of the driving as well as a non-Hermitian Hamiltonian. 
The optimization process is performed by a second-order gradient scheme. The method is applied to a simple one-dimensional model system. 
The results demonstrate a significant enhancement of selected harmonics, with minimized total energy of the driving pulse and controlled permanent ionization probability. 
A successful enhancement of an even harmonic emission is also demonstrated.
\end{abstract}
\maketitle

\tableofcontents

\section{Introduction}

When high intensity light is focused on a dilute gas novel phenomena emerge. One of the most intriguing is High-Harmonic-Generation (HHG), where a nonlinear response of the atomic system leads to emission in very high multiples of carrier frequency of the driving laser \cite{mcpherson1987studies,ferray88,li1989multiple}.
HHG is a crucial step in generating attosecond pulses \cite{mauritsson06,brabec2000intense,paul2001observation,corkum2007attosecond}. 
In addition, HHG has become a light source in extreme UV or soft X-ray for novel spectroscopic applications 
\cite{hassan2016optical,worner2011conical,luzon2017single,bhattacherjee2018ultrafast}. 

HHG constitutes a major breakthrough in experimental physics. However, the main drawback is the low efficiency of the process (typically $\sim 10^{-5}$).
Another major drawback is the energetic requirements, where very high power lasers are required to initiate the process (typically in the range of $\sim 10^{14}-10^{15}W/cm^2$).
In addition, the HHG process involves ionization. The permanent liberation of the electronic density into the macroscopic medium results in the production of plasma, which is responsible to severe experimental problems, such as dispersion and absorption of the emitted radiation.

The low yield of the HHG process has motivated novel approaches to enhance the process. One major direction is the application of \emph{quantum coherent control} for the optimization of the HHG process. In the present study, we establish and explore
a quantum Optimal Control Theory (OCT) scheme to enhance the emission of a specific harmonic output. The minimization of the energetic requirements and restriction of permanent ionization are also addressed.

In general, the optimization of a physical process can be achieved by two approaches:
\begin{enumerate}
	\item \emph{Rational optimization}, based on the intuitive application of physical insights;
	\item \emph{Search procedure optimization}, based on a computational optimization procedure.
\end{enumerate}

Unravelling the mechanism of the process is a step toward a rational optimization.
The three-step-mechanism  was the first successful HHG model \cite{corkum93}. 
This is a single electron
semiclassical model: At the peak of the radiation pulse, the bound electron tunnels out to free space.
In turn, the field switches sign and the electron is launched back to the nucleus where it recombines and emits 
a high energy photon \cite{lewenstein94,IvanovTutorial}.

This simple semiclassical theory has been successful in predicting control approaches able to enhance the yield.
For example, control of the absolute phase of a few cycle pulse enhances the HHG yield. 
The model has inspired the use of polarization and polarization shaping
to control the HHG process \cite{smirnova2009high,fleischer2014spin,neufeld2017high}.

An important aspect ignored by the semiclassical three-step model 
is the importance of quantum effects in HHG, such as interference of the electronic wavefunction, or interference of pathways \cite{rice1992new}.

Other models for HHG have been proposed which were applicable to earlier experiments
which used higher driving frequency \cite{mcpherson1987studies,Knight93,cerjan1993efficient,bental93}. 


Contrary to rational optimization, the search procedure optimization approach does not require previous physical insights.
Machine learning enables optimization of a process without a physical model \cite{mackay2003information}. 
HHG has also been optimized using learning algorithms based on simulations \cite{chu2001optimization}.
Experimental approaches to the optimization of HHG were based on 
open loop learning algorithms \cite{villoresi2004optimization,murnane04,HHGcol,johnson2018high}. 
This is a successful pragmatic approach, however with the disadvantage that physical insight can only be gained 
after extensive analysis.

Quantum optimal control theory (OCT) was developed as a computational scheme able to obtain a control field for a given task \cite{Rabitz,k67}. The control task is formulated as a \emph{maximization problem} of a functional object, based on variational calculus. 
The first control tasks were the control of a chemical yield \cite{k67}. Other tasks emerged \cite{glaser2015training}, 
such as cooling molecular internal degrees of freedom \cite{aroch2018optimizing} or generating quantum gates \cite{k193}.
A major effort has been devoted to create efficient algorithms to solve OCT problems \cite{somloi1993controlled,ohtsuki2004generalized,eitan2011optimal,Degani,schirmer2011efficient,reich2012monotonically}.
These methods are based on iterative algorithms. The largest computational effort in the computational optimization process is devoted to solving the time-dependent Schr\"odinger equation with a time-dependent control field.
The methods differ by their update scheme of the control field from iteration to iteration.
The search process of OCT is significantly more efficient than learning algorithm approaches, since it is based explicitly or implicitly on \emph{gradient information}, which is lacking in learning algorithms.

{(\emph{Terminological remark}: The present paper addresses the optimization of the HHG phenomenon; however, many of the discussed details apply also to the more general problem of optimization of harmonic generation (HG), including low-order harmonic generation phenomena. Other details are unique to the subproblem of optimization of HHG\@. The term HG refers to the general problem, while the term HHG refers specifically to the particular problem addressed in the paper.)}

Control of HG poses a severe challenge on OCT\@.
One difficulty lies in the fact that OCT is naturally formulated in the time domain, while the control requirements are defined in the frequency domain.
An additional difficulty originates in the fact that frequency requirements extend over a duration of time. This results in a uniqueness of the OCT formulation:
While the common OCT formulation addresses targets at a given time-point, the target of HG is extended in time. In OCT such targets require an inhomogeneous 
source term in the Schr\"odinger equation \cite{serban05,gross07,k236}. The treatment of inhomogeneity poses a numerical challenge.


The earliest study of an emission spectrum target dates back to 2008 \cite{RamanASOCT}. 
However, this study is confined to the context of control of coherent anti-Stokes Raman scattering (CARS) spectrum.

Our previous work \cite{thesis,OCTHG} first addressed the general problem of theoretical optimization of HG\@. 
It mainly focused on the optimization of low-order HG {phenomena, such as HG based on a resonance-mediated-absorption mechanism in molecular model systems}. The optimization method was demonstrated to efficiently enhance the emission in selected frequencies. In~\cite{OCTHG} the method was demonstrated also to a HHG problem. However, the chosen optimization problem was not a typical HHG process.
The selected target frequency was within the Bohr frequencies of the system, and thus the optimized physical process was based on the excitation of one of the \emph{characteristic frequencies of the system}. In contrast, in the typical HHG process the spectrum consists of multiples of the fundamental frequency of the \emph{laser source}, where the system plays the role of an up-conversion medium for the incoming frequency. Thus, the physical process is fundamentally different.

Another problem in~\cite{thesis,OCTHG} was the chosen restriction imposed on the source field spectrum. 
Only an upper bound on the control frequency was imposed. 
As a result, the control field included low frequencies which currently cannot be produced in high intensities. 
A more realistic restriction is a frequency band centered around
the carrier frequency of the laser source (see Sec.~\ref{sec:results}).

The present study addresses the particular problem of HHG control, based on the principles developed in our previous work for the general HG problem. However, the HHG problem presents several unique challenges which require a special treatment.

The accurate numerical simulation of the HHG dynamics presents a computational challenge. 
The HHG process is characterized by extreme physical conditions: 
The central atomic or molecular potential has a Coulomb character, which is steep by nature; the process involves ionization, 
where the dynamics of liberated electron in the continuum extends to large spatial distances from the parent ion; 
the dynamics of the liberated electron under the influence of the driving field is characterized by a strongly accelerated motion. 
These extremes require large computational resources for a reliable and accurate description.

Another aspect of the numerical difficulty lies in the fact that HHG is a small amplitude effect. 
Hence, even tiny numerical artefacts lead to large distortions of the high-harmonic spectrum, 
where the magnitude of the spurious effect may be comparable to the physical effects. 
This necessitates the use of highly accurate tools for the description of the HHG process. 
In particular, a highly accurate solution of the time-dependent Schr\"odinger equation is required, which necessitates an appropriate solver. 
Another major challenge lies in the realization of absorbing boundary conditions. Imperfection in their absorption capabilities 
will lead to large spurious effects \cite{krause92,EsryCAP}. 

The \emph{optimization} procedure involves additional challenges. 
The complexity of the physical situation is reflected in the optimization hypersurface, which becomes considerably more complex than that of the simpler low-order HG problems. This requires the use of more sophisticated optimization tools. 
Another problem is that the severity of the numerical artefacts can be enhanced by the optimization process, which tends to maximize them at the expense of real physical effects.


An issue which requires special treatment in the optimization of HHG is the requirement of prevention of permanent ionization. 
While ionization is an integral part of the typical HHG mechanism, the magnitude of the electronic probability 
liberated into the macroscopic medium should be minimized. This requirement has to be reflected in the OCT formulation.



Several studies dealing with theoretical optimization of HHG have been recently published \cite{RasanenHHG,DFTHHG,DombiGeneticHHG,Jin2016,SPO_HHG}.
Ref.~\cite{RasanenHHG} employs a maximization term similar to the one employed in our previous work, as well as in~\cite{RamanASOCT}. 
The study aims at the extension of the cutoff frequency. However, as noted by the authors, the available band of the source includes low frequencies (as in~\cite{OCTHG}), 
which increase the ponderomotive energy and thus extend the cutoff. The low frequencies dominate the spectra of the optimized pulses. 
Thus, the control achievement presented in this study is actually the maximization of the emission in a region which is below the cutoff frequency of the dominating frequency in the pulse. A demonstration of the extension of the cutoff without the extension of the available band to lower frequencies has not been achieved to date.

Ref.~\cite{DFTHHG} utilizes the principles presented in~\cite{thesis,OCTHG} to the development of an OCT formulation in the framework of time-dependent Density-Functional-Theory (TDDFT). This theory enables an optimization of HHG 
in multi-electron systems. Unlike in~\cite{thesis,OCTHG,RasanenHHG}, the system is controlled by the variation of the profile of a \emph{slowly varying envelope} of a fixed carrier frequency.

Refs.~\cite{DombiGeneticHHG,Jin2016,SPO_HHG} employ genetic algorithm optimization schemes, an approach which has already been employed in both theoretical and experimental works, as was mentioned above. 

The contribution of the present study lies in the thorough treatment of the numerical aspects of the problem, as well as in the theoretical treatment of the restriction of permanent ionization, a topic which has not been addressed in other studies.

The aim of this paper is to establish a comprehensive approach based on OCT to
address the target of optimizing a single emission line in HHG\@. The ultimate goal is to identify
novel mechanisms based on interference phenomena which can achieve this goal. Before this task can be achieved,
the technical issues in adopting OCT to HHG have to be addressed and solved. This is the main topic of this paper. The discussion of the mechanisms of optimized fields remains to be addressed in a future publication.


The paper is organized as follows: In Sec.~\ref{sec:theory}, we develop the theoretical OCT formulation; in Sec.~\ref{sec:num}, the numerical tools are described; Sec.~\ref{sec:results} demonstrates the method in the maximization of selected harmonics in a simple model system; the paper is concluded in Sec.~\ref{sec:conclusion}.

\section{Theory}\label{sec:theory}

Our basic OCT formulation of the general HG problem
is reviewed in Sec.~\ref{ssec:OCTHG}. A more detailed description can be found in~\cite{thesis,OCTHG}. 
This formulation establishes the foundations of the method.
In Secs.~\ref{ssec:ion}, \ref{ssec:boundary}, the basic formulation is augmented by the required elements for the HHG problem.

\subsection{The basic OCT formulation of harmonic generation}\label{ssec:OCTHG}

We consider the optimization of a laser pulse defined in the time-interval \text{$t\in[0,T]$}. We assume  a linear polarization in the $x$ direction. The temporal profile of the pulse is defined by the time-dependent electric field  $\epsilon(t)$, where the effect of the magnetic field is ignored.

The basic requirements of the HG optimization consist of two elements:
\begin{enumerate}
	\item The spectrum of the driving field has to be restricted to the spectral band available by the laser source.
	\item The emission of the system has to be maximized at the target frequency band.
\end{enumerate}
The two requirements can be treated separately.

Our control requirements are \emph{spectral}, which are naturally expressed in the \emph{frequency domain}. 
However, the \emph{quantum dynamics} is naturally expressed in the \emph{time domain}. 
These two descriptions are related by a \emph{spectral transform}. 
For this study we employed the \emph{cosine transform}. It was preferred over other spectral transforms (the Fourier transform and the sine transform) due to the properties of its discrete version, 
the \emph{discrete cosine transform} (DCT) (see Appendix~\ref{ssec:dct}),
as will be clarified in Sec.~\ref{ssec:boundary} and Appendix~\ref{app:target}. 

The operation of the cosine transform on an arbitrary function $g(t)$ will be denoted by the symbol 
$\mathcal{C}$, and the transformed function by $\bar{g}(\omega)$:
\begin{equation}\label{eq:costrans}
	\bar{g}(\omega) \equiv \mathcal{C}[g(t)] \equiv \sqrt{\frac{2}{\pi}}\int_0^\infty g(t)\cos(\omega t)\,dt
\end{equation}
The inverse cosine transform will be denoted by $\mathcal{C}^{-1}$:
\begin{equation}\label{eq:invcostrans}
	\mathcal{C}^{-1}[\bar{g}(\omega)] \equiv \sqrt{\frac{2}{\pi}}\int_0^\infty \bar{g}(\omega)\cos(\omega t)\,d\omega = g(t)
\end{equation}
The cosine-transform is its own inverse. It has the important property of being an \emph{orthogonal transformation}.

{In practice, the upper integration limit of the transform does not extend to infinity. The problem is discretized in both time and frequency; this restricts the upper limits of integration of both the direct and the inverse transforms, as follows. The upper integration limit of the inverse transform is always limited by the sampling frequency of the temporal signal, which determines the maximal available frequency, denoted here as $\Omega$. Similarly, the length of the represented time-interval is determined by the density of the discretized $\omega$ grid. The theory is defined within the time-interval \text{$t\in[0,T]$}; hence, the density of the $\omega$ grid is naturally chosen to represent a time-interval of length $T$. Thus, the direct transform is replaced by a \emph{finite-time transform}, with an upper limit $T$.} 

We begin from the treatment of the maximization of the emission in the target band, which is the second control requirement mentioned above. The emission spectrum is generated by the dipole dynamics. This, in turn, is determined by the observables related to the dipole dynamics, e.\,g., the dipole operator, the momentum operator, or the dipole acceleration operator (see Appendix~\ref{app:target}). The emission spectrum consists of the same frequencies contained in the spectra of the time-dependent expectation values of these observables. Thus, the emission spectrum can be controlled by controlling the spectrum of one of these observables. In what follows, we shall address the general problem of the control of the spectrum of an arbitrary Hermitian observable. The HG problem becomes a particular case of the general problem.

Let us denote the controlled observable by $\operator{O}$. The $\exval{\operator{O}}\!(t)$ spectrum becomes
\begin{equation}
	\Ow \equiv \mathcal{C}\left[\exval{\operator{O}}\!(t)\right] = \sqrt{\frac{2}{\pi}}\int_0^T \exval{\operator{O}}\!(t)\cos(\omega t)\,dt
\end{equation}
The problem of maximization of the magnitude of $\Ow$ in the target frequency band can be translated into the maximization of the following functional term:
\begin{align}
	& J_{max} \equiv \frac{1}{2}\int_0^\Omega \fO\overline{\exval{\operator{O}}}^2\!(\omega)\,d\omega, & \label{eq:JmaxO}\\
	& \fO \geq 0, \qquad \max\left[\fO\right] = 1 \label{eq:fO}
\end{align}
where $\fO$ is a \emph{filter function} which represents the target frequency band, \ie it has pronounced values only in the target spectral region (note that the normalization chosen here is different from \cite{thesis,OCTHG}). 
A target maximization term of this type has been employed previously in~\cite{RamanASOCT}.

We proceed to the treatment of the restriction of the driving field spectrum, the first requirement mentioned above. {The function $\epsw$ defines the spectral representation of the time-dependent driving field $\epsilon(t)$ as follows:
\begin{equation}
	\epsilon(t) = \mathcal{C}^{-1}[\epsw] = \sqrt{\frac{2}{\pi}}\int_0^\Omega \epsw\cos(\omega t)\,d\omega, \qquad\qquad t\in [0,T]
\end{equation}
}
The spectral restriction of the driving field is achieved by a frequency-dependent \emph{penalty term} 
inserted into the maximization functional:
\begin{align}
	& J_{energy} \equiv -\alpha\int_0^\Omega\frac{1}{\feps}\bar{\epsilon}^2(\omega)\,d\omega, \label{eq:Jenergy} \\
	& \feps > 0, \qquad \max\left[\feps\right] = 1 \label{eq:feps}\\
	& \alpha>0
\end{align}
where $\feps$ is another \emph{filter function}, which represents the available laser source band, \ie it has pronounced values in the frequency band available to the control field (note that the normalization chosen here is different from \cite{thesis,OCTHG}). 
$\alpha$ is an adjustable penalty factor. $J_{energy}$ penalizes the undesirable frequency regions, 
and thus restricts the band of the optimized $\epsw$.

$J_{energy}$ has also the role of imposing a restriction on the \emph{total energy} of the pulse. By \eqref{eq:feps} we have:
\begin{equation}\label{eq:nepsw_ineq}
	\int_0^\Omega\frac{1}{\feps}\bar{\epsilon}^2(\omega)\,d\omega \geq \int_0^\Omega\bar{\epsilon}^2(\omega)\,d\omega
\end{equation}
{The RHS represents the \emph{norm} of $\epsw$. When $\epsw$ is transformed to the time-domain, its norm is preserved by the orthogonality property of the inverse cosine-transform.} Thus, the magnitude of the RHS becomes equivalent to the \emph{fluence} of the driving field,
\begin{equation}\label{eq:fluence}
	\Phi\left[\epsilon(t)\right] \equiv \int_0^T\epsilon^2(t)\,dt
\end{equation}
Consequently, the fluence becomes the lower limit of the magnitude of the LHS of \eqref{eq:nepsw_ineq}. This means that the magnitude of the fluence is also penalized by $J_{energy}$. Since the fluence is proportional to the total energy of the pulse, 
$J_{energy}$ also restricts the total energy.

The energetic restriction is particularly important in the context of HHG, which is typically a highly inefficient process energetically. 
The simultaneous maximization of the emission amplitude by $J_{max}$ and minimization of the driving field amplitude by 
$J_{energy}$ enhances the efficiency of the process.

It is convenient to define (Cf.~Eq.~\eqref{eq:Jenergy}):
\begin{equation}\label{eq:tfeps}
	\tfeps \equiv \frac{\feps}{\alpha}
\end{equation}

Penalization of the undesirable frequency components of the field has been suggested previously in~\cite{Degani}. 
However, Ref.~\cite{Degani} employs a time-domain formulation, while the present formulation is in the frequency domain, which has considerable advantages (see~\cite[Chapter 3]{thesis}, where the relation to other methods of restricting the driving field spectrum \cite{gross07,Skinner,Degani} was discussed).

The \emph{dynamical requirements} are imposed by a \emph{constraint} to the optimization problem. 
The constraint is introduced into the optimization formulation by the \emph{Lagrange-multiplier method}.

The dynamics is governed by the time-dependent Schr\"odinger equation, subject to a given initial condition:
\begin{equation}\label{eq:Schr}
	\deriv{\ket{\psi(t)}}{t} = -i\operator{H}(t)\ket{\psi(t)}, \qquad \ket{\psi(0)} = \ket{\psi_0}
\end{equation}
(Atomic units are used throughout, thus we set: $\hbar=1$.) The time-dependent Hamiltonian is given by
\begin{equation}
	\operator{H}(t) = \operator{H}_0 - \operator{\mu}_x\epsilon(t)
\end{equation}
where $\operator{H}_0$ is the unperturbed Hamiltonian, and $-\operator{\mu}_x\epsilon(t)$ represents the interaction of the $x$ component of the dipole with the $x$ polarized driving field. The dipole approximation is employed. The Schr\"odinger equation becomes a time-dependent constraint to the optimization problem. 
The maximization functional is modified by the introduction of a corresponding Lagrange-multiplier term (see Appendix~\ref{app:der} for more details):
\begin{equation}\label{eq:JSchr}
	J_{Schr} = -2\Real{\int_0^T\bracketsO{\chi(t)}{\deriv{}{t}+i\operator H(t)}{\psi(t)}\,dt}
\end{equation}

The full maximization functional of the basic HG problem becomes
\begin{equation}
	J[\epsw] = J_{max} + J_{energy} + J_{Schr}
\end{equation}

After imposing the extremum conditions, we obtain the following set of Euler-Lagrange equations (see Appendix~\ref{app:der} for the full derivation):
\begin{align}
	&\deriv{\ket{\psi(t)}}{t} = -i\operator{H}(t)\ket{\psi(t)}, & 	\ket{\psi(0)} = \ket{\psi_0} \label{eq:ELpsi} \\ 
	&\deriv{\ket{\chi(t)}}{t} = -i\operator{H}(t)\ket{\chi(t)} - \mathcal{C}^{-1}\left[\fO\Ow\right]\operator{O}\ket{\psi(t)}, & \ket{\chi(T)} = 0 \label{eq:ELchi} \\
	& \epsw = \feps\mathcal{C}[\eta(t)], \qquad\qquad \eta(t) \equiv -\frac{\Imag{\bracketsO{\chi(t)}{\operator{\mu}_x}{\psi(t)}}}{\alpha}\label{eq:ELepsw} \\
	& \epsilon(t) = \mathcal{C}^{-1}[\epsw] = \mathcal{C}^{-1}\left\lbrace \feps\mathcal{C}[\eta(t)]\right\rbrace \label{eq:ELepst}
\end{align}

The  expression for $\eta(t)$ is identical to that obtained for the driving field $\epsilon(t)$ without the spectral restriction (see~\cite{gross07,thesis,OCTHG}). Thus, Eq.~\eqref{eq:ELepst} can be interpreted as the unrestricted field, filtered by the filter function $\feps$. A complete filtration of undesirable spectral regions is achieved in the limit \text{$\feps\longrightarrow 0$}. 
Although $J_{energy}$ is  rigorously undefined for \text{$\feps=0$}, in practice $\feps$ can be set to $0$ achieving  
a complete filtration of undesirable regions.

The advantage of the current formulation for the spectral restriction of the driving field lies in the flexibility of Eq.~\eqref{eq:ELepsw}, where the freedom in the choice of $\feps$ can have a prominent effect on the profile of the optimized pulse. 
Smooth filtration can be achieved by choosing a smooth filter function. In addition, $\feps$ can be chosen so as to provide an envelope shape to the spectral profile of the field (limitations of this practice are discussed in Sec.~\ref{sec:results}). Finally, in certain experiments the system is irradiated by several sources, which differ both in the spectral band and the available intensity; an appropriate choice of $\feps$ can address these scenarios.

General comments on the presented OCT formulation are in order.

The form of $J_{max}$ enables a \emph{selective enhancement} of spectral emission regions. 
However, it should the noted that there is an ambiguity in the term ``selectivity'', which can be interpreted in two different ways:
\begin{enumerate}
	\item Maximization of the yield of the selected target, regardless of the appearance of ``by-products'' (in our case, the emission in other spectral regions);
	\item Simultaneous maximization of the yield of the selected target, and minimization of the yield of any ``by-product'' 
	(in our case, suppression of other harmonics).
\end{enumerate}
The formulation presented here is aimed at selectivity of the first type. 
However, several other studies target the selectivity of the second type, and the control requirements include suppression of undesirable harmonics (see, for example, \cite{HHGcol}). 
In principle, the suppression of harmonics can be achieved by modifying the definition of $\fO$ (Eq.~\eqref{eq:fO}) to allow negative values. Then the $J_{max}$ term can be used to penalize undesirable spectral emission regions. 
Note that this changes the interpretation of $\fO$ as a filter function. However, this option has not been investigated yet.

Selectivity was the aim of the earliest quantum coherent control studies, which addressed the selective enhancement of the yield in chemical reactions. However, it should be noted that there is a fundamental difference between the optimization of chemical reactions and optimization of HHG\@.
In chemical reactions, the sum of the yields of all products is always $100\%$. 
Consequently, the by products are always produced at the expense of the yield of the desired chemical product. 
Thus, there is no ambiguity in the term of selectivity in the chemical context. In contrary, in the present context, 
the high-harmonic products never sum to $100\%$ of the energetic yield. 
The efficiency of the HHG process is very low---typically, the total energy of the emission is orders of magnitude lower than the invested energy of the incoming pulse.
Therefore, the appearance of high-harmonic ``by products'' (typically, neighbouring harmonics; see Sec.~\ref{sec:results}) is unnecessarily at the expense of the desired harmonics.
On the contrary, the suppression of other harmonics inserts an additional requirement into the control problem, which may be at the expense of the maximization target.

Both $J_{energy}$ and $J_{Schr}$ set constraints on the optimization problem. However, the two terms are fundamentally different in nature. 
The Schr\"odinger equation constraint is a ``hard'' constraint, which is well defined, and should strictly not be violated. 
In contrast, $J_{energy}$ represents softer requirements of desirable trends---the energy should be kept as small as possible, and the spectrum of the pulse should gradually decay in undesirable regions.  The requirements represented by a penalty term are sometimes referred to as ``soft constraints''. 
In what follows, the additional requirements of HHG will be realized by both hard and soft constraint terms. 


\subsection{The target operator}\label{ssec:target}

The target operator $\operator{O}$ in the present study is chosen as the \emph{stationary acceleration operator} of the dipole, which will be defined in what follows. The \emph{dipole acceleration operator} is defined as
\begin{equation}\label{eq:op_acc}
	\operator{\ddot\mu}_x = \operator{\ddX} = -\deriv{V\!\left(\operator{X}\right)}{\operator X}
\end{equation}
where $V(x)$ is the potential. Note that atomic units are used, and the electron has a unit mass and charge. The expression in the RHS of Eq.~\eqref{eq:op_acc} represents the operator expression of the classical force. The potential can be divided into a stationary part and a time-dependent part:
\begin{equation}
	V\left(\operator{X}\right) = V_0\left(\operator{X}\right) - \operator{X}\epsilon(t)
\end{equation}
Accordingly, the acceleration operator can also be divided into a stationary part and a time-dependent part:
\begin{equation}
	\operator{\ddX} = -\deriv{V_0\left(\operator{X}\right)}{\operator{X}} + \epsilon(t)
\end{equation}
The \emph{stationary acceleration operator} is defined by the stationary part of the acceleration operator:
\begin{equation}
	\opC \equiv -\deriv{V_0\left(\operator{X}\right)}{\operator X}
\end{equation}
The time-dependent part, $\epsilon(t)$, contributes only large linear response components to the emission spectrum (as will be explained in Appendix~\ref{app:target}), and thus is not of interest in the present context \cite{DFTHHG}. Moreover, it is advantageous to eliminate these components from the spectrum due to numerical considerations (see Appendix~\ref{app:target}). Accordingly, we set:
\begin{equation}\label{eq:OC}
	\operator{O} \equiv \opC
\end{equation}

A thorough discussion of the considerations in the choice of the target operator is given in Appendix~\ref{app:target}.

With the choice \eqref{eq:OC} of the target operator, the maximization term $J_{max}$ becomes (Cf.\ Eq.~\eqref{eq:JmaxO}):
\begin{equation}\label{eq:JmaxC}
	J_{max} \equiv \frac{1}{2}\int_0^\Omega \fC\overline{\exval{\operator{C}}}^2\!(\omega)\,d\omega, \qquad \fC \geq 0
\end{equation}
where $\fC$ is the target filter function for the $\evC$ spectrum.

\subsection{Restriction of permanent ionization}\label{ssec:ion}

The HHG process involves partial ionization of the system. Part of the amplitude of the liberated electron reverts to the parent ion, and participates in the generation of high harmonics via the overlap with the ionic core. However, part of the electronic amplitude does not return, and the system becomes \emph{permanently ionized}. 
This results in the production of \emph{plasma} in the medium, which is a source of experimental problems (see, for example, \cite{MacroscopicHHG}). 
Therefore, it is highly desirable to reduce the permanent ionization in the process. This task can be achieved by incorporating  an additional control requirement in the OCT formalism.

In our previous work \cite{OCTHG} the restriction of permanent ionization was realized by imposing a penalty on spatial regions which are beyond a chosen spatial threshold. The problem with this formulation is that it is suitable for a complete elimination of permanent ionization. However, we found that this requirement is incompatible for typical HHG problems; some permanent ionization has to be allowed to enable the appearance of significant HHG effects. In principle, the magnitude of the penalty can be reduced to allow some permanent ionization. The difficulty is that it becomes very intricate to quantify the allowed permanent ionization probability in this method.

In the present work we used a different formulation. The penalty is imposed on the permanent ionization itself. We assume that \emph{absorbing boundaries} are employed to eliminate the outgoing amplitude at the edges of the grid.
The penalty on permanent ionization can be formulated by the following penalty term:
\begin{align}
	& J_{ion} \equiv \sigma\left(\brackets{\psi(T)}{\psi(T)}\right), \label{eq:Jion}\\
	&\sigma(y)\leq 0, \label{eq:sigma_leq0}\\
	&\sigma(1) = 0 \label{eq:sigma1}
\end{align}
where $\sigma(y)$ is a \emph{monotonically increasing function}, which has the role of a penalty function. 
When absorbing boundaries are employed, the effective Hamiltonian becomes non-Hermitian.
As a result, the norm of $\ket{\psi(t)}$ is not conserved. The magnitude of $\brackets{\psi(t)}{\psi(t)}$ is gradually decreasing during the process, as the electron's probability is being absorbed by the boundaries. $\brackets{\psi(T)}{\psi(T)}$ represents the survival probability in the process, which can be identified as the probability which is not permanently ionized (see Appendix~\ref{app:plasma}).
Condition~\eqref{eq:sigma_leq0} implies that the lost electronic density is penalized. The requirement that $\sigma(y)$ is a monotonically increasing function implies that the magnitude of the penalty increases with the permanent ionization probability. Condition~\eqref{eq:sigma1} ensures that $J$ remains unaltered in the limit of zero permanent ionization probability.

The main advantage of this formulation lies in the flexibility of the form of $\sigma(y)$, which can be adjusted to represent the complexity of the control requirement. 
As has already been mentioned, some permanent ionization should be allowed; this can be reflected by the definition of the maximal allowed permanent ionization probability, or, equivalently, the \emph{minimal allowed survival probability}. The magnitude of the penalty in the ``allowed'' $\brackets{\psi(T)}{\psi(T)}$ region should be close to 0. As $\brackets{\psi(T)}{\psi(T)}$ approaches the minimal allowed survival probability from the upper limit, the magnitude of the penalty should increase rapidly. 
This can be realized by a sigmoid profile of $\sigma(y)$ (for an example see Fig.~\ref{fig:sigma} in Sec.~\ref{sec:results}).

A further discussion on the interpretation of $J_{ion}$ is given in Appendix~\ref{app:plasma}.
 
\subsection{Imposing boundary conditions on the driving field}\label{ssec:boundary}

A realistic pulse shape must be a continuous function. Rapid jumps in $\epsilon(t)$ cannot be produced in laboratory conditions. 
Very rapid variations in $\epsilon(t)$  will be referred to as ``discontinuities''.

In the basic formulation of Sec.~\ref{ssec:OCTHG}, the continuity of $\epsilon(t)$ \emph{within} the time-interval of the problem, \text{$t\in(0,T)$}, is ensured by the limitation of the pulse to a low frequency regime. However, at the boundaries of the time-domain, \text{$t=0$}, \text{$t=T$}, we encounter a problem: The physical value of $\epsilon(t)$ outside the time-domain \text{$t\in[0,T]$} is zero by definition. However, in practice, $\epsilon(t)$ is constructed by a discrete spectral transform (see Appendix~\ref{ssec:dct}), in which the $\omega$ sampling is discretized; discrete spectral transforms represent \emph{infinite periodic patterns}, which are non-zero outside \text{$t\in[0,T]$}. Thus, the \emph{physical} $\epsilon(t)$ is defined by
\begin{equation}\label{eq:epst_phys}
	\epsilon(t) \equiv
	\begin{cases}
		\mathcal{C}^{-1}[\epsw] & 0\leq t\leq T \\
		0 & t<0,\ t>T
	\end{cases}
\end{equation}
The restriction of the frequency band by $J_{energy}$ prevents the appearance of discontinuities in the infinite periodic function; it does not prevent the possibility of discontinuities in Eq.~\eqref{eq:epst_phys} at the boundaries of the time-interval of the problem, \text{$t=0$}, \text{$t=T$}.

The problem can be solved by formulation of additional control requirements; the driving pulse has to satisfy the following \emph{boundary conditions}:
\begin{align}
	& \epsilon(0) = 0 \label{eq:eps0}\\
	& \epsilon(T) = 0 \label{eq:epsT}
\end{align}
The boundary conditions can be treated as additional \emph{constraints} to the optimization problem. These can be realized by the Lagrange-multiplier method. Eqs.~\eqref{eq:eps0}, \eqref{eq:epsT}, are treated as \emph{constraint equations}. Accordingly, the following Lagrange-multiplier terms are added to $J$:
\begin{align}
	& \Jepsz \equiv -\sqrt{2\pi}\lambda_0\epsilon(0) = -2\lambda_0\int_0^\Omega\epsw \cos(0)\,d\omega = -2\lambda_0\int_0^\Omega\epsw\,d\omega \label{eq:Jeps0}\\
	& \JepsT \equiv -\sqrt{2\pi}\lambda_T\epsilon(T) = -2\lambda_T\int_0^\Omega\epsw \cos(\omega T)\,d\omega \label{eq:JepsT}
\end{align}
where $\sqrt{2\pi}\lambda_0$, $\sqrt{2\pi}\lambda_T$, are the Lagrange-multipliers of the constraint equations \eqref{eq:eps0}, \eqref{eq:epsT}, respectively.

Prevention of discontinuities in $\epsilon(t)$ has a fundamental theoretical importance. 
If discontinuities are present at the boundaries, the spectrum of the physical $\epsilon(t)$ contains very high frequencies. This has considerable physical significance in the context of HHG, since very energetic photons are contained in the driving pulse. The HHG process is an up-conversion process, in which low energy photons in the driving pulse are converted by the system into high energy photons. The presence of highly energetic photons in the driving pulse completely changes the physical picture, and leads to spurious HHG effects.

A dynamical analysis of these spurious effects enables a more thorough understanding of their physical origin. A sudden change in the Hamiltonian leads to the violation of the \emph{adiabatic approximation}. We found that non-adiabatic processes are a prerequisite  
to the generation of high harmonics.  Typically, this is achieved  by extremely high intensities. A discontinuity in the field generates 
spurious non-adiabatic transitions in much lower intensities. This topic requires a dedicated paper.

While this effect is typically small in magnitude, it is of considerable importance in the context of HHG, which is governed by small amplitude effects. In our previous work \cite{OCTHG} the topic of the boundary conditions of the field was ignored (as in many OCT works). Thus, the physical significance of the results presented in the HHG problem of Ref.~\cite{OCTHG} is quite limited.

Actually, conditions~\eqref{eq:eps0}, \eqref{eq:epsT}, are insufficient to prevent the appearance of highly energetic components in the physical $\epsilon(t)$ spectrum; discontinuities in the time derivatives,
\begin{equation*}
	\nderiv{\epsilon(t)}{t}{n}, \qquad n\geq 1
\end{equation*}
are also a source of high frequency components in the spectrum, which can be questionable both experimentally and theoretically (this was ignored, e.\,g., in~\cite{DFTHHG}). We found that a discontinuity in the first derivative is responsible for significant spurious HHG effects. However, the effect of a discontinuity in the second derivative on the high-harmonic spectrum was found to be negligible. 
Thus, the following boundary conditions to $\epsilon(t)$ are added:
\begin{align}
	& \derat{\epsilon(t)}{t}{t=0} = 0 \label{eq:Deps0}\\
	& \derat{\epsilon(t)}{t}{t=T} = 0 \label{eq:DepsT}
\end{align}
These boundary conditions are satisfied automatically by the DCT, in which $\epsilon(t)$ is spanned by a cosine series. This is an important advantage of the DCT over the discrete Fourier transform, in which the derivative boundary conditions define additional constraints in the optimization problem.

\subsection{The full OCT formulation of HHG control}

The total maximization functional is the sum of all the individual terms:
\begin{equation}
	J \equiv J_{max} + J_{ion} + J_{energy} + J_{\epsilon(0)} + J_{\epsilon(T)} + J_{Schr}
\end{equation}
where the different components are given by Eqs.~\eqref{eq:JmaxC}, \eqref{eq:Jion}, \eqref{eq:Jenergy}, \eqref{eq:Jeps0}, \eqref{eq:JepsT}, \eqref{eq:JSchr}.

The extremum conditions yield the following set of Euler-Lagrange equations:
\begin{align}
	\deriv{\ket{\psi(t)}}{t} &= -i\operator{H}(t)\ket{\psi(t)}, \nonumber \\
	& \ket{\psi(0)} = \ket{\psi_0} \label{eq:Schr_full} \\
	\deriv{\ket{\chi(t)}}{t} &= -i\operator{H}^\dagger(t)\ket{\chi(t)} - \mathcal{C}^{-1}\left[\fC\Cw\right]\opC\ket{\psi(t)}, \nonumber \\
	&\ket{\chi(T)} = \sigma'\left(\brackets{\psi(T)}{\psi(T)}\right)\ket{\psi(T)}\label{eq:ihSchr} \\
	\operator{H}(t) &= \operator{H}_0 - \operator{X}\epsilon(t) \label{eq:H}\\
	\epsilon(t) &= \mathcal{C}^{-1}[\epsw] \\
	\epsw &= \epswu - \tfeps[\lambda_0 + \lambda_T\cos(\omega T)] \label{eq:epsw_full}\\
	&\epswu \equiv \feps\mathcal{C}[\eta(t)], \qquad\qquad \eta(t) \equiv -\frac{\Imag{\bracketsO{\chi(t)}{\operator{X}}{\psi(t)}}}{\alpha} \label{eq:epswunc_full}\\
	&\epsunc(t) = \mathcal{C}^{-1}[\epswu] \label{eq:epsunc}\\
	&\lambda_0 = \frac{\epsunc(0)d - \epsunc(T)b}{ad - b^2} \label{eq:lambda0_full} \\
	&\lambda_T = \frac{\epsunc(T)a - \epsunc(0)b}{ad - b^2} \label{eq:lambdaT_full}\\
	& a \equiv \mathcal{C}^{-1}\left[\tfeps\right]\!\biggm|_{t=0} \\
	& b \equiv \mathcal{C}^{-1}\left[\tfeps\right]\!\biggm|_{t=T} \\
	& d \equiv \mathcal{C}^{-1}\left[\tfeps\cos(\omega T)\right]\!\biggm|_{t=T}\label{eq:d}
\end{align}
The derivation of the equations is given in Appendix~\ref{app:der}. These equations form the base for the optimization
procedure.

\section{Numerical methods}\label{sec:num}

Optimal control equations are solved by iterative forward backward propagation of the Schr\"odinger equation with an
update scheme to change the control field from iteration to iteration. The control of HG is a particularly difficult problem
and therefore the standard procedures that have been developed for OCT do not apply. The solution of the
Schr\"odiner equation is complicated due to explicit time dependence of the control Hamiltonian. An additional difficulty is that 
due to absorbing boundaries the Hamiltonian is non-Hermitian. High accuracy is required since the HHG phenomenon is generated
from a minor fraction of the wavefunction.

\subsection{Optimization procedure}\label{ssec:optim}

The optimization procedure is based on a second-order gradient method (quasi-Newton)---the BFGS method (see~\cite[Chapter~3]{Fletcher} and references therein).

The BFGS method replaces the relaxation process employed in our previous work (see in detail \cite[Chapter~3.2.3]{thesis}). The relaxation process was found to be successful in the optimization of low-order HG processes. However, we found it to be rather slow in the optimization of the HHG process. This can be attributed to the complexity of the physical situation, 
which results in an optimization hypersurface which is more complex than that of the simpler HG problems. 
This necessitates the use of second-order information in the optimization (the Hessian), which can be approximated by a quasi-Newton method.

We found that the BFGS method yields a drastic improvement compared to the relaxation process. 
However, this required adjustments of several details of the implementation to the present problem. 
The implementation details are described in Appendix~\ref{app:BFGS}.

\subsection{Dynamics}\label{ssec:dynamics}

The propagation with an explicit time dependent Hamiltonian is solved by 
a new highly accurate and efficient algorithm \cite{k273,SemiGlobal}. 
The algorithm is based on a \emph{semi-global propagation approach}, which is governed by multiple considerations which are both local in time, and global in time. The propagation is performed in relatively large time-steps, where each time-interval is approximated globally as a unified unit by an interpolation based approach. 
The application of the algorithm to the physical situation of HHG has already been described in detail in \cite[Sec.~4]{SemiGlobal}.

The Fourier grid method \cite{FourierGrid,k56} is employed for the Hamiltonian operation.

Absorbing boundary conditions are employed in order to prevent a wraparound of the wave-function at the boundaries of the grid. The absorbing boundaries are implemented by a complex absorbing potential (see \cite{CAP} for a comprehensive review). This implies that the Hamiltonian becomes non-Hermitian. The propagation method was extended to the treatment of non-Hermitian dynamics in Ref.~\cite{SemiGlobal}.

The choice of the complex absorbing potential is of utmost importance in the present setting. 
It has already been recognized in the early HHG simulations (see~\cite{krause92}) that reflection of the wave-function from the absorbing boundaries leads to large distortions of the high-harmonic spectrum. Different absorbing potentials vary in their absorption capabilities. 
Reliable HHG simulations require absorbing potentials which reduce the reflection and transmission of the wave-function to extremely low rates. Attempts to locate absorbing potentials with this property by inspection have failed \cite{krause92}.

This issue is crucial in the context of optimization. The optimization process cannot distinguish between a physical effect and a numerical artefact. As a result, it might tend to amplify the magnitude of a numerical artefact instead of maximizing a real physical effect.

In the present work, we employed an \emph{optimization procedure} for the construction of the absorbing potential. 
The optimization is performed  to locate an absorbing potential with minimal reflection and transmission. 
The method relies on the principles presented in~\cite{squareBarriers}, but with several necessary modifications. 
The real part of the absorbing potential is constructed from a finite cosine series. The imaginary part is constructed from another finite cosine series, where the imaginary potential is given by squaring the cosine series and adding a minus sign. This prevents the presence of positive imaginary components in the potential, which is a source of numerical instability in the propagation process.
The optimization parameters are the cosine coefficients. The reflection and transmission amplitudes are obtained 
by static scattering calculations (see \cite{staticScattering}). Further details are available in~\cite[Sec.~4.2]{SemiGlobal}. 
However, a full description of the method has not been published to date.

The absorbing potential is available in the Supplemental Material of the paper as a \texttt{MATLAB} data file.

\section{Results}\label{sec:results}

The optimization of selected harmonics in the range of the thirteenth to the seventeenth harmonic of the fundamental of a Ti-sapphire laser is demonstrated in a simple single electron model system. The selected target frequencies are above the ionization threshold.
(\emph{Remark}: Atomic units are used throughout.)

\subsection{The system}

\begin{figure}
	\centering \includegraphics[width=3in]{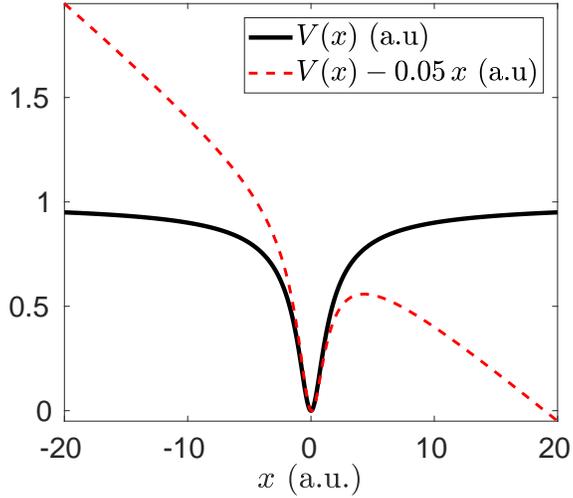}
	\caption{(Color online) The truncated Coulomb potential (Eq.~\eqref{eq:Vatom}, black), and the distorted potential under the influence of a strong field (dashed red, gray).}\label{fig:Vatom}	
\end{figure}

Our system is a simplified one-dimensional model of an atomic system. A ``one-dimensional electron'' is placed in a central potential of a Coulombic nature, which represents the parent atom potential. The central potential has the form of a \emph{truncated Coulomb potential} (see Fig.~\ref{fig:Vatom}):
\begin{equation}\label{eq:Vatom}
	V(x) = 1 - \frac{1}{\sqrt{x^2 + 1}}
\end{equation}
This form eliminates the singularity of the Coulomb potential at \text{$x=0$}.

The one-dimensional model has been extensively studied in the context of intense laser atomic physics in general, and HHG in particular. It is by no means a realistic atomic model; nevertheless, it preserves the fundamental features of the high-harmonic spectrum \cite{krause92}. The model constitutes the simplest system for which the HHG phenomenon can be observed. The simplicity of the model is a considerable advantage for the study of HHG\@. The realistic HHG process is extremely complicated and rich with physical effects of secondary importance, such as multielectron effects \cite{IvanovTutorial,DFTHHG}, {effects of higher dimensionality (\eg} spreading of the electron in transversal directions to the polarization of the driving field; {nonlinear interaction between spatial degrees of freedom)}, and macroscopic propagation effects \cite{MacroscopicHHG}. The current simplified model refines the most fundamental elements of the HHG process, which can contribute to their study and understanding.

The time-dependent \emph{physical} Hamiltonian becomes
\begin{equation}
	\operator{H}(t) = \frac{\operator{P}^2}{2} + 1 - \frac{1}{\sqrt{\operator{X}^2 + 1}} - \operator{X}\epsilon(t)
	\label{eq:ham}
\end{equation}
This Hamiltonian is supplemented by a complex potential to account for the absorbing boundaries. In addition, the potential is modified such that the classical force induced by the physical potential is smoothly ``turned off'' near the absorbing boundaries, as described in detail in \cite[Sec.~4]{SemiGlobal}.

The fundamental Bohr frequency of the model system, \text{$\omega_{1,0} = 0.395\au$}, is similar to that of the hydrogen atom ($0.375\au$) or the argon atom ($0.424\au$).

The spatial domain is \text{$x\in[-240,240)$}. We use an equidistant grid, with 768 points. The distance between adjacent grid points becomes $0.625\au$. The absorbing potential extends over $40\au$ at each boundary of the $x$ grid (see~\cite[Sec.~4.2]{SemiGlobal}). Thus, the actual {\em physical domain} is \text{$x\in[-200, 200]$}.

\subsection{General details of the control problems}\label{ssec:details}

\begin{figure}
	\centering \includegraphics[width=3in]{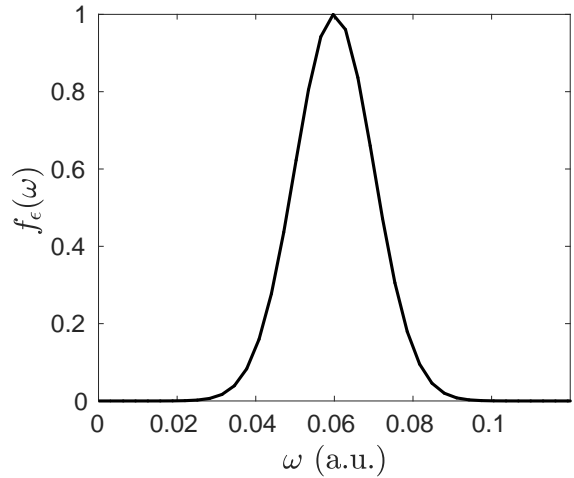}
	\caption{The driving field filter function}\label{fig:feps}
\end{figure}

The initial state $\ket{\psi_0}$ in all problems is the ground state of the stationary Hamiltonian.
The time interval allocated to the process is \text{$t\in[0,1000\au]$}, which corresponds to a pulse duration of $24.2_{fs}$.
The driving field filter function has a Gaussian profile (see Fig.~\ref{fig:feps}): 
\begin{equation}
	\feps = \exp\left[-\frac{(\omega - 0.06)^2}{2\cdot 0.01^2}\right]
\end{equation}
The Gaussian is centred at \text{$\omega=0.06\au$}, which corresponds to a wavelength \text{$\lambda = 760_{nm}$}---similar to the central wavelength of the Titanium-Sapphire laser. Let us denote:
\begin{equation}
	\omega_0 = 0.06\au
\end{equation}
$\omega_0$ will be referred to as the \emph{fundamental frequency} of the laser source.

The filter function of the target frequency has also a Gaussian profile:
\begin{equation}\label{eq:fmun}
	\fC = \exp\left[-\frac{(\omega - n\omega_0)^2}{2\cdot 0.01^2}\right]
\end{equation}
where $n$ denotes the harmonic order of the target frequency. $n$ varies with the specific optimization problem.

The total energy penalty factor is \text{$\alpha=2\times 10^{-6}$}.

\begin{figure}
	\centering \includegraphics[width=3in]{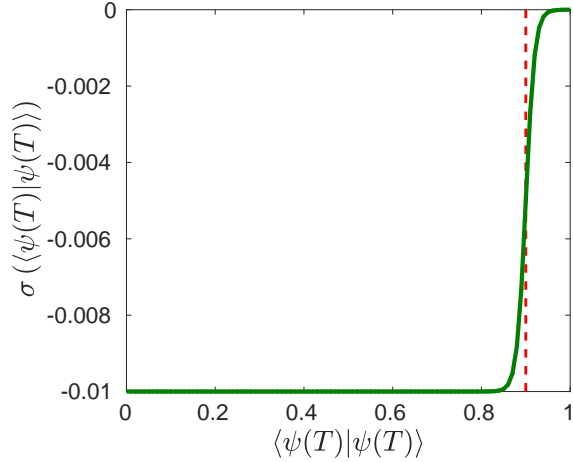}
	\caption{(Color online) The ionization penalty function; the minimal allowed survival probability (0.9) is marked by a vertical dashed red line.}\label{fig:sigma}
\end{figure}

The ionization penalty function is (see Fig.~\ref{fig:sigma})
\begin{equation}
	\sigma(y) = 5\times 10^{-3}\left\lbrace\tanh\left[50\left(y - 0.9\right)\right] -\tanh(5)\right\rbrace
\end{equation}
$\sigma(y)$ is chosen so as to restrict the maximal allowed permanent ionization to $10\%$ of the probability.

The initial guess for the driving field is constructed from an  unconstrained function $\bar{\epsilon}_{unc}^{(0)}(\omega)$, substituted into Eqs.~\eqref{eq:epsw_full}, \eqref{eq:epsunc}-\eqref{eq:lambdaT_full} (as explained in Appendix~\ref{sssec:iguess}). The following unconstrained field has been used for all problems:
\begin{equation}
	\bar{\epsilon}_{unc}^{(0)}(\omega) = 5\,\exp\left[-\frac{(\omega - 0.06)^2}{2\cdot 0.01^2}\right]\,\sin\left[\frac{(\omega - 0.06)\pi}{0.015}\right]
\end{equation}

The termination condition is given by Eq.~\eqref{eq:termination}.

\subsection{Maximization of selected harmonics}

The maximization of the 13'th harmonic is the first target to be studied. 
The target filter function is given by the substitution of \text{$n=13$} into Eq.~\eqref{eq:fmun}. The target frequency, \text{$\omega=0.78\au$}, is the first odd-harmonic frequency which is above the ionization threshold frequency, $0.67\au$.

\begin{figure}
	\centering \includegraphics[width=3in]{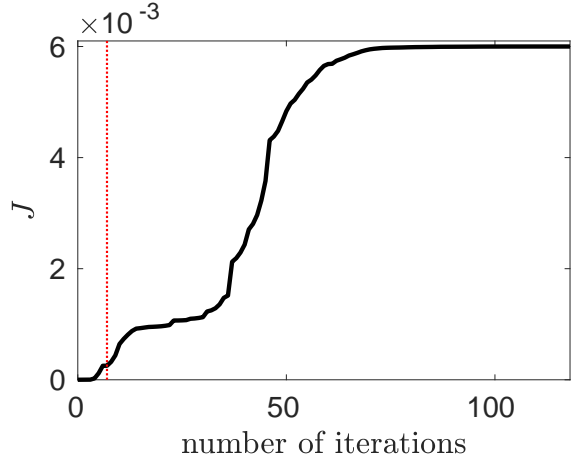}
	\caption{(Color online) The convergence curve for the optimization of the 13'th harmonic; the process is divided into two stages by a doted vertical red (gray) line---before and after the reset of the Hessian {(see Appendix~\ref{sssec:practical})}.}\label{fig:conv}
\end{figure}

The inverse-Hessian approximation was reset after 7 iterations (see Appendix~\ref{sssec:practical}), when the termination condition~\eqref{eq:termination} was first matched. The convergence curve is shown in Fig.~\ref{fig:conv}, where the optimization process is divided in two stages---before and after the reset of the Hessian.

\begin{figure}
	\centering \includegraphics[width=3in]{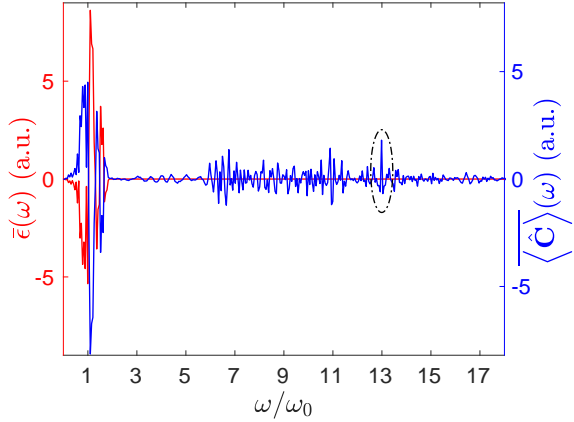}
	\caption{(Color online) The spectra of the driving field (red, gray) and the stationary acceleration expectation (blue, dark gray) Vs.\ the harmonic order, for the 13'th harmonic problem; the response at the target frequency is marked.}\label{fig:epsC13}
\end{figure}

The spectra of the optimized driving field and the stationary acceleration expectation are plotted in Fig.~\ref{fig:epsC13} Vs.\ the harmonic order. The driving field is successfully restricted to the region of the fundamental frequency of the source. The response at the 13'th harmonic is marked. The enhanced emission at the target frequency is apparent.

Several other important peaks are present in the emission spectrum: There is a large linear response around $\omega_0$; a large peak is present at the fundamental Bohr frequency of the system, which equals $6.6\omega_0$; a significant response is observed at the 11'th harmonic, which is the neighbouring odd-harmonic of the target harmonic. The significant response at neighbouring harmonics is typical
since no suppression of other harmonics was included in the control requirements (see Sec.~\ref{ssec:OCTHG}).

\begin{figure}
	\centering \includegraphics[width=3in]{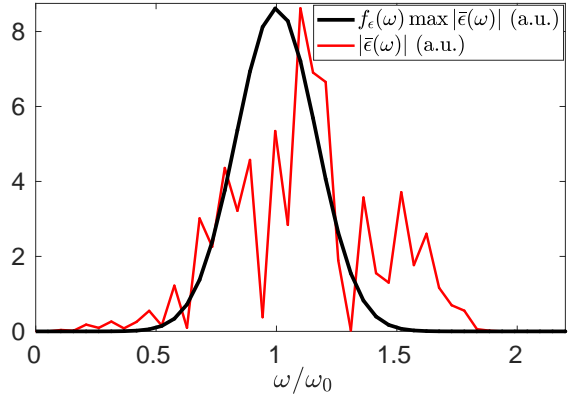}
	\caption{(Color online) A comparison of the forms of the filter function and the driving field spectrum of the 13'th harmonic problem; the absolute value of the driving field spectrum, $\abs{\epsw}$ (red, gray), and the filter function $\feps$ normalized to the peak of the driving field spectrum (black), are plotted Vs.\ the harmonic order. It is shown that there is a significant deviation from the filter function envelope, in particular at the blue side of the spectrum. An additional general blue shift can be also observed.}\label{fig:epsw13feps}
\end{figure}

Let us examine more carefully the spectral restriction of the driving field frequency by the filter function $\feps$. In Fig.~\ref{fig:epsw13feps}, we compare the magnitude of $\epsw$ with the form of $\feps$. It can be observed that the general form of $\epsw$ is influenced by the Gaussian envelope shape imposed by $\feps$. However, it can be also observed that there is a significant deviation from the filter function envelope. The decay rate of $\epsw$ is slower than that of $\feps$, in particular at the blue side of the spectrum. A blue shift can be observed also at the central region of the emission spectrum.
It can be found that the vast majority of the contribution to the magnitude of $J_{energy}$ originates from the wings of the driving field spectral profile, in particular at the blue side. 
This implies that the extension of the available spectrum from the laser source has a fundamental role in the enhancement of the HHG process, in particular the blue extension.
This statement can be verified by the variation of the standard-deviation of $\feps$. A more thorough physical analysis is left for a future publication.

Let us compare the harmonic yield produced by the optimized pulse at the target frequency with that of a reference pulse. Our reference pulse is constructed from a periodic wave with frequency $\omega_0$, which is constrained to the required boundary conditions of the present problem. We start from a periodic waveform, confined to a finite time-duration:
\begin{equation}
	\epsilon_{harmonic}(t;\beta) \equiv \beta\sin[\omega_0(t-500)], \qquad t\in[0, 1000] 
\end{equation} 
This form is used to construct a field which satisfies the cosine series boundary conditions, \eqref{eq:Deps0}, \eqref{eq:DepsT}:
\begin{equation}
	\epsilon_{ref,unc}(t;\beta) = \mathcal{C}^{-1}\left\lbrace\feps\mathcal{C}\left[\epsilon_{harmonic}(t;\beta)\right]\right\rbrace
\end{equation}
Finally, $\epsilon_{ref,unc}(t;\beta)$ is substituted in Eqs.~\eqref{eq:epsw_full}, \eqref{eq:epsunc}-\eqref{eq:lambdaT_full}, in order to obtain a new pulse $\epsilon_{ref}(t;\beta)$ which satisfies also the zero boundary conditions of $\epsilon(t)$, \eqref{eq:eps0}, \eqref{eq:epsT} (as explained in Appendix~\ref{app:der}).

The optimized pulse is compared to $\epsilon_{ref}(t;\beta)$ with two different values of $\beta$. The two values correspond to normalization of $\epsilon_{ref}(t;\beta)$ to two different physical properties of the optimized pulse:
\begin{enumerate}
	\item The fluence $\Phi\left[\epsilon(t)\right]$ (Eq.~\eqref{eq:fluence}), or equivalently, the total energy;
	\item The peak intensity, or equivalently, $\max\abs{\epsilon(t)}$.
\end{enumerate}
The two choices of $\beta$ will be marked as $\beta_1$, $\beta_2$, respectively.

\begin{figure}
	\centering \includegraphics[width=3in]{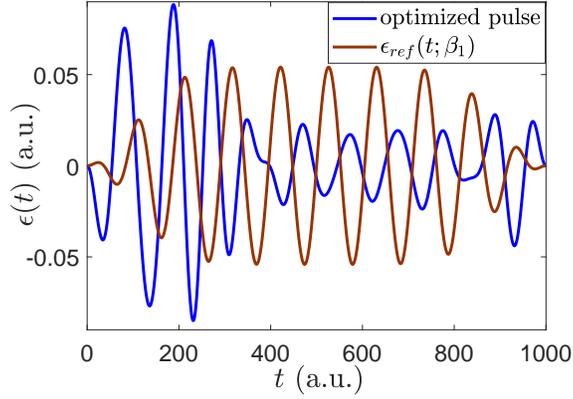}
	\caption{(Color online) The temporal profiles of the optimized pulse (blue, dark gray) and the reference pulse normalized to the same total energy (brown, gray).}\label{fig:epst13ref}
\end{figure}

The temporal profiles of the optimized pulse and $\epsilon_{ref}(t;\beta_1)$ are plotted in Fig.~\ref{fig:epst13ref}.

\begin{figure}
	\centering \includegraphics[width=3in]{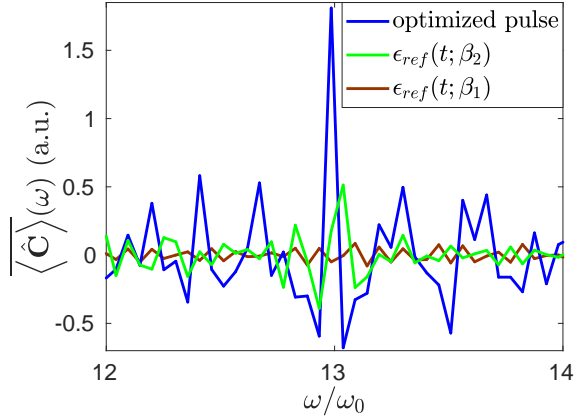}
	\caption{(Color online) The stationary acceleration expectation spectra of the optimized pulse (blue, dark gray), the reference pulse normalized to the same peak intensity (green, light gray) and the reference pulse normalized to the same total energy (brown, gray); the region of the target harmonic of the optimized pulse (the 13'th harmonic) is shown.}\label{fig:Cw13ref}
\end{figure}

The response of the different pulses at the 13'th harmonic is compared in Fig.~\ref{fig:Cw13ref}. It can be observed that $\epsilon_{ref}(t;\beta_1)$ does not induce a significant response. The response is significantly enhanced under the influence of $\epsilon_{ref}(t;\beta_2)$. 
However, the response of $\epsilon_{ref}(t;\beta_2)$ is still not comparable to that of the optimized pulse.

In Table~\ref{tab:ref} we compare for the three pulses the survival probability at the end of the process, the fluence, and the $J_{max}$ value (as defined in Eq.~\eqref{eq:JmaxC}). The ionization probability of the optimized pulse is successfully restricted to the allowed rate, defined by $\sigma(\brackets{\psi(T)}{\psi(T)})$ (less than $10\%$). The ionization probability of $\epsilon_{ref}(t;\beta_1)$ is very small, which implies that the response of the system to the exerted field is not significant. The ionization probability of $\epsilon_{ref}(t;\beta_2)$ is close to $22\%$, which is significantly higher than the optimized pulse, and beyond our control requirements. The fluence of $\epsilon_{ref}(t;\beta_2)$ is also considerably larger. It is shown that the optimized pulse induces a much larger response in the target frequency, with considerably lower ionization probability and total energy.

\begin{table}
\begin{equation*}
	\renewcommand{\arraystretch}{1.5}
	\begin{array}{|c||c|c|c|}
		\hline
		\text{\textbf{Pulse}} & \brackets{\psi(T)}{\psi(T)} & \Phi\left[\epsilon(t)\right]\text{ (a.u.)} & J_{max}\text{ (a.u.)} \\ \hline \hline
		\text{optimized}   		  & 0.926 & 0.992 & 6.97\times 10^{-3}\\ \hline
		\epsilon_{ref}(t;\beta_1) & 0.999 & 0.992 & 3.87\times 10^{-5}\\ \hline
		\epsilon_{ref}(t;\beta_2) & 0.783 & 2.65  & 8.73\times 10^{-4}\\ \hline
	\end{array}
\end{equation*}
\caption{The survival probability at \text{$t=T$}, the fluence, and the $J_{max}$ value of the optimized pulse and the two reference pulses.}\label{tab:ref}
\end{table}

Other target harmonics were optimized: The 15'th, 17'th and 14'th harmonics.

The targeting of an even harmonic (the 14'th) is used to test the limits of the control opportunities. It is well known that the typical HHG spectrum consists of odd harmonics. This property of the HHG spectrum can be referred to as the \emph{selection rules} of the process. It is shown in \cite{BenTal_symmetry,ofir98} that the HHG selection rules originate in the symmetry properties of the problem. The derivation of the selection rules is based on the \emph{Floquet formalism}, which is defined for a CW field. Nevertheless, a quasi-periodic pulse of finite duration can be approximated by the Floquet formalism. {The symmetry of the problem in the idealized Floquet formalism is combined from spatial symmetry properties (the central potential of the atom is symmetric under inversion; the dipole operator is anti-symmetric under inversion) and temporal symmetry properties (the periodic harmonic waveform is anti-symmetric under a temporal shift of a half period; see \cite{BenTal_symmetry,ofir98} for a fuller explanation).}

However, the assumptions underlying the derivation of the selection rules in \cite{BenTal_symmetry,ofir98} may not hold in an optimized pulse. We shall mention several important issues:
\begin{enumerate}
	\item The {temporal} symmetry properties of the harmonic waveform assumed in \cite{BenTal_symmetry,ofir98} can be broken in an optimized field.
	\item It is assumed in \cite{BenTal_symmetry,ofir98} that the system is found in a single Floquet state. This assumption needn't be valid if the system is appropriately controlled to be in a superposition of such states.
	\item The assumption that the system can be approximately described by the Floquet formalism may totally collapse in an optimized pulse, which may considerably deviate from a quasi-periodic template.
\end{enumerate}
All these issues are related to the \emph{breaking of symmetry} of the problem. 
The question is to which extent this broken symmetry can lead to the enhancement of the ``forbidden'' even harmonics.

During the optimization processes of the 14'th and 15'th harmonics, it was necessary to reset the Hessian approximation, 
after the termination condition \eqref{eq:termination} was first matched.

\begin{figure}
	\begin{center}
		\begin{tabular}{c}
			\includegraphics[width=2.9in]{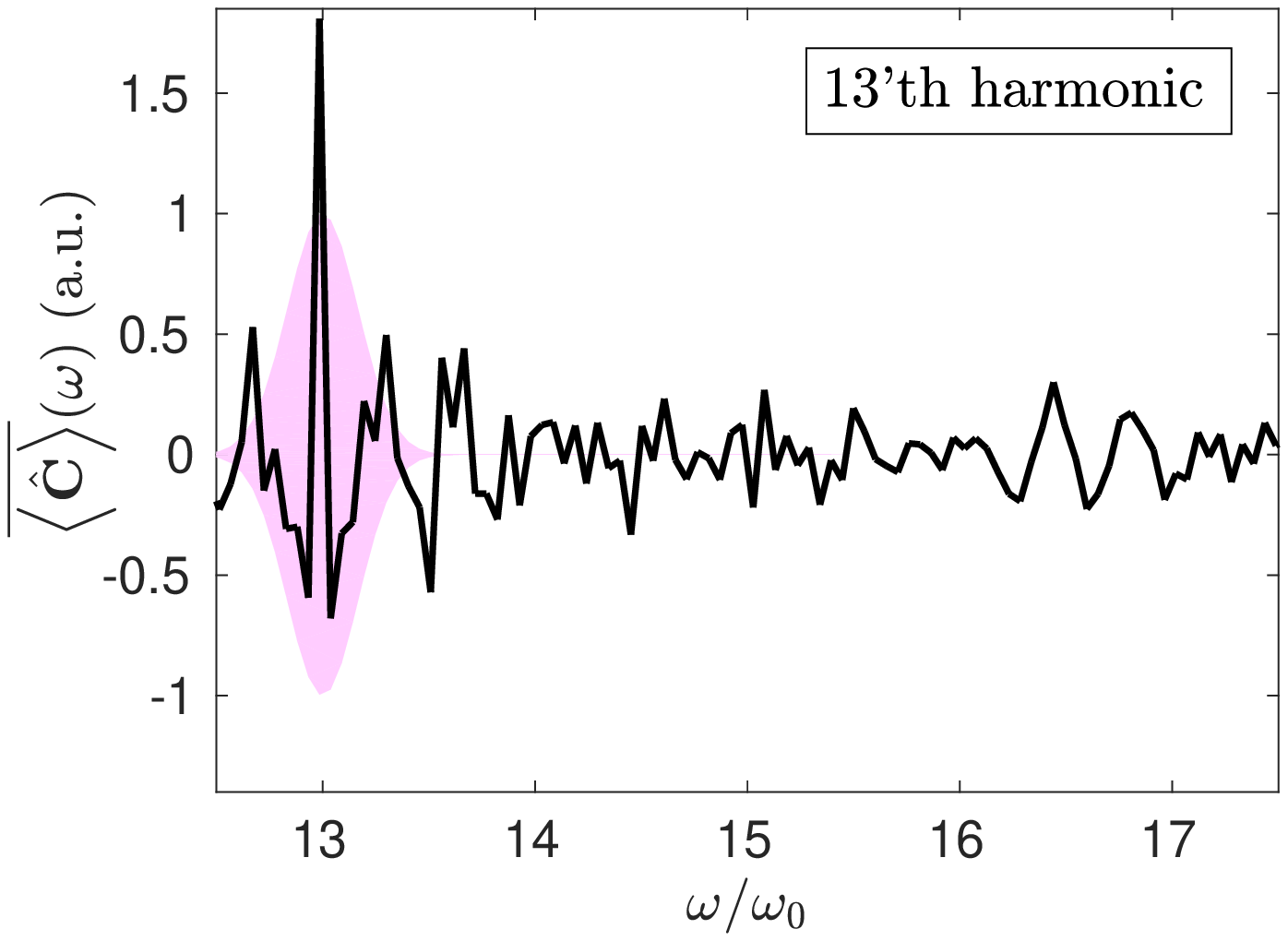} \\
			\includegraphics[width=2.9in]{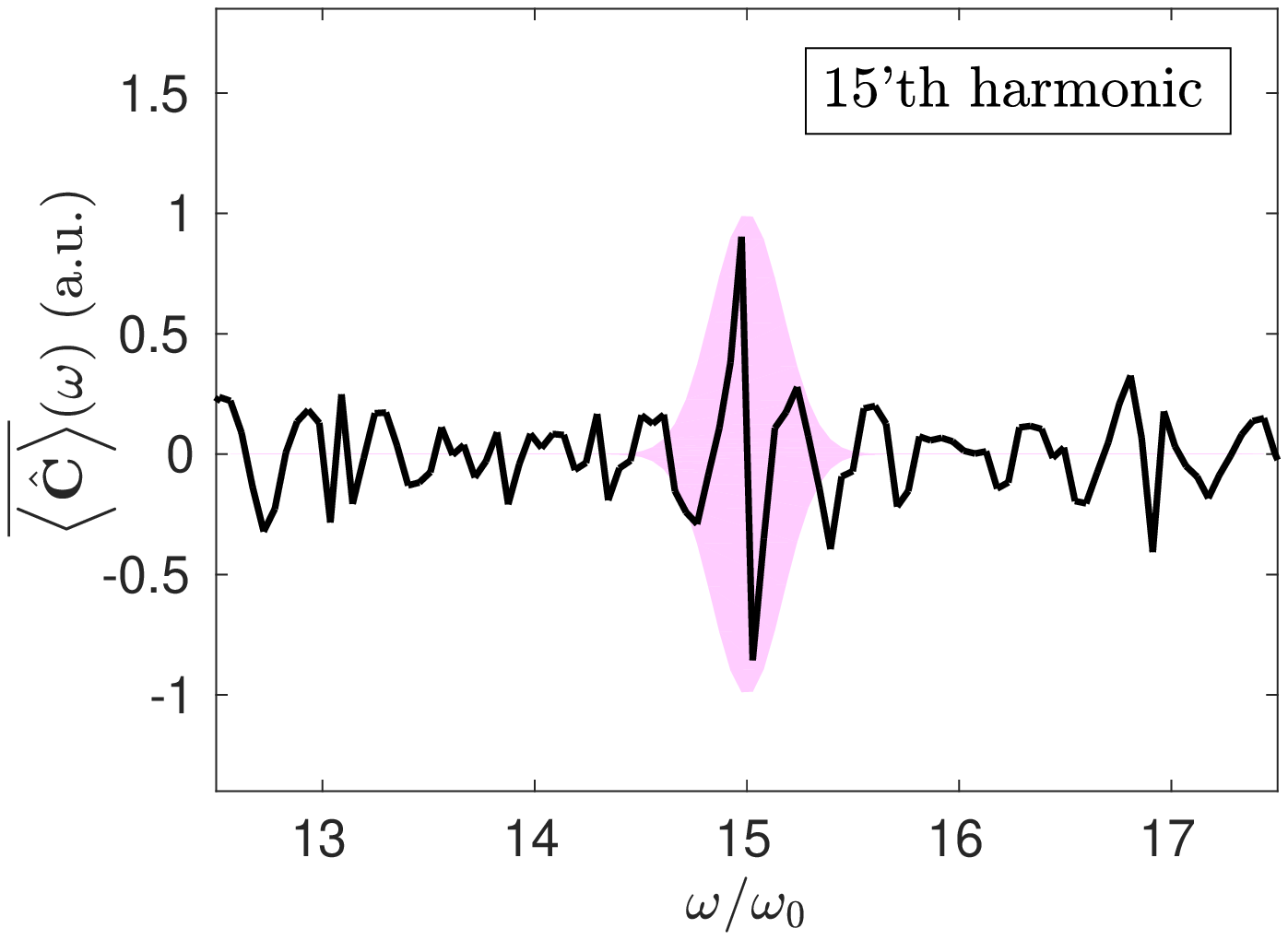} \\
			\includegraphics[width=2.9in]{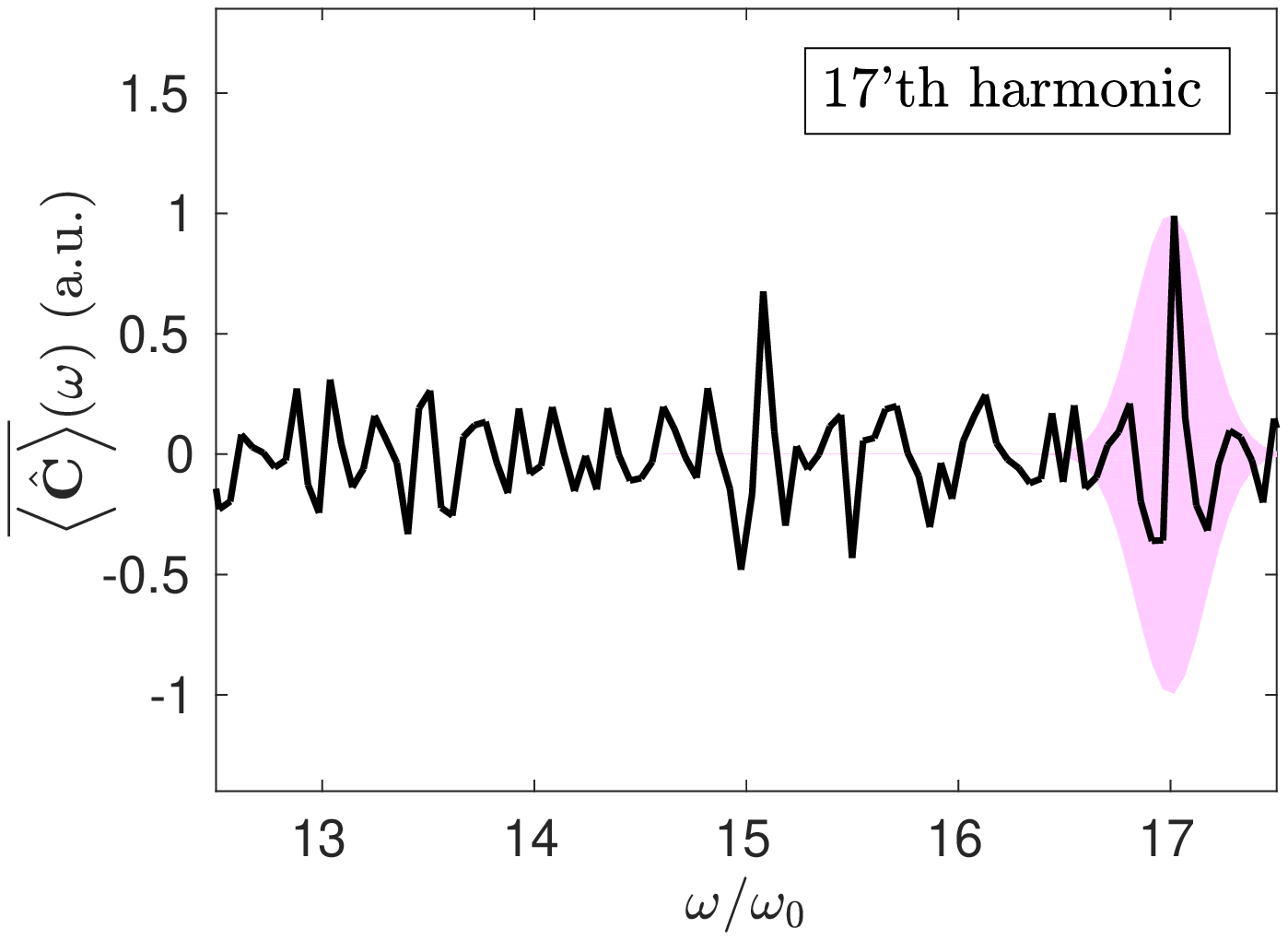} \\
			\includegraphics[width=2.9in]{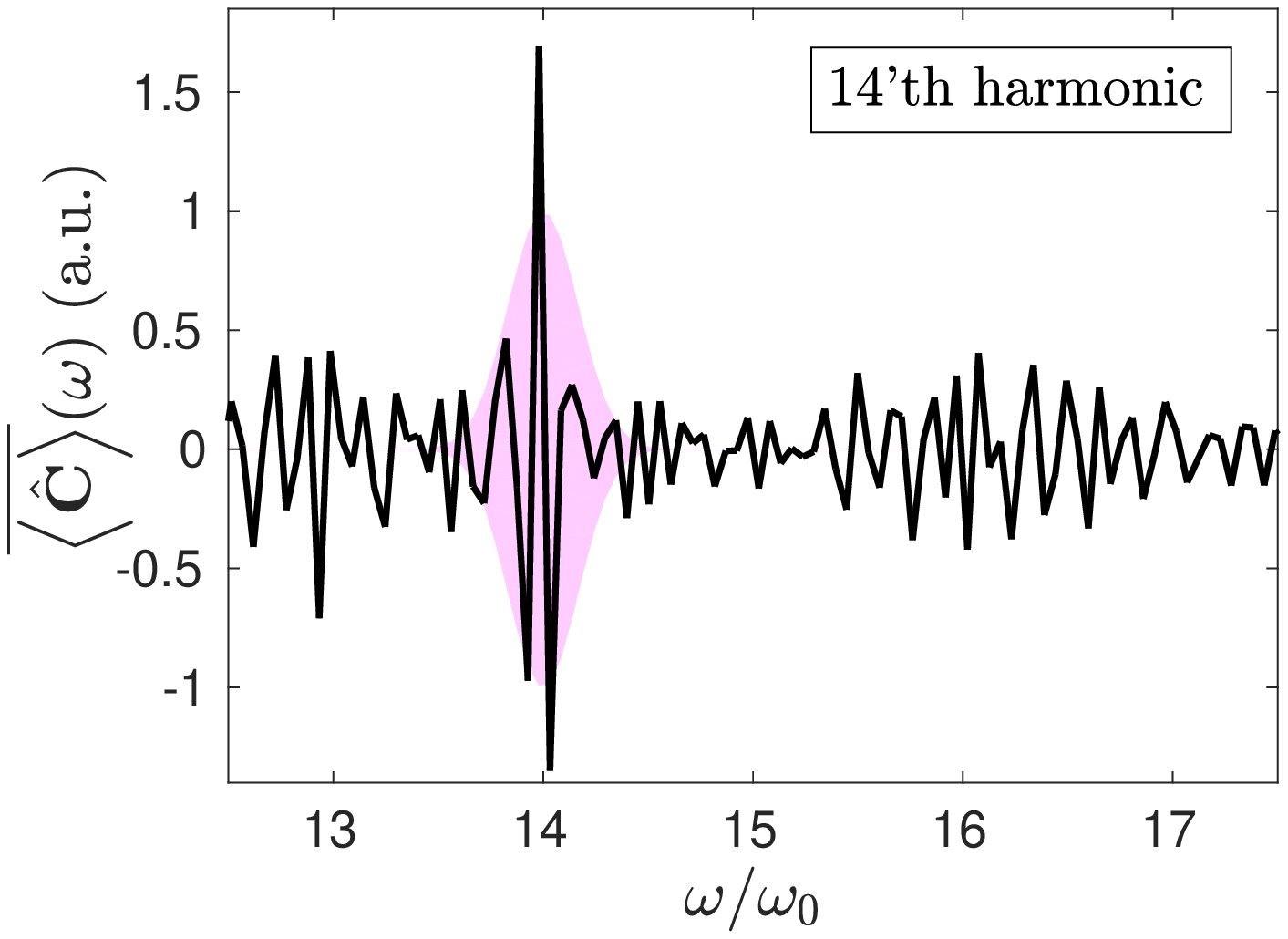}
		\end{tabular}
	\end{center}
	\caption{(Color online) The stationary acceleration expectation spectra of the various target harmonic problems in the region of interest; the profile of the target filter function is marked in light magenta (gray) for each problem.}\label{fig:harmonicsCw}
\end{figure}

In Fig.~\ref{fig:harmonicsCw}, we present the response of the system in the region of interest for the four problems (including the 13'th harmonic problem). 
It is shown that the response is selectively enhanced in all the required frequencies.


\begin{figure}
	\centering \includegraphics[width=3in]{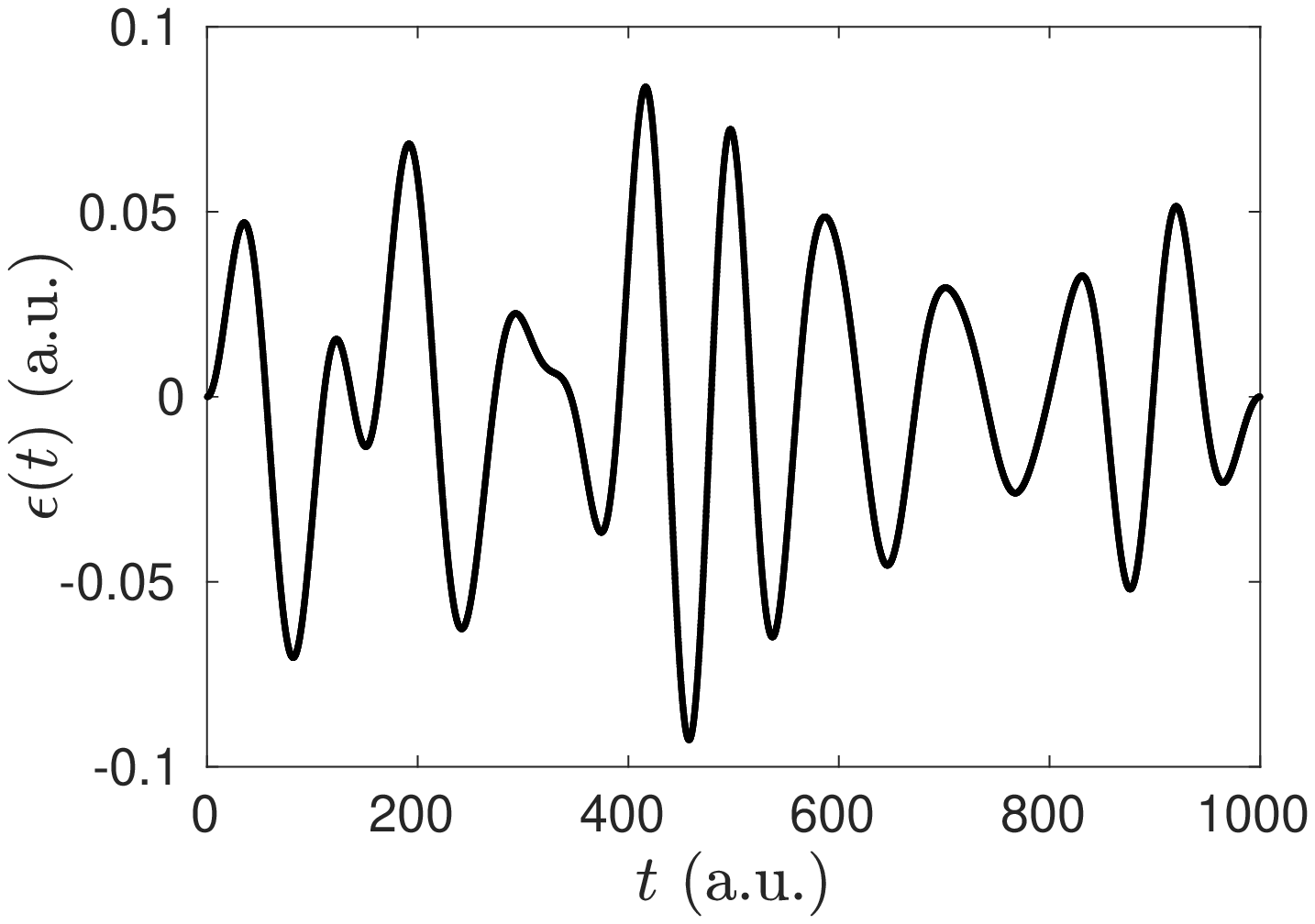}
	\caption{The temporal profile of the optimized pulse in the 14'th harmonic problem.}\label{fig:epst14}
\end{figure}

Remarkably, the enhancement of the response at the even 14'th harmonic target is as successful compared to the other odd harmonic targets. In Fig.~\ref{fig:epst14}, the temporal profile of the optimized pulse is shown. The field seems to be considerably more complex than the field of the optimized pulse of the 13'th harmonic problem, presented in Fig.~\ref{fig:epst13ref}. The reason could be the necessity of breaking symmetry in the even harmonic problem. It can be readily observed that the optimized field does not satisfy the symmetry properties of a harmonic waveform. 
In addition, the Floquet formalism clearly becomes inappropriate for the description of this very complex profile.

As in the spectrum of the 13'th harmonic problem, significant response in neighbouring harmonics appears also in the other spectra presented in Fig.~\ref{fig:harmonicsCw}. Interestingly, for the odd harmonic targets, only odd neighbouring harmonics are observed. 
This implies that the symmetry properties of HHG inferred from a CW field, remain relevant for the description of the process induced by the optimized pulses. 
In contrary, in the 14'th harmonic problem we can observe the appearance of both an odd harmonic (the 13'th) and an even one (the 16'th). 
This could be expected by the break of symmetry which allows the appearance of the target frequency.

In Table~\ref{tab:harmonics_ion} we present the survival probability at the end of the process for all our target frequencies. The ionization probability in all problems is successfully restricted to less than $10\%$, and lies in the range of \text{$6\!-\!8\;\%$}.

\begin{table}
\begin{equation*}
	\renewcommand{\arraystretch}{1.5}
	\begin{array}{|c||c|}
		\hline
		n  & \brackets{\psi(T)}{\psi(T)} \\ \hline \hline
		13 & 0.926 \\ \hline
		14 & 0.923 \\ \hline
		15 & 0.932 \\ \hline
		17 & 0.937 \\ \hline
	\end{array}
\end{equation*}
\caption{The survival probability at \text{$t=T$} of the optimized pulses for the different target frequencies.}\label{tab:harmonics_ion}
\end{table}

The validity of the approximations introduced by the employment of absorbing boundary conditions has been tested for all problems, as described in Appendix~\ref{app:absorbing}.

The source \texttt{MATLAB} codes of the results are available in the Supplemental Material of the paper, along with relevant data files.

\section{Conclusion}\label{sec:conclusion}

In the present study, an optimization method of HHG has been developed in the framework of quantum OCT\@. 
The target was a specific emission frequency. 
Several restrictions have been imposed in the OCT formulation as ``hard'' and ``soft'' constraints. 
In particular, the restriction of permanent ionization has been addressed 
by its formulation as a soft constraint. This requirement has not been addressed in previous theoretical studies.

Special emphasis was given to the numerical implementation. 
The simulation of the dynamics was performed by highly accurate methods, 
which is crucial to achieve a reliable description of the HHG process. 
The solver of the explicitly time dependent Schr\"odinger equation 
was performed by a new highly accurate approach \cite{k273,SemiGlobal}. 
A new optimization method was employed for the construction of the complex-absorbing-potential, used for the realization of the absorbing boundary conditions. The complexity of the HHG optimization problem required the employment of a second-order gradient method for the optimization process. Several details of the implementation to the present problem required special care.

The results demonstrated significant selective enhancement of the harmonic yield, with simultaneous minimization of the total energy of the driving pulse and control of the permanent ionization probability. The violation of the high harmonic selection rules has also been demonstrated.

\begin{figure}
	\begin{center}
		\begin{tabular}{c}
			\includegraphics[width=3in]{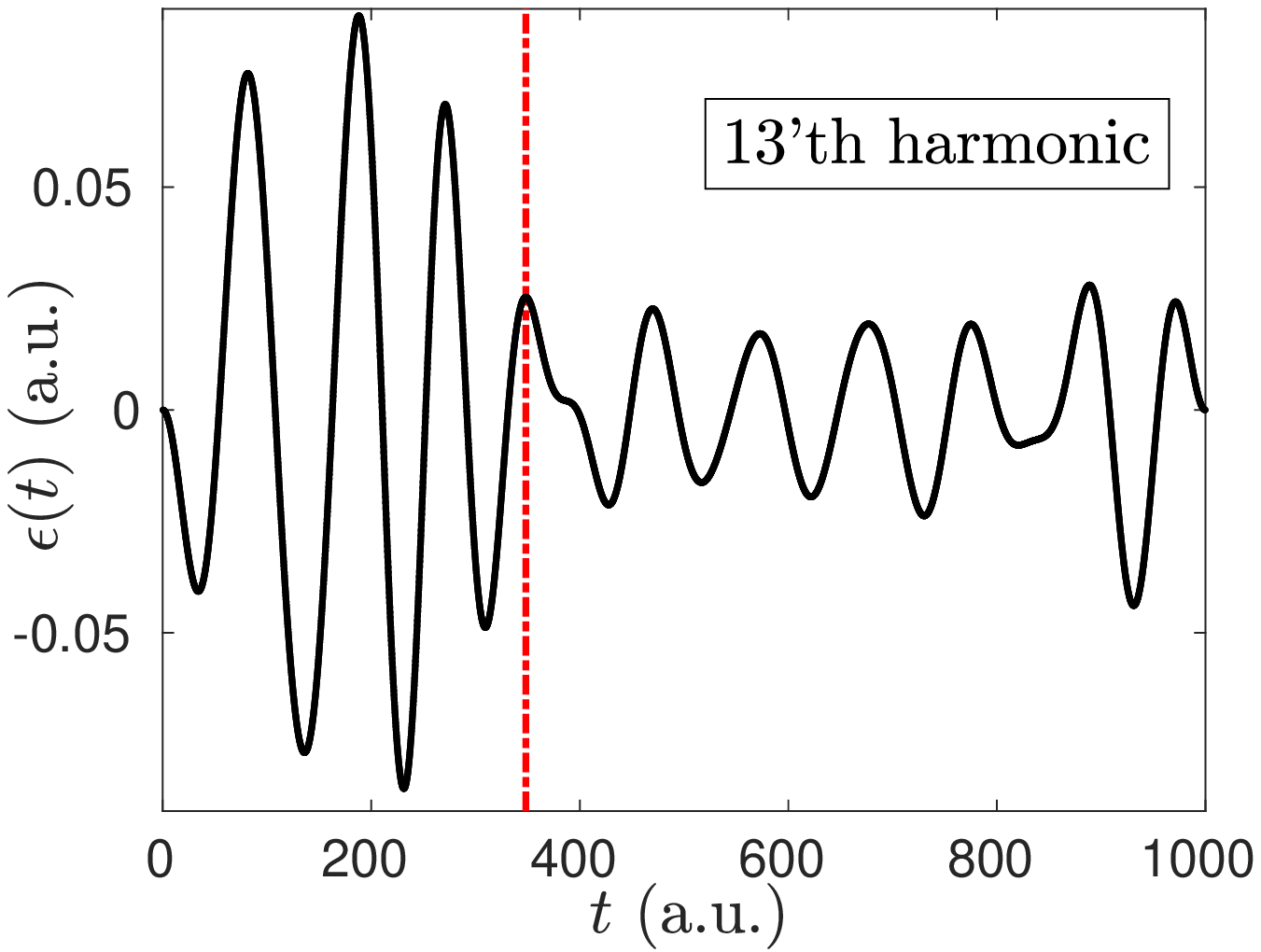}\\
			\includegraphics[width=3in]{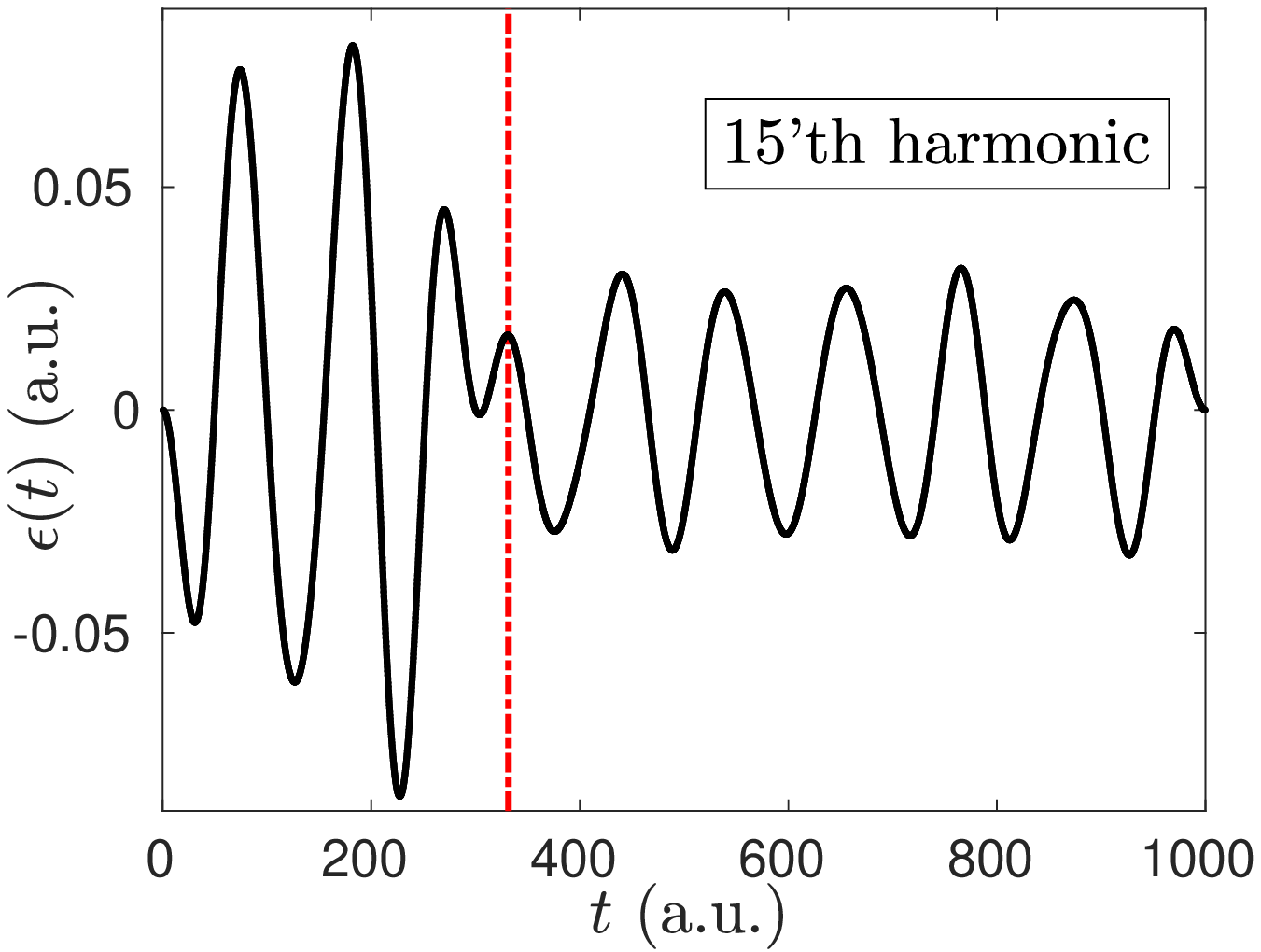} \\
			\includegraphics[width=3in]{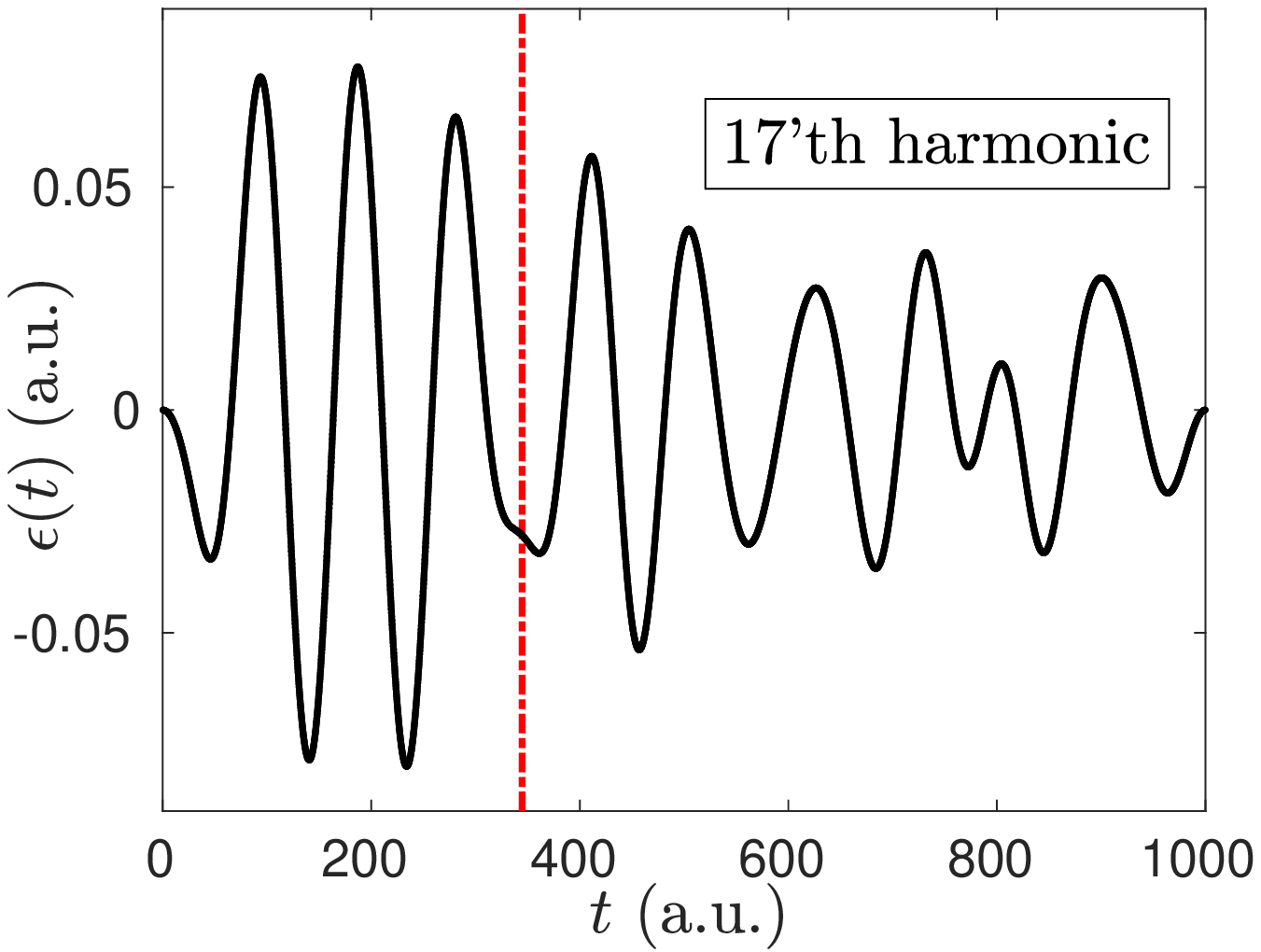}
		\end{tabular}
	\end{center}
	\caption{(Color online) The temporal profiles of the optimized pulses in the odd-harmonic problems; the process can be divided into two stages, where the first stage is characterized by non-periodic patterns, higher intensities and a tendency to higher frequencies, while the second one is characterized by quasi-periodic patterns, with lower intensities and frequencies around the fundamental frequency of the source. The division into two stages is marked by a vertical dashed red (gray) line.}\label{fig:epst_div}
\end{figure}

The present paper is devoted to the control aspects of the optimization method. The physical interpretation of optimized fields requires a separate thorough discussion, and thus is beyond the framework of the present paper. Nevertheless, we shall briefly mention a particular insight into the mechanisms underlying the 
optimized fields in the odd harmonic problems of Sec.~\ref{sec:results}. 
We found that the optimized HHG processes consist of two stages:
\begin{enumerate}
	\item Liberation of the electron from the adiabatic regime; it was found that adiabaticity imposes a barrier on the initiation of the HHG process 
	(as has already been mentioned in Sec.~\ref{ssec:boundary}).
	\item Generation of harmonics by quasi-periodic patterns of the driving fields.
\end{enumerate}
{The two stage pattern is demonstrated in Fig.~\ref{fig:epst_div}; the temporal profiles of the optimized fields in the odd harmonic problem are plotted, where the division into two stages in each optimized field is marked by a vertical dashed red line (it should be noted that the transition from the first stage to the second one does not take place in a well-defined time-point, and thus there is some arbitrariness in this marking). The first stage is characterized by non-periodic patterns, with higher intensities and a tendency to higher frequencies, while the second one is characterized by quasi-periodic patterns, with lower intensities and frequencies around the fundamental frequency of the source. Both the high intensities and the high frequencies in the first stage contribute to the deviation from the adiabatic regime.} A fuller discussion is left for a future publication.

We hope that the present optimization method will contribute to a fuller understanding of physics of the HHG process in general, 
and of optimal pulses for controlling an emission line in particular.

\begin{acknowledgments}
We thank Daniel Strasser, Roi Baer and Nimrod Moiseyev for useful discussions.
This research was supported by the Israel Science Foundation, Grant No.\ 2244/14.
\end{acknowledgments}

\appendix

\section{The choice of the target operator}\label{app:target}

The target operator $\operator{O}$ has been chosen as the stationary acceleration operator $\opC$ (see Sec.~\ref{ssec:target}). The present appendix discusses the considerations which lead to this choice.

As an introduction, we shall present several obvious candidates for $\operator{O}$ which are related to the dipole motion. For simplicity, we consider the dipole of a single electron. Only the $x$ component is considered, in accordance with the polarization of the driving field. The following observables are directly related to the dipole motion:
\begin{enumerate}
	\item The dipole operator:
	\begin{equation}
		\operator{\mu}_x = \operator{X}
	\end{equation}		
	\item The dipole velocity operator:
	\begin{equation}
		\operator{\dot\mu}_x = \operator{\dot X} = \operator{P}_x
	\end{equation}	
	\item The dipole acceleration operator (see Sec.~\ref{ssec:target}):
	\begin{equation}\label{eq:op_acc1}
		\operator{\ddot\mu}_x = \operator{\ddX} = -\deriv{V\!\left(\operator{X}\right)}{\operator X}
	\end{equation}
\end{enumerate}
Note that atomic units are used, and the electron has a unit mass.

The spectra of the three observables can be related mathematically. The simplest relations are between the \emph{Fourier} spectral functions. The Fourier spectral representation of a temporal signal is traditionally defined by its \emph{inverse Fourier transform}. Let us use the following notation for the inverse Fourier transform of an arbitrary function $g(t)$:
\begin{equation}
	g^f(\omega) \equiv \mathcal{F}^{-1}\left[g(t)\right] \equiv \frac{1}{\sqrt{2\pi}}\int_{-\infty}^\infty g(t)\exp(i\omega t)\,dt 
\end{equation}
$g(t)$ is restored from $g^f(\omega)$ by the direct Fourier transform:
\begin{equation}
	g(t) = \mathcal{F}\left[g^f(\omega)\right] \equiv \frac{1}{\sqrt{2\pi}}\int_{-\infty}^\infty g^f(\omega)\exp(-i\omega t)\,d\omega 
\end{equation}
The three Fourier spectral functions are related by
\begin{equation}\label{eq:Fourier_relations}
	{\exval{\operator{\ddot\mu}_x}\!}^f(\omega) = -i\omega{\exval{\operator{\dot\mu}_x}\!}^f(\omega) = -\omega^2{\exval{\operator{\mu}_x}}^f(\omega)
\end{equation}
The relations \eqref{eq:Fourier_relations} can be derived from the direct Fourier expression for the dipole expectation,
\begin{equation}\label{eq:invFmu}
	\exval{\operator{\mu}_x}\!(t) = \frac{1}{\sqrt{2\pi}}\int_{-\infty}^\infty {\exval{\operator{\mu}_x}}^f(\omega)\exp(-i\omega t)\,d\omega
\end{equation}
by considering the single and the double differentiation of the equation w.r.t.\ $t$. It can be seen from \eqref{eq:Fourier_relations} that the three spectra \emph{contain the same spectral regions}. However, the higher frequency regions are emphasized in higher derivatives of $\operator{\mu}_x$. The three spectral functions also differ by a phase shift.

The \emph{cosine transform} spectra of the dipole and the acceleration are related by
\begin{equation}
	\overline{\exval{\operator{\ddot\mu}_x}}(\omega) = -\omega^2\muwx
\end{equation}
The relation to $\overline{\exval{\operator{\dot\mu}_x}}(\omega)$ is more complex. However, the general picture remains similar, where the higher frequencies are emphasized in higher derivatives.

The choice of the target operator $\operator{O}$ requires discussions on two different grounds:
\begin{enumerate}
	\item In the \emph{physical ground}, we first have to identify the observable associated as the source of emission.
	\item In the \emph{numerical ground}, we should identify the observable which is the preferable one from a numerical viewpoint. The spectrum of this observable can be related mathematically to the emission spectrum, which may be associated with another observable. 
\end{enumerate}

We begin from the physical discussion. In classical electrodynamics an accelerated charge emits radiation  proportional to the \emph{acceleration} of the charge. 
Therefore, it has been assumed that the high-harmonic spectrum is proportional to the dipole acceleration spectrum. 
However, in 2008 it was claimed in~\cite{Diestler08} that the emission high-harmonic spectrum is proportional to the \emph{velocity} spectrum. 
Currently, there was a debate on this topic in the literature (see~\cite{xvaHHG,xvaHHGcomment,xvaHHGreply}). 
In recent publications there is still no consensus on this point.


The present study focuses on the optimization of selected harmonics (see Sec.~\ref{sec:results}). Therefore, the target band extends over a small interval in the spectrum. 
Hence, the $\omega$ factor difference between the acceleration and velocity spectra does not change significantly over the target region, and thus does not make an important difference between the spectra. Therefore, the physical question is not of fundamental importance for our results.

However, the numerical question has considerable importance. First, we have to introduce the numerical problem of concern. Spectra produced by discrete spectral transforms almost always contain background of numerical origin. This numerical background might hinder small magnitude physical effects. There are three important sources of background:
\begin{enumerate}
	\item \emph{Discretization}: The discrete spectral transforms are based on discrete basis functions, with a discrete frequency sampling. If the signal contains a frequency which is not present in the discrete set, the transformed signal contains peaks at the nearest frequencies, but with a long ``tail'', which induces background noise throughout the spectrum. The magnitude of the background depends on the magnitude of the peak, and the distance of the represented frequency from the nearest frequency in the discrete set. In the discrete cosine and sine transforms, the frequency sampling is twice as dense as the Fourier sampling; however, the phase flexibility of the discrete Fourier transform is lost, with a fixed phase for each frequency component. The effect of mismatch of phase of the signal with the basis functions has the same effect as that of frequency mismatch.
	\item \emph{Truncation error}: The discrete time-sampling results in a truncation of the spectral series at a cutoff frequency $\Omega$. The truncation error induces the effect of \emph{aliasing}.
	\item \emph{Boundary effects}: Each spectral transform is adjusted to certain boundary conditions of the signal. $\evO$ is not expected to satisfy any of these boundary conditions at the final time $T$. This results in discontinuities at the boundary (a discrete spectral series represents an \emph{infinite periodic signal}, which is defined also for \text{$t>T$}). A discontinuity contains frequencies from the entire spectrum, and induces noise throughout the spectrum (see \cite[Chapter 3]{thesis}). The DCT has the advantage that the discontinuity is in the \emph{derivative} of the signal, and not in the function value; this reduces drastically the boundary effects compared to the discrete Fourier and sine transforms. However, the effect is not completely eliminated.
\end{enumerate}

In our previous study \cite{thesis,OCTHG} we targeted the dipole spectrum, $\muwx$. The problem with this choice is that the spectrum is scaled by $1/\omega$ compared to the velocity spectrum, or $1/\omega^2$ compared to the acceleration spectrum. This led to difficulties in the optimization of higher frequencies, which were hindered by the background of the spectrum. The problem disappeared when we adopted the choice of the acceleration operator spectrum as the optimization target, as in \cite{RasanenHHG,DFTHHG}. In the acceleration spectrum, the magnitude of the physical information in higher frequencies is amplified compared to the other spectra discussed here. Hence, it becomes the preferable choice for harmonic generation problems. Conversely, in down conversion optimization problems the dipole spectrum is expected to be preferable.

In principle, if necessary, it is possible to obtain further amplification of higher frequencies; the target operator can be chosen as a time-derivative of the dipole of a higher order than the second. This was not required in the present study.

The acceleration spectrum still contains a large linear response component, which induces background throughout the spectrum. A further improvement can be achieved with some insight into the acceleration spectrum. As was mentioned in Sec.~\ref{ssec:target}, the acceleration operator can be divided into a stationary part and a time-dependent part:
\begin{equation}
	\operator{\ddX} = \opC + \epsilon(t)
\end{equation}
By linearity of the spectral transform, the acceleration spectrum can also be divided into two parts:
\begin{equation}
	\ddXw = \Cw + \epsw 
\end{equation}
The second term of the RHS is the driving field spectrum, which contains only \emph{linear response} in the low frequency regime. It is irrelevant to the region of interest in the spectrum---the high harmonic regime. Thus, the part of interest in the acceleration spectrum is the \emph{stationary acceleration} only \cite{DFTHHG}. Accordingly, we can set:
\begin{equation}\label{eq:OC1}
	\operator{O} \equiv \opC
\end{equation}
The omission of the time-dependent part of the acceleration drastically reduces the magnitude of the linear response peaks in the spectrum. Consequently, the background is considerably reduced. This reveals another advantage of the acceleration spectrum over the dipole or velocity spectra---the majority of the linear response can be isolated and eliminated from the spectrum.

\section{The derivation of the Euler-Lagrange equations}\label{app:der}

For the sake of clarity, let us first summarize the different components of the maximization functional. The full maximization functional is
\begin{equation}
	J \equiv J_{max} + J_{ion} + J_{energy} + J_{\epsilon(0)} + J_{\epsilon(T)} + J_{Schr}
\end{equation}
The different terms in the functional can be classified as follows:
\begin{enumerate}
	\item A maximization term,
	\begin{equation}
		J_{max} \equiv \frac{1}{2}\int_0^\Omega \fC\overline{\exval{\operator{C}}}^2\!(\omega)\,d\omega, \qquad \fC \geq 0, \qquad \max\left[\fC\right] = 1
	\end{equation}
	where
	\begin{equation}
		\Cw \equiv \sqrt{\frac{2}{\pi}}\int_0^T \exval{\operator{C}}\!(t)\cos(\omega t)\,dt \label{eq:Cwd}
	\end{equation}
	\item Two penalty terms, which represent ``soft constraints'':
	\begin{enumerate}
		\item The constraint on the ionization probability is represented by
		\begin{equation}
			J_{ion} \equiv \sigma\left(\brackets{\psi(T)}{\psi(T)}\right), \qquad \sigma(y)\leq 0, \qquad \sigma(1) = 0 \label{eq:Jiond}
		\end{equation}
		\item The following term represents the constraint on the energy of the incident field, as well as the restriction of its spectrum:
		\begin{equation}
			J_{energy} \equiv -\int_0^\Omega\frac{1}{\tfeps}\bar{\epsilon}^2(\omega)\,d\omega, \qquad \tfeps>0 \label{eq:Jenergyd}
		\end{equation}
		{where $\epsw$ defines the spectral representation of $\epsilon(t)$ as follows:
		\begin{equation}
			\epsilon(t) = \sqrt{\frac{2}{\pi}}\int_0^\Omega \epsw\cos(\omega t)\,d\omega
		\end{equation}
		}
	\end{enumerate}
	\item Three Lagrange-multiplier terms, which represent three ``hard'' constraints:
	\begin{enumerate}
		\item The constraint of the zero boundary condition on $\epsilon(t)$ at \text{$t=0$},
		\begin{equation} \label{eq:eps0d}
			\epsilon(0) = 0,
		\end{equation}
		is represented by the following Lagrange-multiplier term:
		\begin{equation} \label{eq:Jeps0d}
			\Jepsz \equiv -\sqrt{2\pi}\lambda_0\epsilon(0) = -2\lambda_0\int_0^\Omega\epsw \cos(0)\,d\omega = -2\lambda_0\int_0^\Omega\epsw\,d\omega
		\end{equation}
		\item The constraint of the zero boundary condition on $\epsilon(t)$ at \text{$T=0$},
		\begin{equation} \label{eq:epsTd}
			\epsilon(T) = 0,
		\end{equation} 
		is represented by the following Lagrange-multiplier term:
		\begin{equation} \label{eq:JepsTd}
			\JepsT \equiv -\sqrt{2\pi}\lambda_T\epsilon(T) = -2\lambda_T\int_0^\Omega\epsw \cos(\omega T)\,d\omega
		\end{equation}
		\item The $J_{Schr}$ term represents the Schr\"odinger equation constraint on $\ket{\psi(t)}$,
		\begin{equation}
			\deriv{\ket{\psi(t)}}{t} = -i\operator{H}(t)\ket{\psi(t)}	\label{eq:Schrd}	
		\end{equation}
		The Schr\"odinger equation relates $\ket{\psi(t)}$ to $\epsw$ via the time-dependent Hamiltonian,
		\begin{equation}
			\operator{H}(t) = \operator{H}_0 - \operator{X}\epsilon(t) = \operator{H}_0 - \operator{X}\left(\sqrt{\frac{2}{\pi}}\int_0^\Omega \bar{\epsilon}(\omega)\cos(\omega t)\,d\omega\right)\label{eq:Hd}
		\end{equation}
		The Schr\"odinger equation constraint has to be imposed at each time-point of the time-interval \text{$t\in[0,T]$}.
		
		In order to formulate the constraint correctly, we should note that the vector $\ket{\psi(t)}$ is a complex entity. Hence, each of its components consists of two degrees of freedom in each time-point---the real and imaginary parts. The Schr\"odinger equation, being a complex equation, constrains both degrees of freedom. Each of these constraints requires a distinct Lagrange-multiplier term. It is convenient to treat $\ket{\psi(t)}$ and $\bra{\psi(t)}$ as independent variables, in order to represent the two degrees of freedom of the complex state vector. The constraint equation on $\ket{\psi(t)}$ is \eqref{eq:Schrd}. The constraint equation on $\bra{\psi(t)}$ is its adjoint equation,
		\begin{equation}
			\deriv{\bra{\psi(t)}}{t} = i\bra{\psi(t)}\operator{H}^\dagger(t) \label{eq:conjSchrd}
		\end{equation}
		These constraint equations are subject to the following initial conditions, respectively:
		\begin{align}
			& \ket{\psi(0)} = \ket{\psi_0} \label{eq:initcond}\\
			& \bra{\psi(0)} = \bra{\psi_0} \label{eq:initcond_bra}
		\end{align}
		The constraint equations, \eqref{eq:Schrd}, \eqref{eq:conjSchrd}, together with the initial conditions, \eqref{eq:initcond}, \eqref{eq:initcond_bra}, ensure that
		\begin{equation}
			\bra{\psi(t)} = \ket{\psi(t)}^\dagger
		\end{equation}
		in all $t$. Assuming this, only Eqs.~\eqref{eq:Schrd}, \eqref{eq:initcond} are required for the performance of the computations in practice.
		
		The resulting Lagrange-multiplier term is a continuous summation of the constraint terms over $t$:
		\begin{align}
			&J_{Schr} \equiv \nonumber \\
			&-\left[\int_0^T\bracketsk{\chi(t)}{\deriv{\psi(t)}{t}+i\operator H(t)\psi(t)}\,dt + \int_0^T\bracketsb{\deriv{\psi(t)}{t}+i\operator H(t)\psi(t)}{\chi(t)}\,dt\right]  \label{eq:JSchrd}
		\end{align}
		In this stage, $\ket{\chi(t)}$ and $\bra{\chi(t)}$ are treated as \emph{independent Lagrange-multiplier functions}. Nevertheless, the symmetry of the problem under the adjoint operation suggests that
		\begin{equation}
			\bra{\chi(t)} = \ket{\chi(t)}^\dagger \label{eq:chibraket}
		\end{equation}		
		This justifies the use of the same letter $\chi$ for both Lagrange-multipliers. The full justification to \eqref{eq:chibraket} will be given in what follows. For the sake of brevity, we shall rely on Eq.~\eqref{eq:chibraket} before it was fully justified, and write $J_{Schr}$ as
		\begin{equation}
			J_{Schr} = -2\Real{\int_0^T\bracketsO{\chi(t)}{\deriv{}{t}+i\operator H(t)}{\psi(t)}\,dt}
		\end{equation}
	\end{enumerate}
\end{enumerate}


The extremum conditions are:
\begin{align}
	&\fnlderiv{J}{\bar{\epsilon}(\omega)} = 0 \label{eq:dJdepswd}\\
	&\fnlderiv{J}{\ket{\psi(t)}} = 0 \label{eq:dJdpsitkd} \\ 
	&\fnlderiv{J}{\bra{\psi(t)}} = 0 \label{eq:dJdpsitbd} \\	
	&\fnlderiv{J}{\ket{\psi(T)}} = 0 \label{eq:dJdpsiTkd} \\ 
	&\fnlderiv{J}{\bra{\psi(T)}} = 0 \label{eq:dJdpsiTbd}
\end{align}

$J_{Schr}$ is more easily handled after integrating by part the following expression:
\[
	\int_0^T\bracketsk{\chi(t)}{\deriv{\psi(t)}{t}}\,dt
\]
We obtain:
\begin{equation}
	J_{Schr} = -2\Real\left[\bracketsk{\chi(T)}{\psi(T)}-\bracketsk{\chi(0)}{\psi(0)} - \int_0^T\bracketsb{\left(\deriv{}{t}+i\operator H^\dagger(t)\right)\chi(t)}{\psi(t)}\,dt\right] \label{eq:JSchrp}
\end{equation}

We start from condition \eqref{eq:dJdepswd}. First, we shall derive the equation for a simpler problem, in which the boundary condition constraints on the field (Eqs.~\eqref{eq:eps0d}, \eqref{eq:epsTd}) are excluded. Then, we shall derive the equation for the full problem, which is conveniently expressed in the terms of the simpler problem expression for the field.

In the unconstrained problem, the $\Jepsz$ and $\JepsT$ terms (Eqs.~\eqref{eq:Jeps0d}, \eqref{eq:JepsTd}, respectively) are excluded from $J$. The LHS of Eq.~\eqref{eq:dJdepswd} becomes:
\begin{equation}
	\fnlderiv{J}{\epsw} = \fnlderiv{J_{energy}}{\epsw} + \fnlderiv{J_{Schr}}{\epsw} \label{eq:dJdepsunc}
\end{equation}
where
\begin{align}
	\fnlderiv{J_{energy}}{\epsw} &= -\frac{2\epsw}{\tfeps} \label{eq:dJenergydeps}\\
	\fnlderiv{J_{Schr}}{\epsw} &= 2\,\Real\left[-i\int_0^T\bracketsO{\chi(t)}{\fnlderiv{\operator{H}(t)}{\epsw}}{\psi(t)}\,dt\right] \nonumber \\
	&=-2\,\Imag\left[\sqrt{\frac{2}{\pi}}\int_0^T\bracketsO{\chi(t)}{\operator{X}}{\psi(t)}\cos(\omega t)\,dt\right] \nonumber \\
	&=-2\,\Imag\left\lbrace \mathcal{C}\left[\bracketsO{\chi(t)}{\operator{X}}{\psi(t)}\right]\right\rbrace \nonumber \\
	&=  2\,\mathcal{C}\left[-\Imag{\bracketsO{\chi(t)}{\operator{X}}{\psi(t)}}\right] \label{eq:dJSchrdeps}
\end{align}
The resulting gradient expression is
\begin{equation}
	\fnlderiv{J}{\epsw} = 2\left(-\frac{\epsw}{\tfeps} + \mathcal{C}\left[-\Imag{\bracketsO{\chi(t)}{\operator{X}}{\psi(t)}}\right]\right)
\end{equation}
The extremum condition \eqref{eq:dJdepswd} yields the following expression for $\epsw$:
\begin{equation}\label{eq:epswresultunc}
	\epsw = \tfeps\mathcal{C}\left[-\Imag{\bracketsO{\chi(t)}{\operator{X}}{\psi(t)}}\right] 
\end{equation}
The field in the time-domain is
\begin{equation}
	\epsilon(t) = \mathcal{C}^{-1}\left\lbrace\tfeps\mathcal{C}\left[-\Imag{\bracketsO{\chi(t)}{\operator{X}}{\psi(t)}}\right]\right\rbrace
\end{equation}

Now we turn to the solution of the full constrained problem. The LHS of Eq.~\eqref{eq:dJdepswd} is modified to
\begin{equation}
	\fnlderiv{J}{\epsw} = \fnlderiv{J_{energy}}{\epsw} + \fnlderiv{\Jepsz}{\epsw} + \fnlderiv{\JepsT}{\epsw} + \fnlderiv{J_{Schr}}{\epsw} \label{eq:dJdeps}
\end{equation}
where
\begin{align}
	& \fnlderiv{\Jepsz}{\epsw} =  -2\lambda_0\\
	& \fnlderiv{\JepsT}{\epsw} =  -2\lambda_T\cos(\omega T)
\end{align}
The gradient expression is modified to
\begin{equation} \label{eq:grad_functional}
	\fnlderiv{J}{\epsw} = 2\left(-\frac{\epsw}{\tfeps} + \mathcal{C}\left[-\Imag{\bracketsO{\chi(t)}{\operator{X}}{\psi(t)}}\right] - \lambda_0 - \lambda_T\cos(\omega T)\right)
\end{equation}
Eq.~\eqref{eq:dJdepswd} yields the following modified expression for $\epsw$:
\begin{equation}\label{eq:epswresult}
	\epsw = \tfeps\mathcal{C}\left[-\Imag{\bracketsO{\chi(t)}{\operator{X}}{\psi(t)}}\right] - \tfeps[\lambda_0 + \lambda_T\cos(\omega T)]
\end{equation}
The first term in the RHS of Eq.~\eqref{eq:epswresult} is recognized as the same expression as $\epsw$ in the unconstrained problem, Eq.~\eqref{eq:epswresultunc}. Let us denote:
\begin{equation} \label{eq:epswunc}
	\epswu \equiv \tfeps\mathcal{C}\left[-\Imag{\bracketsO{\chi(t)}{\operator{X}}{\psi(t)}}\right]
\end{equation}
We can rewrite the expression to $\epsw$ in the constrained problem as
\begin{equation}\label{eq:epswresult2}
	\epsw = \epswu - \tfeps[\lambda_0 + \lambda_T\cos(\omega T)]
\end{equation}
Now, in order to find explicit expressions to $\lambda_0$ and $\lambda_T$, we should enforce the boundary constraints, \eqref{eq:eps0d}, \eqref{eq:epsTd}, on the solution \eqref{eq:epswresult2}. We start from the constraint \eqref{eq:eps0d}. It can be rewritten in the terms of $\epsw$ as follows:
\begin{equation} \label{eq:eps0epsw}
	\epsilon(0) = \sqrt{\frac{2}{\pi}}\int_0^\Omega\epsw\cos(0)\,d\omega = \sqrt{\frac{2}{\pi}}\int_0^\Omega\epsw\,d\omega = 0
\end{equation}
Let us substitute the solution \eqref{eq:epswresult2} in the constraint equation:
\begin{equation} \label{eq:lambdas1}
	\epsunc(0) - \left(\lambda_0\,\mathcal{C}^{-1}\left[\tfeps\right]\!\biggm|_{t=0} + \lambda_T\,\mathcal{C}^{-1}\left[\tfeps\right]\!\biggm|_{t=T}\right) = 0
\end{equation}
The constraint \eqref{eq:epsTd} can be written in the terms of $\epsw$ as follows:
\begin{equation} \label{eq:epsTepsw}
	\epsilon(T) = \sqrt{\frac{2}{\pi}}\int_0^\Omega\epsw\cos(\omega T)\,d\omega = 0
\end{equation}
Substituting Eq.~\eqref{eq:epswresult2} in Eq.~\eqref{eq:epsTepsw} we obtain:
\begin{equation} \label{eq:lambdas2}
	\epsunc(T) - \left(\lambda_0\,\mathcal{C}^{-1}\left[\tfeps\right]\!\biggm|_{t=T} + \lambda_T\,\mathcal{C}^{-1}\left[\tfeps\cos(\omega T)\right]\!\biggm|_{t=T}\right) = 0
\end{equation}
Eqs.~\eqref{eq:lambdas1}, \eqref{eq:lambdas2}, constitute a two equation system of the two variables $\lambda_0$, $\lambda_T$. Let us denote, for convenience:
\begin{align}
	& a \equiv \mathcal{C}^{-1}\left[\tfeps\right]\!\biggm|_{t=0} \label{eq:adefd}\\
	& b \equiv \mathcal{C}^{-1}\left[\tfeps\right]\!\biggm|_{t=T} \label{eq:bdefd}\\
	& d \equiv \mathcal{C}^{-1}\left[\tfeps\cos(\omega T)\right]\!\biggm|_{t=T} \label{eq:ddefd}
\end{align}
The system of equations can be written in a matrix-vector form in the following way:
\begin{equation}
	\begin{bmatrix}
		a & b \\
		b & d
	\end{bmatrix}
	\begin{bmatrix}
		\lambda_0 \\
		\lambda_T
	\end{bmatrix}
	=
	\begin{bmatrix}
		\epsunc(0) \\
		\epsunc(T)
	\end{bmatrix}
\end{equation}
Let us define:
\begin{equation}
	M \equiv
	\begin{bmatrix}
		a & b \\
		b & d
	\end{bmatrix}
\end{equation}
The solution of this system is
\begin{align}
	&\lambda_0 = \frac{\epsunc(0)d - \epsunc(T)b}{\det M} \label{eq:lambda0d}\\
	&\lambda_T = \frac{\epsunc(T)a - \epsunc(0)b}{\det M} \label{eq:lambdaTd}
\end{align}
where \text{$\det M = ad-b^2$} is the determinant of $M$.

Note that the derivation of Eqs.~\eqref{eq:epswresult2}, \eqref{eq:lambda0d}, \eqref{eq:lambdaTd}, does not require any knowledge of the forms of the other terms in the functional ($J_{max}$ and $J_{ion}$), which do not have an explicit dependence on $\epsw$. Thus, the same technique of imposing frequency restrictions and boundary constraints can be used directly for various control problems, without altering the expression of the field.

Eq.~\eqref{eq:epswresult2} can be utilized to produce a field which satisfies the boundary condition constraints, \eqref{eq:eps0d}, \eqref{eq:epsTd}, from an arbitrary unconstrained field. The given unconstrained field can be substituted in Eq.~\eqref{eq:epswresult2} as $\epswu$, even though it was not produced by Eq.~\eqref{eq:epswunc}. The derivation of the expressions of the $\lambda$'s does not rely on any previous knowledge on $\epswu$. Thus, it holds in general for any unconstrained field. If, furthermore, the given $\epswu$ is restricted to the required frequency region which is represented by $\tfeps$, then $\epsw$ becomes also restricted to the required frequency region. This technique can be used to produce a guess field which satisfies both the boundary conditions and the frequency requirements.

We proceed to the condition \eqref{eq:dJdpsitkd}. Let us derive the LHS of this equation:
\begin{align}
	\fnlderiv{J}{\ket{\psi(t)}} &= \fnlderiv{J_{max}}{\ket{\psi(t)}} + \fnlderiv{J_{Schr}}{\ket{\psi(t)}} \label{eq:dJdpsicomp}\\
	\fnlderiv{J_{max}}{\ket{\psi(t)}} &= \frac{1}{2}\int_0^\Omega \fC\deriv{\overline{\exval{\operator{C}}}^2\!(\omega)}{\Cw}\,\fnlderiv{\Cw}{\exval{\opC}\!(t)}\,\fnlderiv{\exval{\opC}\!(t)}{\ket{\psi(t)}}\,d\omega \nonumber \\
	&= \sqrt{\frac{2}{\pi}}\int_0^\Omega \fC\Cw\cos(\omega t)\,d\omega \bra{\psi(t)}\opC \nonumber \\
	&= \mathcal{C}^{-1}\left[\fC\Cw\right]\bra{\psi(t)}\opC \label{eq:dJmaxdpsi}\\
	\fnlderiv{J_{Schr}}{\ket{\psi(t)}} = & \deriv{\bra{\chi(t)}}{t} + \bra{i\operator{H}^\dagger(t)\chi(t)} \label{eq:dJSchrdpsi}
\end{align}
Eq.~\eqref{eq:dJdpsitkd} becomes
\begin{equation}\label{eq:ihSchrbra}
	\deriv{\bra{\chi(t)}}{t} =  -\bra{i\operator{H}^\dagger(t)\chi(t)} - \bra{\psi(t)}\mathcal{C}^{-1}\left[\fC\Cw\right]\opC
\end{equation}
Following analogous steps, the condition~\eqref{eq:dJdpsitbd} yields:
\begin{equation}\label{eq:ihSchrket}
	\deriv{\ket{\chi(t)}}{t} = -i\operator{H}^\dagger(t)\ket{\chi(t)} - \mathcal{C}^{-1}\left[\fC\Cw\right]\opC\ket{\psi(t)}
\end{equation}
It can be seen that $\bra{\chi(t)}$ is subject to the adjoint evolution equation of $\ket{\chi(t)}$.

Let us derive the LHS of the condition \eqref{eq:dJdpsiTkd}:
\begin{align}
	\fnlderiv{J}{\ket{\psi(T)}} &= \fnlderiv{J_{ion}}{\ket{\psi(T)}} + \fnlderiv{J_{Schr}}{\ket{\psi(T)}} \\
	\fnlderiv{J_{ion}}{\ket{\psi(T)}} &= \deriv{\left[\sigma\left(\brackets{\psi(T)}{\psi(T)}\right)\right]}{\brackets{\psi(T)}{\psi(T)}}\, \fnlderiv{\brackets{\psi(T)}{\psi(T)}}{\ket{\psi(T)}} \nonumber \\
	&= \sigma'\left(\brackets{\psi(T)}{\psi(T)}\right)\bra{\psi(T)} \\
	\fnlderiv{J_{Schr}}{\ket{\psi(T)}} &= -\bra{\chi(T)}
\end{align}
Eq.~\eqref{eq:dJdpsiTkd} becomes
\begin{equation}\label{eq:chiTbra}
	\bra{\chi(T)} = \sigma'\left(\brackets{\psi(T)}{\psi(T)}\right)\bra{\psi(T)}
\end{equation}
Following analogous steps, the condition~\eqref{eq:dJdpsiTbd} yields:
\begin{equation}\label{eq:chiTket}
	\ket{\chi(T)} = \sigma'\left(\brackets{\psi(T)}{\psi(T)}\right)\ket{\psi(T)}	
\end{equation}
From Eqs.~\eqref{eq:chiTbra}, \eqref{eq:chiTket}, we have:
\begin{equation}\label{eq:chiTbraket}
	\bra{\chi(T)} = \ket{\chi(T)}^\dagger
\end{equation}
The evolution equations \eqref{eq:ihSchrbra}, \eqref{eq:ihSchrket}, together with Eq.~\eqref{eq:chiTbraket}, lead conclusively to Eq.~\eqref{eq:chibraket}, as expected. Assuming this, only Eqs.~\eqref{eq:ihSchrket}, \eqref{eq:chiTket} are required to perform the computations in practice.

We collect the equations which define $\epsw$, $\ket{\psi(t)}$ and $\ket{\chi(t)}$, \ie Eqs.~\eqref{eq:Schrd}, \eqref{eq:initcond}, \eqref{eq:ihSchrket}, \eqref{eq:chiTket}, \eqref{eq:Hd}, \eqref{eq:epswresult2}, \eqref{eq:epswunc}, \eqref{eq:lambda0d}, \eqref{eq:lambdaTd}, \eqref{eq:adefd}, \eqref{eq:bdefd}, \eqref{eq:ddefd}: 
\begin{align}
	\deriv{\ket{\psi(t)}}{t} &= -i\operator{H}(t)\ket{\psi(t)}, \nonumber \\
	& \ket{\psi(0)} = \ket{\psi_0} \label{eq:Schr_sum} \\
	\deriv{\ket{\chi(t)}}{t} &= -i\operator{H}^\dagger(t)\ket{\chi(t)} - \mathcal{C}^{-1}\left[\fC\Cw\right]\opC\ket{\psi(t)}, \nonumber \\
	&\ket{\chi(T)} = \sigma'\left(\brackets{\psi(T)}{\psi(T)}\right)\ket{\psi(T)}\label{eq:ihSchr_sum} \\
	\operator{H}(t) &= \operator{H}_0 - \operator{X}\epsilon(t) \label{eq:Hsum}\\
	\epsilon(t) &= \mathcal{C}^{-1}[\epsw] \label{eq:epst_sum}\\
	\epsw &= \epswu - \tfeps[\lambda_0 + \lambda_T\cos(\omega T)] \label{eq:epsw_sum}\\
	&\epswu \equiv \tfeps\mathcal{C}\left[-\Imag{\bracketsO{\chi(t)}{\operator{X}}{\psi(t)}}\right] \label{eq:epswunc_sum}\\
	&\epsunc(t) = \mathcal{C}^{-1}[\epswu] \label{eq:epsunc_sum}\\
	&\lambda_0 = \frac{\epsunc(0)d - \epsunc(T)b}{ad - b^2} \\
	&\lambda_T = \frac{\epsunc(T)a - \epsunc(0)b}{ad - b^2} \\
	& a \equiv \mathcal{C}^{-1}\left[\tfeps\right]\!\biggm|_{t=0} \\
	& b \equiv \mathcal{C}^{-1}\left[\tfeps\right]\!\biggm|_{t=T} \\
	& d \equiv \mathcal{C}^{-1}\left[\tfeps\cos(\omega T)\right]\!\biggm|_{t=T}\label{eq:dsum}
\end{align}
These constitute the Euler-Lagrange equations of the problem.

\section{Numerical details}\label{app:num_details}

\subsection{Discretization of the cosine transform}\label{ssec:dct}

In practice, the cosine transform \eqref{eq:costrans} is replaced by a 
\emph{discrete-cosine-transform} (DCT). There are several types of DCT's. We used a boundary including DCT, which is sometimes referred to as the \emph{DCT of the first type} (DCT-1). The DCT-1 of a time-dependent function, $g(t)$, sampled at \text{$N_t + 1$} equidistant time points,
\[
	t_k, \qquad k=0,1,\ldots,N_t
\]
is defined as:
\begin{align}
	& \bar{g}(\omega_j) = \sqrt{\frac{2}{N_t}}\sum_{k=0}^{N_t}\frac{1}{h_k} g(t_k)\cos\left(\frac{jk\pi}{N_t}\right), & &j=0,1,\ldots,N_t \label{eq:DCT}\\
	& h_k =
	\begin{cases}
		2 & \qquad k=0 \text{ or } k=N_t \\
		1 & \qquad 1\leq k\leq N_t-1
	\end{cases} \label{eq:hdef}
\end{align}
The inverse transform is defined by
\begin{align}
	& g(t_k) = \sqrt{\frac{2}{N_t}}\sum_{j=0}^{N_t}\frac{1}{h_j} \bar{g}(\omega_j)\cos\left(\frac{jk\pi}{N_t}\right), & & k=0,1,\ldots,N_t \label{eq:iDCT}
\end{align}
It can be observed from Eqs.~\eqref{eq:DCT}, \eqref{eq:iDCT} that DCT-1 is its own inverse.

The argument of the cosine function in Eqs.~\eqref{eq:DCT}, \eqref{eq:iDCT} is equivalent to a discretized variant of the continuous argument $\omega t$ in the continuous transform \eqref{eq:costrans}, where the variables are discretized as follows:
\begin{align}
	& t_k = k\frac{T}{N_t} &  k=0,1,\ldots,N_t \label{eq:t_samp}\\
	& \omega_j = j\frac{\pi}{T} & j=0,1,\ldots,N_t \label{eq:w_samp}
\end{align}

In order to be consistent with the continuous formulation, the direct transform was multiplied by the factor $T/\sqrt{N_t\pi}$:
\begin{equation}\label{eq:DCTdirect}
	\bar{g}(\omega_j) = \sqrt{\frac{2}{\pi}}\frac{T}{N_t}\sum_{k=0}^{N_t}\frac{1}{h_k} g(t_k)\cos\left(\frac{jk\pi}{N_t}\right)
\end{equation}
where $T/N_t$ replaces $dt$ in the continuous integral form (Cf.~Eq.~\eqref{eq:t_samp}). The inverse transform was divided by the same factor:
\begin{align}
	g(t_k) &= \frac{\sqrt{2\pi}}{T}\sum_{j=0}^{N_t}\frac{1}{h_j} \bar{g}(\omega_j)\cos\left(\frac{jk\pi}{N_t}\right) \nonumber \\
	&= \sqrt{\frac{2}{\pi}}\frac{\pi}{T}\sum_{j=0}^{N_t}\frac{1}{h_j} \bar{g}(\omega_j)\cos\left(\frac{jk\pi}{N_t}\right) \label{eq:DCTinv}
\end{align}
where $\pi/T$ replaces $d\omega$ in the continuous integral form (Cf.~Eq.~\eqref{eq:w_samp}). The consistency with the continuous formulation has an important advantage: When using Eq.~\eqref{eq:DCT} as is, the definition of the spectral function $\bar{g}(\omega)$, represented by the transform, varies with the sampling frequency. The consistency with the continuous form makes $\bar{g}(\omega)$ independent of the sampling. 

The choice of this type of DCT is important for the following reason: The boundary constraints, \eqref{eq:eps0d}, \eqref{eq:epsTd}, are imposed at the boundaries of the time-domain, \ie at \text{$t=0$} and \text{$t=T$}. Consequently, $\epsunc(t)$ and the inverse cosine transform in Eqs.~\eqref{eq:lambdas1}, \eqref{eq:lambdas2}, are evaluated at the time-domain boundaries. Therefore, it is important that the discrete transform will include both boundaries. Moreover, the Euler-Lagrange equations define a \emph{two-point boundary value problem}: $\ket{\psi(t)}$ is propagated from the initial condition at \text{$t=0$}, and $\ket{\chi(t)}$ is backward propagated from the final condition at \text{$t=T$}. Thus, the boundary including DCT is compatible with the control equations.

In this context, we shall give a practical remark on the computation of $d$ in Eqs.~\eqref{eq:lambda0d}, \eqref{eq:lambdaTd}. $d$ is computed in the DCT-1 form as
\begin{align}
	d &= \frac{\sqrt{2\pi}}{T}\sum_{j=0}^{N_t}\frac{1}{h_j}\tilde{f}_\epsilon(\omega_j)\cos^2(\omega_j T) & \omega_j = j\frac{\pi}{T} \nonumber \\
	&= \frac{\sqrt{2\pi}}{T}\sum_{j=0}^{N_t}\frac{1}{h_j}\tilde{f}_\epsilon\!\left(\frac{j\pi}{T}\right)\cos^2(j\pi) \nonumber \\			
	&= \frac{\sqrt{2\pi}}{T}\sum_{j=0}^{N_t}\frac{1}{h_j}\tilde{f}_\epsilon\!\left(\frac{j\pi}{T}\right)(-1)^2 \nonumber \\
		&= \frac{\sqrt{2\pi}}{T}\sum_{j=0}^{N_t}\frac{1}{h_j}\tilde{f}_\epsilon\!\left(\frac{j\pi}{T}\right)
\end{align}
which is equivalent to the discretized expression to $a$. Thus, Eq.~\eqref{eq:lambda0d} can be computed as
\begin{equation}
	\lambda_0 = \frac{\epsunc(0)a - \epsunc(T)b}{\det M}
\end{equation}
where $\det M$ can be computed as
\begin{equation}\label{eq:detMsimp}
	\det M = a^2 - b^2
\end{equation}
The $\det M$ in Eq.~\eqref{eq:lambdaTd} is also computed by Eq.~\eqref{eq:detMsimp}.

\subsection{Details of the optimization procedure}\label{app:BFGS}

As was mentioned in Sec.~\ref{sec:num}, the optimization procedure is based on the BFGS method (see, for example, \cite[Chapter~3]{Fletcher}). The BFGS method is a universal method of optimization which is well established and widespread. However, several details are crucial for a successful implementation of the method in the present context. Commercial programs might completely fail when they are used without certain modifications.

\subsubsection{Discretization of the optimization space}

The optimization is performed in a discrete variable space. Thus, we first have to reformulate the problem by discrete means in order to obtain the necessary expressions for the optimization process.

The optimized function, $\epsw$, is discretized as follows:
\begin{equation}\label{eq:epsw_discrete}
	\epsw\longrightarrow
	\epswv \equiv
	\begin{bmatrix}
		\bar\epsilon(\omega_0)\\
		\bar\epsilon(\omega_1)\\
		\vdots\\
		\bar\epsilon(\omega_{N_t})
	\end{bmatrix}\qquad
	\omega_j = j\frac{\pi}{T} 
\end{equation}
The $\omega$ sampling is the same as the DCT sampling (Cf.~Eq.~\eqref{eq:w_samp}). The functional has to be expressed by the terms of the discrete optimization space:
\begin{equation}
	J\left[\epsw\right]\longrightarrow J^d\left[\epswv\right]
\end{equation}
where $J^d$ is a discretized variant of $J$. The $\epsw$ dependence in $J$ is reformulated by discrete means; $J_{energy}$ is replaced by
\begin{equation}\label{eq:Jenergy_discrete}
	J_{energy}^d = -\sum_{j=0}^{N_t}\frac{1}{\tfepsj}\bar\epsilon^2(\omega_j)w_j, \qquad w_j = \frac{1}{h_j}\frac{\pi}{T}
\end{equation}
where the $h_j$'s are defined by Eq.~\eqref{eq:hdef}. The $w_j$'s are the \emph{integration weights}. The $\epsw$ dependence of $\epsilon(t)$ in $\Jepsz$, $\JepsT$ and $J_{Schr}$ is expressed by the inverse DCT expression (Cf.~Eq.~\eqref{eq:DCTinv}):
\begin{align}
	\epsilon(t) &= \sqrt{\frac{2}{\pi}}\frac{\pi}{T}\sum_{j=0}^{N_t}\frac{1}{h_j} \bar{\epsilon}(\omega_j)\cos\left(\omega_j t\right) \nonumber \\
	&= \sqrt{\frac{2}{\pi}}\sum_{j=0}^{N_t}\bar{\epsilon}(\omega_j)\cos\left(\omega_j t\right)w_j \label{eq:epst_discrete}
\end{align}
The resulting gradient expression is
\begin{equation}\label{eq:grad_discrete}
	\deriv{J^d}{\bar{\epsilon}(\omega_j)} = w_j\fnlderat{J}{\epsw}{\omega=\omega_j}
\end{equation}
where $\fnlderiv{J}{\epsw}$ is given by Eq.~\eqref{eq:grad_functional}.

\subsubsection{Definition of the numerical optimization problem}\label{ssec:BFGSdef}

Typically, the space of $\epswv$ is much larger than what is required in practice. The reason lies in the form of Eq.~\eqref{eq:epswresult}---it can be easily seen that $\epsw$ attains negligible values in the regions in which the filter function $\feps$ is negligible. Hence, we can neglect the terms of $\epswv$ in these regions and fix them to zero in advance. Only the terms in the non-negligible regions participate in the optimization process. This results in a dramatic decrease of the dimension of the optimization space. This practice is important also for another reason---it solves the problem of division by zero in the first term in the RHS of Eq.~\eqref{eq:grad_functional}.

Let us denote the vector of the $\omega_j$ values which participate in the optimization by $\vomr$. In our calculations, we have chosen $\vomr$ as to satisfy the following condition:
\begin{equation}
	f_{\epsilon}(\vomr)\geq 2.22\times 10^{-16}
\end{equation}
The value in the RHS represents the machine precision of double type variables.

For convenience, let us define the solution vector in the reduced space:
\begin{equation}
	\epswr \equiv \bar\epsilon(\vomr)
\end{equation}

We naturally formulated our optimization problem as a \emph{maximization} problem. This is also the common convention in quantum OCT texts for the vast majority of the optimization problems. However, the BFGS method is traditionally formulated as a \emph{minimization} process, which is the common convention in the optimization literature in general. In order to apply the BFGS method to our problem, we have to use a uniform formulation. There exist two options:
\begin{enumerate}
	\item The BFGS method can be reformulated as a maximization process.
	\item We can perform a minimization of $-J^d$ instead of the maximization of $J^d$.
\end{enumerate}
We adopt the second option, in order to prevent confusion, and to enable a direct use of existing BFGS codes.

In summary, our optimization problem is defined as the minimization of the objective $-J^d$ with respect to the solution vector $\epswr$.

We denote the gradient vector of the \emph{minimization problem} as $\vg$. Its general term is given by
\begin{align}
	-\deriv{J^d}{\bar{\epsilon}(\omega_j)} &= -w_j\fnlderat{J}{\epsw}{\omega=\omega_j} \nonumber \\
	&= 2w_j\left(\frac{\epswj}{\tfepsj} - \mathcal{C}\left[-\Imag{\bracketsO{\chi(t)}{\operator{X}}{\psi(t)}}\right]\biggm|_{\omega=\omega_j} + \lambda_0 + \lambda_T\cos(\omega_j T)\right) \label{eq:grad_optim}
\end{align}
$\vg$ can be expressed by a vector notation in the following way:
\begin{equation}\label{eq:vgrad}
	\vg = 2\vw\circ\left(\epswr\oslash\tilde{f}_{\epsilon}(\vomr) - \mathcal{C}\left[-\Imag{\bracketsO{\chi(t)}{\operator{X}}{\psi(t)}}\right]\biggm|_{\vomega=\vomr} + \lambda_0 + \lambda_T\cos(\vomr T)\right)
\end{equation}
where $\vw$ is the weight vector. $\circ$ denotes the Hadamard product (elementwise product) and $\oslash$ the Hadamard division.

The approximated Hessian matrix will be denoted by $S$.


\subsubsection{Iterative formulation}

The optimization process is mathematically described by an iterative \emph{update rule} for the solution vector $\epswr$. This requires the formulation of the optimization equations by iterative means.

First, we have to index the iteration number of the mathematical objects involved in the iterative process. The iteration index will be denoted by a superscript in parentheses. For example, the solution vector in the $k$'th iteration is denoted by $\epswr^{(k)}$. $\epswr^{(0)}$ denotes the initial guess solution.

$\epsilon^{(k)}(t)$ is given by the substitution of $\epswr^{(k)}$ into Eq.~\eqref{eq:epst_discrete}. $\ket{\psi^{(k)}(t)}$ and $\ket{\chi^{(k)}(t)}$ are given by Eqs.~\eqref{eq:Schr_sum}--\eqref{eq:Hsum} with the substitution of $\epsilon^{(k)}(t)$ into the Hamiltonian (Eq.~\eqref{eq:Hsum}). $\lambda_0^{(k)}$ and $\lambda_T^{(k)}$ are given by Eqs.~\eqref{eq:epswunc_sum}--\eqref{eq:dsum}, where Eq.~\eqref{eq:epswunc_sum} is computed by the substitution of $\ket{\psi^{(k)}(t)}$ and $\ket{\chi^{(k)}(t)}$.

The objective of the minimization process in the $k$'th iteration is given by $-J^{d\,(k)}$, where $J^{d\,(k)}$ is computed by $\epswr^{(k)}$ and $\ket{\psi^{(k)}(t)}$. Note that the Lagrange-multiplier terms, $\Jepsz$, $\JepsT$ and $J_{Schr}$, are identically zero; thus, the Lagrange-multipliers do not participate in the computation of $J^{d\,(k)}$.

The gradient in the $k$'th iteration is given by the following expression (Cf.~Eq.~\eqref{eq:vgrad}):
\begin{equation}\label{eq:vgradk}
	\vg^{(k)} = 2\vw\circ\left(\epswr^{(k)}\oslash\tilde{f}_{\epsilon}(\vomr) - \mathcal{C}\left[-\Imag{\bracketsO{\chi^{(k)}(t)}{\operator{X}}{\psi^{(k)}(t)}}\right]\biggm|_{\vomega=\vomr} + \lambda_0^{(k)} + \lambda_T^{(k)}\cos(\vomr T)\right)
\end{equation}


Let us define:
\begin{align}
	\vdelta^{(k)} &\equiv \epswr^{(k+1)} - \epswr^{(k)} \\
	\vgamma^{(k)} &\equiv \vg^{(k+1)} - \vg^{(k)}
\end{align}
The BFGS update rule for the approximated inverse-Hessian is
\begin{equation}\label{eq:invHess_update}
	{S^{(k+1)}}^{-1} = {S^{(k)}}^{-1} + \left(1 + \frac{\vgamma^{(k)\,T}{S^{(k)}}^{-1}\vgamma^{(k)}}{\vdelta^{(k)\,T}\vgamma^{(k)}}\right) \frac{\vdelta^{(k)}\vdelta^{(k)\,T}}{\vdelta^{(k)\,T}\vgamma^{(k)}} - \frac{\vdelta^{(k)}\vgamma^{(k)\,T}{S^{(k)}}^{-1} + {S^{(k)}}^{-1}\vgamma^{(k)}\vdelta^{(k)\,T}}{\vdelta^{(k)\,T}\vgamma^{(k)}}
\end{equation}
The initial inverse-Hessian guess, ${S^{(0)}}^{-1}$, is supplied by the user.

The direction of search in the $k$'th iteration will be denoted by $\vp^{(k)}$. It is given by the general quasi-Newton expression:
\begin{equation}\label{eq:pk}
	\vp^{(k)} = -{S^{(k)}}^{-1}\vg^{(k)}
\end{equation}

The update rule for the solution vector is
\begin{equation}\label{eq:update}
	\epswr^{(k+1)} = \epswr^{(k)} + \kappa^{(k)}\vp^{(k)}, \qquad \kappa^{(k)}>0
\end{equation}
where $\kappa^{(k)}$ is determined by the \emph{line-search} procedure. The line-search requires the knowledge of $\epswr^{(k)}$, $\vp^{(k)}$, $-J^{d\,(k)}$ and $\vg^{(k)}$. It also requires a procedure which returns $-J^d$ and $\vg$ for any solution vector $\epswr$ which is a candidate for $\epswr^{(k+1)}$.

Note that the only Euler-Lagrange equation which does not participate in the iterative calculation is the equation for the solution $\epsw$ \eqref{eq:epsw_sum}. Nevertheless, the solution equation has been utilized in the considerations which led to the reduction of the optimization space (Sec.~\ref{ssec:BFGSdef}). In what follows, it turns out that the solution equation form is actually contained implicitly in the optimization procedure.

The termination condition of the iterative process is chosen as
\begin{equation}\label{eq:termination}
	\frac{\norm{\epswr^{(k)} - \epswr^{(k-1)}}}{\norm{\epswr^{(k)}}}\leq 10^{-4}
\end{equation}

%
%

\subsubsection{Initial guess for the solution vector and the Hessian}\label{sssec:iguess}

The update equation \eqref{eq:update} presents a problem: The RHS consists of two terms. The first is the previous solution $\epswr^{(k)}$, which is assumed to satisfy the boundary constraints, \eqref{eq:eps0d} and \eqref{eq:epsTd}. However, the second term $-\kappa^{(k)}{S^{(k)}}^{-1}\vg^{(k)}$ does not satisfy the boundary constraints in general. As a result, the update rule does not conserve the boundary constraints.

Another property which is not conserved by the update rule is the restriction of the driving field spectrum. The direction of search $\vp^{(k)}$ is not confined, in general, to the allowed frequency region. As an illustration of this statement, let us take \text{$S^{(k)}=I$}, where $I$ is the identity matrix. The identity matrix is usually chosen as the initial guess for the Hessian in the BFGS process. With this choice, the direction of search becomes \text{$\vp^{(k)}=-\vg^{(k)}$}, as in the first order gradient method. It can be readily observed from Eq.~\eqref{eq:grad_optim} that the gradient direction does not satisfy the spectral restriction. The terms $\lambda_0$, $\lambda_T\cos(\omega_j T)$ obviously have significant values throughout the spectrum of $\omega_j$ values. Actually, the different terms in parenthesis in Eq.~\eqref{eq:grad_optim} can be recognized as the same terms as in both sides of Eq.~\eqref{eq:epswresult}, \emph{divided by $\tfeps$}. The division by the filter function simply \emph{cancels the filtration} which is present in Eq.~\eqref{eq:epswresult}. Thus, the gradient expression is not restricted to the required  spectral region, represented by $\feps$.

When $\vp^{(k)}$ is not spectrally restricted, the optimization process cannot proceed; very large penalties are put on the ``forbidden'' frequency regions, and thus the result cannot improve unless $\kappa^{(k)}$ is extremely small. Consequently, the optimization process becomes highly ineffective and impractical. Moreover, the update of $\epswr$ is often so small that it does not change beyond the roundoff error regime.

Note that the conservation problems of the update rule originate from the form \eqref{eq:update}, which determines $\epswr$ independently of the Euler-Lagrange equation for the solution, \eqref{eq:epsw_sum}. These problems do not exist in optimization methods in which the update rule is based directly on the solution equation. An example of a method of this class is the relaxation process proposed in our previous work (Ref.~\cite[Chapter~3.2.3]{thesis}, \cite{OCTHG}). An update rule of this type satisfies automatically the boundary and spectral restrictions which are present in Eq.~\eqref{eq:epsw_sum}.

The conservation problems in the BFGS process can be solved by an appropriate choice of the \emph{initial guess solution} $\epswr^{(0)}$ and the \emph{initial Hessian} $S^{(0)}$. First, $\epswr^{(0)}$ should satisfy the boundary conditions and the spectral restrictions. A guess solution which satisfies both requirements can be constructed by the following steps:
\begin{enumerate}
	\item Choose a spectral function $\bar{\epsilon}_{unc}^{(0)}(\omega)$ which satisfies the spectral restrictions, but unnecessarily the boundary constraints.
	\item Obtain a constrained spectral function $\bar{\epsilon}^{(0)}(\omega)$ by substituting $\bar{\epsilon}_{unc}^{(0)}(\omega)$ into Eqs.~\eqref{eq:epsw_sum}, \eqref{eq:epsunc_sum}--\eqref{eq:dsum} (the justification to this practice is explained in Appendix~\ref{app:der}).
	\item $\epswr^{(0)}$ is simply given by the terms of $\bar{\epsilon}^{(0)}(\omega)$ which participate in the optimization procedure.
\end{enumerate}
Note that if \text{$\bar{\epsilon}_{unc}^{(0)}(\omega)\propto\tfeps$}, then Eq.~\eqref{eq:epsw_sum} yields \text{$\bar{\epsilon}^{(0)}(\omega)\equiv 0$}. This initial field is not very useful---there seems to be a local minimum or a saddle point in the zero field solution. Consequently, the optimization procedure cannot proceed from this point in the optimization space.

$S^{(0)}$ should be chosen as the Hessian of $-J_{energy}^d$ (see Eq.~\eqref{eq:Jenergy_discrete}). The general term of the Hessian is given by
\begin{equation}\label{eq:Hess0term}
	-\pderivdd{J_{energy}^d}{\epswj}{\bar{\epsilon}(\omega_l)} = \delta_{jl}\frac{2w_j}{\fepsj}
\end{equation}
Thus, $S^{(0)}$ is a \emph{diagonal matrix}. It can be written in a matrix notation in the following way:
\begin{equation}\label{eq:Hess0}
	S^{(0)} = -\nabla_{\vomr}(\nabla_{\vomr}J_{energy}^d)^T = 2\,\diag\left[\vw\oslash\tilde{f}_{\epsilon}(\vomr)\right]
\end{equation}
The initial inverse-Hessian matrix is given by
\begin{equation}\label{eq:invHess0}
	{S^{(0)}}^{-1} = \frac{1}{2}\,\diag\left[\tilde{f}_{\epsilon}(\vomr)\oslash\vw\right]
\end{equation}
This choice of $S^{(0)}$ yields the following initial direction of search:
\begin{align}
	\vp^{(0)} &= -{S^{(0)}}^{-1}\vg^{(0)} \nonumber \\
	&= -\epswr^{(0)} + \tilde{f}_{\epsilon}(\vomr)\circ\left\lbrace\mathcal{C}\left[-\Imag{\bracketsO{\chi^{(0)}(t)}{\operator{X}}{\psi^{(0)}(t)}}\right]\biggm|_{\vomega=\vomr} - \left[\lambda_0^{(0)} + \lambda_T^{(0)}\cos(\vomr T)\right]\right\rbrace \label{eq:p0}
\end{align}
Let us write the function version of the second term in the RHS:
\begin{equation*}
	\tfeps\left\lbrace\mathcal{C}\left[-\Imag{\bracketsO{\chi^{(0)}(t)}{\operator{X}}{\psi^{(0)}(t)}}\right] - \left[\lambda_0^{(0)} + \lambda_T^{(0)}\cos(\omega T)\right]\right\rbrace
\end{equation*}
This expression can be recognized as the expression for $\epsw$ from the solution Euler-Lagrange equation (see Eqs.~\eqref{eq:epsw_sum}, \eqref{eq:epswunc_sum}), with the substitution of $\ket{\psi^{(0)}(t)}$, $\ket{\chi^{(0)}(t)}$, $\lambda_0^{(0)}$ and $\lambda_T^{(0)}$. Let us denote:
\begin{align}
	&\epswELk \equiv \tfeps\left\lbrace\mathcal{C}\left[-\Imag{\bracketsO{\chi^{(k)}(t)}{\operator{X}}{\psi^{(k)}(t)}}\right] - \left[\lambda_0^{(k)} + \lambda_T^{(k)}\cos(\omega T)\right]\right\rbrace \\
	&\epswrEL^{(k)} \equiv \bar{\epsilon}_{EL}^{(k)}(\vomr)
\end{align}
Using this notation, Eq.~\eqref{eq:p0} can be rewritten as
\begin{equation}\label{eq:p0EL}
	\vp^{(0)} = -\epswr^{(0)} + \epswrEL^{(0)}
\end{equation}
$\epswrEL^{(k)}$ satisfies the boundary constraints and spectral restrictions for all $k$. $\epswr^{(0)}$ is also assumed to satisfy both requirements by an appropriate choice, as above. According to Eq.~\eqref{eq:p0EL}, $\vp^{(0)}$ is a linear combination of $\epswr^{(0)}$ and $\epswrEL^{(0)}$. Thus, $\vp^{(0)}$ also satisfies both requirements. Consequently, the update rule for $\epswr^{(1)}$ preserves the boundary and spectral restrictions.  

Note that the relaxation method proposed in our previous work yields the same update direction as that of Eq.~\eqref{eq:p0EL}, generalized to all $k$. It has been shown in~\cite[Chapter~3.2.3]{thesis} that the relaxation method is equivalent to a quasi-Newton method with the same approximated Hessian \eqref{eq:Hess0} for all iterations.

The situation is more complicated in the BFGS method, where the Hessian is updated in each iteration. Nevertheless, the BFGS update rule preserves the required properties for all $k$, with the proper choice of $\epswr^{(0)}$ and $S^{(0)}$, as above. It can be shown that the resulting direction of search $\vp^{(k)}$ is spanned by the following set of vectors:
\begin{equation*}
	\epswr^{(0)},\ \epswrEL^{(0)},\ \epswrEL^{(1)},\ldots,\epswrEL^{(k)}
\end{equation*}
The full justification is left for a future publication. All the vectors in this set satisfy the required properties. Consequently, the required properties are preserved by the resulting update rule.

Thus, it has been shown that by a proper choice of $S^{(0)}$, the BFGS method yields an update rule which is based implicitly on the solution equation \eqref{eq:epsw_sum}.

It is noteworthy that there is an important particular case in which the common choice for the initial Hessian, \text{$S^{(0)}=I$}, conserves the required properties. This situation is characterized by the following conditions:
\begin{enumerate}
	\item $\feps\in\{0,1\}$, \ie $\feps$ can attain the values 0 and 1 only. For instance, $\feps$ is a rectangular function.
	\item The boundary frequencies, \text{$\omega_0=0$} and \text{$\omega_{N_t}=N_t\pi/T$}, are not contained in the allowed frequency region represented by $\feps$.
\end{enumerate}
The issue of conservation of the spectral restriction becomes irrelevant whenever the first condition is satisfied, since the ``forbidden'' spectral regions are completely eliminated from the optimization process in advance. If, in addition, the second condition is satisfied, ${S^{(0)}}^{-1}$ from Eq.~\eqref{eq:invHess0} differs from the identity matrix $I$ just by a constant factor. {This contributes a constant factor to $\vp^{(0)}$, which does not affect its direction. Thus,} the common choice \text{$S^{(0)}=I$} yields an update rule which conserves the boundary conditions as well.

\subsubsection{The line-search}

The line-search procedure is based on the scheme outlined in Ref.~\cite[Chapter~2.6]{Fletcher}. We shall comment on several details in which we deviated from this scheme.

In~\cite[Chapter~2.6]{Fletcher} it is assumed that the computation of the gradient is much more expensive numerically than the computation of the optimized function value. This assumption is reflected in the details of the line-search scheme---the gradient is computed only when it is essential for the search process. This excludes a certain situation in which the gradient information is not essential, although its knowledge improves the search process---the choice of the next point in the search is based on interpolation considerations; in the scheme outlined in~\cite[Chapter~2.6]{Fletcher} it is preferred to employ a quadratic interpolation, rather than a cubic one with the cost of an extra gradient evaluation. In our situation, the economical considerations regarding the gradient computation are irrelevant---the vast majority of the numerical effort consists of the propagation process, and the extra computational effort in the gradient calculation is negligible. Thus, the gradient should be computed whenever it can be utilized for the improvement of the process. Hence, we always employed a cubic interpolation for the choice of the next point in the search.

The conditions for an acceptable point in~\cite[Chapter~2.5]{Fletcher} include the requirement that the point is located below a decreasing ``$\rho$-line'' (the first Goldstein condition). This condition is unnecessary in our case, since it is employed to exclude a situation which is irrelevant in our problem (see~\cite[Page~30]{Fletcher}). In an existing line-search program we can set \text{$\rho=0$}.

It is also unnecessary to include a lower bound for the functional value ($\bar{f}$ in \cite[Chapter~2.6]{Fletcher}) from similar considerations. In an existing program we can set \text{$\bar{f}=-\infty$}.

There is a unique issue in quantum optimal control theory problems which requires a specialized treatment in the context of the line-search scheme. The computation of the functional involves propagation under the solution vector field. Typically, the precise description of the dynamics becomes more demanding numerically as the magnitude of the field increases. Of course, the high numerical requirements cannot be avoided if the optimal field is actually large in magnitude. However, it may occur that the optimization search path passes through fields which are much larger in magnitude than the optimal field. Thus, it is desirable to control the optimization path such that large field regions are avoided.

In the context of the line-search process, it often happens that the bracket located in the bracketing phase is much larger than required. In other words, the search unnecessarily extends ``too far'' in the search direction $\vp$. The corresponding fields might be considerably larger than the optimal field. As a result, the propagation process might become very time-consuming, inaccurate, or unstable.

We addressed this problem in a primitive way---we restricted the peak magnitude of the field which is allowed in the line-search process. Let us denote the maximal allowed magnitude of $\epsilon(t)$ by $\epsilon_{max}$; the updated field in each iteration should satisfy the following condition:
\begin{equation}
	\abs{\epsilon^{(k+1)}(t)} \leq \epsilon_{max}, \qquad t\in[0, T]
\end{equation}
which is equivalent to the following set of conditions:
\begin{align}
	& \epsilon^{(k+1)}(t) \leq \epsilon_{max} \label{eq:epsmax}\\
	& \epsilon^{(k+1)}(t) \geq -\epsilon_{max} \label{eq:minus_epsmax}\\
	& t\in[0, T] \nonumber
\end{align}
These conditions yield a condition on the maximal allowed $\kappa^{(k)}$ in the search, as will be readily shown. First, conditions \eqref{eq:epsmax}, \eqref{eq:minus_epsmax} have to be discretized as follows:
\begin{align}
	& \veps^{(k+1)} \leq \epsilon_{max} \label{eq:epsmax_discrete}\\
	& \veps^{(k+1)} \geq -\epsilon_{max} \label{eq:minus_epsmax_discrete}
\end{align}
where we defined:
\begin{equation}
	\veps \equiv
	\begin{bmatrix}
		\epsilon(t_0)\\
		\epsilon(t_1)\\
		\vdots\\
		\epsilon(t_{N_t})
	\end{bmatrix}
\end{equation}
$\veps^{(k+1)}$ can be expressed in the terms of the discretized solution $\epswv$:
\begin{equation}\label{eq:idct_veps}
	\veps^{(k+1)} = C^{-1}\epswv^{(k+1)}
\end{equation}
where $C$ denotes a matrix which represents the DCT linear transformation defined by Eq.~\eqref{eq:DCTdirect}. Accordingly, $C^{-1}$ represents the inverse DCT transformation defined by Eq.~\eqref{eq:DCTinv}. In order to express $\epswv^{(k+1)}$ by the terms of $\kappa^{(k)}$, the update rule \eqref{eq:update} has to reformulated in the full \text{$N_t+1$} dimensional space in which $\epswv^{(k+1)}$ is defined. Let us define for $\vp^{(k)}$ a corresponding vector $\vp_f^{(k)}$ in the full space. The update rule in the full space becomes
\begin{equation}\label{eq:update_full}
	\epswv^{(k+1)} = \epswv^{(k)} + \kappa^{(k)}\vp_f^{(k)}
\end{equation}
Eqs.~\eqref{eq:idct_veps}, \eqref{eq:update_full} yield:
\begin{equation}\label{eq:update_veps}
	\veps^{(k+1)} = C^{-1}\epswv^{(k)} + \kappa^{(k)}C^{-1}\vp_f^{(k)} = \veps^{(k)} + \kappa^{(k)}\vq^{(k)}
\end{equation}
where
\begin{equation}
	\vq^{(k)} \equiv C^{-1}\vp_f^{(k)}
\end{equation}
It can be easily found that the substitution of Eq.~\eqref{eq:update_veps} into conditions \eqref{eq:epsmax_discrete}, \eqref{eq:minus_epsmax_discrete} yields the following set of \text{$N_t+1$} conditions on $\kappa^{(k)}$:
\begin{equation}\label{eq:kappa_cond}
	\kappa^{(k)} \leq \frac{\epsilon_{max} - \sgn\left(q_{j+1}^{(k)}\right)\epsilon^{(k)}(t_j)}{\abs{q_{j+1}^{(k)}}}, \qquad  0 \leq j \leq N_t
\end{equation}
where $q_{j+1}^{(k)}$ is the $(j+1)$'th entry of $\vq^{(k)}$, which corresponds to the time-point $t_j$.

This practice should be distinguished from optimization with inequality constraints. Conditions \eqref{eq:epsmax}, \eqref{eq:minus_epsmax} do not represent an additional \emph{constraint} on the \emph{optimization problem}; they just impose a restriction on the \emph{search process}.

When condition~\eqref{eq:kappa_cond} is imposed, it may occur that the bracketing phase fails. Usually, this indicates that the chosen $\epsilon_{max}$ is too small for the given problem, and it should be increased. Alternatively, the problem can be altered by the increment of the penalties on energy or ionization. These modifications will result in an optimal field of smaller magnitude.

The parameter values of the line-search procedure are summarized in Table~\ref{tab:ls_params}. We use the notations of \cite[Chapter~2.6]{Fletcher}, with the exception of the parameter $\kappa^{(0)}$, which is defined in the present text.

\begin{table}
\begin{equation*}
	\renewcommand{\arraystretch}{1.5}
	\begin{array}{|c||c|}
		\hline
		\sigma & 0.9 \\ \hline
		\rho & 0 \\ \hline
		\tau_1 & 9 \\ \hline
		\tau_2 & 0.1 \\ \hline
		\tau_3 & 0.5 \\ \hline
		\bar{f} & -\infty \\ \hline
		\kappa^{(0)} & 1 \\ \hline
		\epsilon_{max} & 0.15 \\ \hline
	\end{array}
\end{equation*}
\caption{Line-search parameters}\label{tab:ls_params}
\end{table}


\subsubsection{Practical remarks}\label{sssec:practical}

The choice of the initial guess field is important for the success of the computational method. It is recommended to choose $\epswr^{(0)}$ such that \text{$J^{d\,(0)}>0$}, for the following reason. As has already been mentioned, there seems to be a local {minimum of $-J^d$}, or a saddle point, in the zero field solution, \text{$\epswr \equiv \bvec{0}$}. This solution yields \text{$J^d=0$}. If {\text{$-J^{d\,(0)}>0$}, or equivalently,} \text{$J^{d\,(0)}<0$}, the optimization process has a strong tendency to converge to the zero field solution.

The task of locating a guess solution which yields \text{$J^{d\,(0)}>0$} often becomes non-trivial. The guess field should produce some response in the required region, such that the magnitude of $J_{max}^{(0)}$ is sufficiently large. One option is to use some physical insight in the choice of $\epswr^{(0)}$ such that the magnitude of $J_{max}^{(0)}$ is significant. Another option is to employ a \emph{preparation optimization}---we perform several iterations of another optimization problem, in order to find a field for which $J_{max}^{(0)}$ becomes significant. In the preparation problem, the penalties on energy and ionization are reduced, such that the initial functional value \emph{of the preparation problem} becomes positive.

Occasionally, a quasi-Newton scheme applied to the present problem fails in the approximation of the Hessian. The optimization ``gets stuck'' after several iterations, when the resulting direction of search $\vp^{(k)}$ ceases to be useful. Theoretically, the BFGS update rule of the inverse-Hessian \eqref{eq:invHess_update} conserves the positive definiteness of $S^{-1}$. According to Eq.~\eqref{eq:pk}, this yields a direction of search $\vp^{(k)}$ which is always a descent direction. However, in practice, the magnitude of the negative slope in the $\vp^{(k)}$ direction may be too small to exceed the roundoff error regime. As a result, no improvement can be achieved in a search in the $\vp^{(k)}$ direction.

The origin of the problem lies in the failure of the assumptions underlying the inverse-Hessian update rule \eqref{eq:invHess_update}. The update rule relies on the following quadratic approximation:
\begin{equation}\label{eq:Squad}
	\vgamma^{(k)} \approx S^{(k+1)}\vdelta^{(k)}
\end{equation}
The quadratic approximation is relevant only when the new point $\epswr^{(k+1)}$ is located in the nearest valley in the $\vp^{(k)}$ direction. 
The situation might be different when there exist additional further valleys in the $\vp^{(k)}$ direction. 
In this case, the accepted $\epswr^{(k+1)}$ in the line-search scheme may be located in these further valleys, where approximation \eqref{eq:Squad} does not hold. Consequently, the update rule \eqref{eq:invHess_update} becomes irrelevant for the approximation of the inverse Hessian. After the accumulation of several events of this type, the resulting ${S^{(k)}}^{-1}$ ceases to be useful.

The existence of several valleys in the search direction is typical to a guess field of small magnitude combined with small penalties on energy and ionization. This results in small magnitudes of the penalty terms in the region of $\epswr^{(k)}$ in the optimization space. In general, the penalty terms introduce a ``wall'' or a barrier in the optimization space. This can prevent the appearance of additional valleys. If the penalty terms in the vicinity of $\epswr^{(k)}$ are small in magnitude, they do not have a significant effect on the optimization space in the region of $\epswr^{(k)}$, and additional further valleys may appear.

The problem can be solved by the ``reset'' of the inverse-Hessian approximation when the process ``gets stuck''. The inverse-Hessian is set again to the form of \eqref{eq:invHess0}, and the iterative update of the inverse-Hessian by Eq.~\eqref{eq:invHess_update} is restarted from this point. This is equivalent to starting a new optimization from the final solution of the first optimization.

A failure of the inverse-Hessian approximation can usually be detected before the total failure of the optimization, since it is characterized by an unusual behaviour of the process. We may observe too many iterations in the different phases of the line-search, and certain iterations of the $\epswr$ update with very small improvement. If an unusual behaviour is detected, it is recommended to reset the inverse-Hessian in this stage, instead of waiting until the optimization is completely ``stuck''.

An alternative way to address this problem is to use the form of \eqref{eq:invHess0} as the inverse-Hessian approximation also for \text{$k>0$}, until $\epswr^{(k)}$ and the penalty terms become sufficiently large to prevent the appearance of the problem.

\section{The effect of the absorbing boundaries}\label{app:absorbing}

As was mentioned in Sec.~\ref{ssec:dynamics}, the employment of absorbing boundary conditions might be a primary source of inaccuracy in HHG simulations. In principle, two sources of inaccuracy are introduced when an infinitely long spatial grid is replaced by absorbing boundary conditions:
\begin{enumerate}
	\item \emph{Physical effect}: The absorption of a portion of the wave-function amplitude at the boundaries can have an effect on the dynamics in the central region of the wave-function. The possible physical effects can be classified as \emph{direct} or \emph{indirect} effects:
	\begin{enumerate}
		\item \emph{Direct effect}: The unjustified elimination of an electronic amplitude which is due to revert to the central region in a later stage in the dynamical process;
		\item \emph{Indirect effect}: The wave-function dynamics in all spatial regions is \emph{coupled} by the time-dependent Schr\"odinger equation. Thus, the elimination of the amplitude at the boundaries has some effect on the dynamics in the central region of the wave-function. 
	\end{enumerate}
	\item \emph{Numerical effect}: Imperfection in the absorption capabilities of the absorbing boundaries results in the effects of reflection from the absorbing boundary, or transmission and wraparound of the electronic amplitude.
\end{enumerate}
Both the physical and the numerical effects can be reduced by placing the absorbing boundaries further from the central region of the grid. However, neither of them can be completely eliminated. As was discussed in Sec.~\ref{ssec:dynamics}, the magnitude of the numerical effect strongly depends on the choice of the complex-absorbing-potential. This motivated the employment of the optimization scheme described in Sec.~\ref{ssec:dynamics} for its construction. 

The validity of the approximations introduced by the employment of absorbing boundaries was tested in all problems of Sec.~\ref{sec:results}. As a test of validity, the optimized fields obtained in all problems were used to propagate the same initial condition in a doubled grid, \ie \text{$x\in[-480, 480)$} where the spacing between adjacent points remains unaltered (this test has failed in \cite{krause92}, due to the presence of the numerical effect mentioned above). The resulting $\Cw$ spectra were calculated. We found that there is no significant difference between the spectra of the original grid and the doubled grid. In order to quantify the magnitude of the deviation of the response in the original grid from the doubled grid, we define the relative difference of $J_{max}$:
\begin{equation}\label{eq:relE_Jmax}
	\Delta_{rel} J_{max} \equiv \frac{J_{max} - J_{max,doubled}}{J_{max,doubled}}
\end{equation}
where $J_{max,doubled}$ is the $J_{max}$ calculated for the doubled grid. In Table~\ref{tab:relE}, we present $\Delta_{rel} J_{max}$ for the various optimized harmonics. It can be observed that the magnitude of the relative difference does not exceed the order of $\sim 10^{-3}$. Thus, the effects introduced by the absorbing boundaries are not expected to have a significant effect on the optimization problem.

\begin{table}
\begin{equation*}
	\renewcommand{\arraystretch}{1.5}
	\begin{array}{|c||c|}
		\hline
		n  & \Delta_{rel} J_{max} \\ \hline \hline
		13 & 2.42\times 10^{-3} \\ \hline
		14 & 9.02\times 10^{-4} \\ \hline
		15 & 1.44\times 10^{-3} \\ \hline
		17 & 2.41\times 10^{-5} \\ \hline
	\end{array}
\end{equation*}
\caption{The relative difference of $J_{max}$ from the resulting $J_{max}$ in the doubled grid (Eq.~\eqref{eq:relE_Jmax}) for the various optimized harmonics.}\label{tab:relE}
\end{table}

\section{Prevention of plasma production represented as a control requirement}\label{app:plasma}

The minimization of plasma production in HHG is one of the aims of the current study. In this appendix we shall further discuss its realization by the control formulation outlined in Sec.~\ref{sec:theory}.

There is a fundamental problem in formulating a \emph{control requirement} representing the \emph{physical requirement} of prevention of plasma production. Plasma production is a macroscopic effect, which takes place in the macroscopic medium. However, the treatment of the dynamics in the present study is in the isolated system level. Thus, there is no direct access to the physical effect of plasma production by the current dynamical treatment. Consequently, it is impossible to give a direct formulation of the physical requirement as a control requirement.

As an alternative, it is possible to define the physical requirement of \emph{localization} of the liberated electron around the parent ion. The localization of the electron ensures that it is not liberated into the macroscopic medium, and thus the plasma production is prevented. In more precise means, we can define a radius around the parent ion, where the electronic probability inside the radius is considered as localized. We shall refer to this radius as the \emph{localization radius}. Such a physical requirement is accessible by the dynamical treatment of the isolated system, and thus can be formulated as a control requirement.

The immediate question which arises is how the localization radius should be chosen. It is important to avoid an over-localization of the electron around the parent ion. The ionization and liberation of the electron into large distances in the continuum is an integral part of the HHG mechanism, where the electron gains large energies in its accelerated motion reverted into the parent ion over a large distance. If the chosen localization radius is too small, the feasibility of production of the target harmonic can be completely prevented, or at least reduced.

It should be emphasized that the insertion of an additional control requirement always reduces the optimal yield, since the space of the allowed control opportunities becomes restricted. However, we should distinguish between two types of reduction of the optimal harmonic yield by the localization requirement:
\begin{enumerate}
	\item \emph{Direct reduction}: The spatial restriction imposed on the electron damages directly the opportunities provided by the  \emph{physical mechanism} responsible for HHG, as above. In this case, the localization requirement imposes a direct restriction on the electronic probability which \emph{participates in the HHG process}.
	\item \emph{Indirect reduction}: The HHG process has the \emph{side effect} of liberated electronic probability which does not revert to the parent ion, and consequently, is responsible for plasma production. This part of the electronic probability does not participate in the physical mechanism responsible for the production of high harmonics. The localization requirement restricts this side effect. An indirect consequence of this restriction is the reduction of the control opportunities, where the optimal mechanism should satisfy the additional restriction of the side effect. Thus, the localization requirement restricts directly an electronic probability which \emph{does not participate in the HHG process}, and the probability which participates in the process also becomes restricted as an indirect consequence of this requirement.
\end{enumerate}
When we state that over-localization of the electron should be avoided, we intend to the prevention of the \emph{direct reduction} of the optimal yield. However, the indirect reduction is an unavoidable consequence of the physical requirement of prevention of plasma production.

In what follows, we claim that the considerations in the choice of the localization radius are similar to the considerations of the choice position of the \emph{absorbing boundaries}.

We shall start from the considerations of the choice of the position of the absorbing boundaries. In Appendix~\ref{app:absorbing} we distinguished between two sources of inaccuracy induced by the absorbing boundaries, which were classified as a \emph{physical effect} and a \emph{numerical effect}. Let us assume, for the moment, that the absorption capabilities of the absorbing boundaries are nearly perfect, and the numerical effect is negligible. In this case, the position of the absorbing boundaries is chosen by purely physical considerations; the choice should reduce the physical effect to a negligible magnitude. The term ``negligible'' depends on the required accuracy. This ensures that the absorption of the outgoing electronic amplitude has no significant effect on the physics in the central region of the grid.

The appropriate position of the absorbing boundaries depends on the specific profile of the laser pulse.

The determination of the appropriate position in an optimization problem becomes problematic. The laser pulse profile varies during the optimization process, and thus the position of the absorbing boundaries has to be far enough from the central region to consider all fields which take place in the optimization path. The problem is that these fields are unknown in advance. However, the position of the absorbing boundaries must be determined in advance, since it defines the system, and consequently, the optimization problem.

Nevertheless, an upper limit for the position of the absorbing boundaries can be roughly estimated, since we actually have some idea about important physical properties of the fields in the optimization path---the control requirements restrict the intensity of the field and the available frequency band. The position of the absorbing boundaries can be estimated by comparison to HHG problems with similar physical conditions.

Now we return to the discussion of the choice of the localization radius. In order to avoid the direct reduction of the optimal yield by the localization requirement, the localization requirement should exclude only electronic probability which is far enough from the parent ion to become irrelevant to the physics responsible for the high harmonic production. In other words, the localization radius is chosen such that the electronic probability which is beyond the localization radius has a negligible effect on the dynamics in the central region. This coincides with the appropriate position of the absorbing boundaries. This justifies the formulation of the localization control requirement by $J_{ion}$ (defined in Eq.~\eqref{eq:Jion}), where the absorbed probability is identified as the permanently ionized probability.

Let us consider also the situation in which the imperfection in the absorption capabilities of the absorbing potential cannot be ignored. In this case, the numerical effect of the absorbing boundaries is non-negligible. The position of the absorbing boundaries is not led by physical considerations only, but also by numerical considerations. The common practice is to place the absorbing boundaries further from the central region of the grid, such that the numerical effect becomes negligible. In this case, the correspondence between the localization radius and the position of the absorbing boundaries is lost.

Nevertheless, the method outlined in Sec.~\ref{ssec:ion} for the restriction of permanent ionization can still be applied. The considerations which were outlined here for the choice of the localization radius impose only a lower limit on the localization radius, but not an upper limit. The only problem in the exaggeration of the magnitude of the localization radius is the extra numerical cost of larger grids. Hence, when the use of larger grid is crucial from numerical considerations, the localization radius can still be defined by the position of the absorbing boundaries. However, it should be emphasized that the minimal allowed survival amplitude has to be increased as the distance of the absorbing boundaries from the parent ion is increased. The profile of $\sigma(y)$ has to be altered accordingly. Thus, the current formulation couples between the \emph{numerical realization of the dynamics} and the \emph{optimization problem}.

\bibliography{HHGcontrol}

\begin{thebibliography}{67}%
\makeatletter
\providecommand \@ifxundefined [1]{%
 \@ifx{#1\undefined}
}%
\providecommand \@ifnum [1]{%
 \ifnum #1\expandafter \@firstoftwo
 \else \expandafter \@secondoftwo
 \fi
}%
\providecommand \@ifx [1]{%
 \ifx #1\expandafter \@firstoftwo
 \else \expandafter \@secondoftwo
 \fi
}%
\providecommand \natexlab [1]{#1}%
\providecommand \enquote  [1]{``#1''}%
\providecommand \bibnamefont  [1]{#1}%
\providecommand \bibfnamefont [1]{#1}%
\providecommand \citenamefont [1]{#1}%
\providecommand \href@noop [0]{\@secondoftwo}%
\providecommand \href [0]{\begingroup \@sanitize@url \@href}%
\providecommand \@href[1]{\@@startlink{#1}\@@href}%
\providecommand \@@href[1]{\endgroup#1\@@endlink}%
\providecommand \@sanitize@url [0]{\catcode `\\12\catcode `\$12\catcode
  `\&12\catcode `\#12\catcode `\^12\catcode `\_12\catcode `\%12\relax}%
\providecommand \@@startlink[1]{}%
\providecommand \@@endlink[0]{}%
\providecommand \url  [0]{\begingroup\@sanitize@url \@url }%
\providecommand \@url [1]{\endgroup\@href {#1}{\urlprefix }}%
\providecommand \urlprefix  [0]{URL }%
\providecommand \Eprint [0]{\href }%
\providecommand \doibase [0]{http://dx.doi.org/}%
\providecommand \selectlanguage [0]{\@gobble}%
\providecommand \bibinfo  [0]{\@secondoftwo}%
\providecommand \bibfield  [0]{\@secondoftwo}%
\providecommand \translation [1]{[#1]}%
\providecommand \BibitemOpen [0]{}%
\providecommand \bibitemStop [0]{}%
\providecommand \bibitemNoStop [0]{.\EOS\space}%
\providecommand \EOS [0]{\spacefactor3000\relax}%
\providecommand \BibitemShut  [1]{\csname bibitem#1\endcsname}%
\let\auto@bib@innerbib\@empty
\bibitem [{\citenamefont {McPherson}\ \emph {et~al.}(1987)\citenamefont
  {McPherson}, \citenamefont {Gibson}, \citenamefont {Jara}, \citenamefont
  {Johann}, \citenamefont {Luk}, \citenamefont {McIntyre}, \citenamefont
  {Boyer},\ and\ \citenamefont {Rhodes}}]{mcpherson1987studies}%
  \BibitemOpen
  \bibfield  {author} {\bibinfo {author} {\bibfnamefont {A.}~\bibnamefont
  {McPherson}}, \bibinfo {author} {\bibfnamefont {G.}~\bibnamefont {Gibson}},
  \bibinfo {author} {\bibfnamefont {H.}~\bibnamefont {Jara}}, \bibinfo {author}
  {\bibfnamefont {U.}~\bibnamefont {Johann}}, \bibinfo {author} {\bibfnamefont
  {T.~S.}\ \bibnamefont {Luk}}, \bibinfo {author} {\bibfnamefont
  {I.}~\bibnamefont {McIntyre}}, \bibinfo {author} {\bibfnamefont
  {K.}~\bibnamefont {Boyer}}, \ and\ \bibinfo {author} {\bibfnamefont {C.~K.}\
  \bibnamefont {Rhodes}},\ }\href@noop {} {\bibfield  {journal} {\bibinfo
  {journal} {JOSA B}\ }\textbf {\bibinfo {volume} {4}},\ \bibinfo {pages} {595}
  (\bibinfo {year} {1987})}\BibitemShut {NoStop}%
\bibitem [{\citenamefont {{M. Ferray, A. Lhuillier, X. F. Li, L. A. Lompre, G.
  Mainfray and C. Manus}}(1988)}]{ferray88}%
  \BibitemOpen
  \bibfield  {author} {\bibinfo {author} {\bibnamefont {{M. Ferray, A.
  Lhuillier, X. F. Li, L. A. Lompre, G. Mainfray and C. Manus}}},\ }\href@noop
  {} {\bibfield  {journal} {\bibinfo  {journal} {{J. Phys. B-At. Mol. Opt.
  Phys.}}\ }\textbf {\bibinfo {volume} {{21}}},\ \bibinfo {pages} {L31}
  (\bibinfo {year} {1988})}\BibitemShut {NoStop}%
\bibitem [{\citenamefont {Li}\ \emph {et~al.}(1989)\citenamefont {Li},
  \citenamefont {L'Huillier}, \citenamefont {Ferray}, \citenamefont
  {Lompr\'e},\ and\ \citenamefont {Mainfray}}]{li1989multiple}%
  \BibitemOpen
  \bibfield  {author} {\bibinfo {author} {\bibfnamefont {X.~F.}\ \bibnamefont
  {Li}}, \bibinfo {author} {\bibfnamefont {A.}~\bibnamefont {L'Huillier}},
  \bibinfo {author} {\bibfnamefont {M.}~\bibnamefont {Ferray}}, \bibinfo
  {author} {\bibfnamefont {L.~A.}\ \bibnamefont {Lompr\'e}}, \ and\ \bibinfo
  {author} {\bibfnamefont {G.}~\bibnamefont {Mainfray}},\ }\href {\doibase
  10.1103/PhysRevA.39.5751} {\bibfield  {journal} {\bibinfo  {journal} {Phys.
  Rev. A}\ }\textbf {\bibinfo {volume} {39}},\ \bibinfo {pages} {5751}
  (\bibinfo {year} {1989})}\BibitemShut {NoStop}%
\bibitem [{\citenamefont {Mauritsson}\ \emph {et~al.}(2006)\citenamefont
  {Mauritsson}, \citenamefont {Johnsson}, \citenamefont {Gustafsson},
  \citenamefont {L'Huillier}, \citenamefont {Schafer},\ and\ \citenamefont
  {Gaarde}}]{mauritsson06}%
  \BibitemOpen
  \bibfield  {author} {\bibinfo {author} {\bibfnamefont {J.}~\bibnamefont
  {Mauritsson}}, \bibinfo {author} {\bibfnamefont {P.}~\bibnamefont
  {Johnsson}}, \bibinfo {author} {\bibfnamefont {E.}~\bibnamefont
  {Gustafsson}}, \bibinfo {author} {\bibfnamefont {A.}~\bibnamefont
  {L'Huillier}}, \bibinfo {author} {\bibfnamefont {K.~J.}\ \bibnamefont
  {Schafer}}, \ and\ \bibinfo {author} {\bibfnamefont {M.~B.}\ \bibnamefont
  {Gaarde}},\ }\href {\doibase 10.1103/PhysRevLett.97.013001} {\bibfield
  {journal} {\bibinfo  {journal} {Phys. Rev. Lett.}\ }\textbf {\bibinfo
  {volume} {97}},\ \bibinfo {pages} {013001} (\bibinfo {year}
  {2006})}\BibitemShut {NoStop}%
\bibitem [{\citenamefont {Brabec}\ and\ \citenamefont
  {Krausz}(2000)}]{brabec2000intense}%
  \BibitemOpen
  \bibfield  {author} {\bibinfo {author} {\bibfnamefont {T.}~\bibnamefont
  {Brabec}}\ and\ \bibinfo {author} {\bibfnamefont {F.}~\bibnamefont
  {Krausz}},\ }\href@noop {} {\bibfield  {journal} {\bibinfo  {journal}
  {Reviews of Modern Physics}\ }\textbf {\bibinfo {volume} {72}},\ \bibinfo
  {pages} {545} (\bibinfo {year} {2000})}\BibitemShut {NoStop}%
\bibitem [{\citenamefont {Paul}\ \emph {et~al.}(2001)\citenamefont {Paul},
  \citenamefont {Toma}, \citenamefont {Breger}, \citenamefont {Mullot},
  \citenamefont {Aug{\'e}}, \citenamefont {Balcou}, \citenamefont {Muller},\
  and\ \citenamefont {Agostini}}]{paul2001observation}%
  \BibitemOpen
  \bibfield  {author} {\bibinfo {author} {\bibfnamefont {P.~M.}\ \bibnamefont
  {Paul}}, \bibinfo {author} {\bibfnamefont {E.~S.}\ \bibnamefont {Toma}},
  \bibinfo {author} {\bibfnamefont {P.}~\bibnamefont {Breger}}, \bibinfo
  {author} {\bibfnamefont {G.}~\bibnamefont {Mullot}}, \bibinfo {author}
  {\bibfnamefont {F.}~\bibnamefont {Aug{\'e}}}, \bibinfo {author}
  {\bibfnamefont {P.}~\bibnamefont {Balcou}}, \bibinfo {author} {\bibfnamefont
  {H.~G.}\ \bibnamefont {Muller}}, \ and\ \bibinfo {author} {\bibfnamefont
  {P.}~\bibnamefont {Agostini}},\ }\href {\doibase 10.1126/science.1059413}
  {\bibfield  {journal} {\bibinfo  {journal} {Science}\ }\textbf {\bibinfo
  {volume} {292}},\ \bibinfo {pages} {1689} (\bibinfo {year} {2001})},\ \Eprint
  {http://arxiv.org/abs/http://science.sciencemag.org/content/292/5522/1689.full.pdf}
  {http://science.sciencemag.org/content/292/5522/1689.full.pdf} \BibitemShut
  {NoStop}%
\bibitem [{\citenamefont {Corkum}\ and\ \citenamefont
  {Krausz}(2007)}]{corkum2007attosecond}%
  \BibitemOpen
  \bibfield  {author} {\bibinfo {author} {\bibfnamefont {P.~{\'a}.}\
  \bibnamefont {Corkum}}\ and\ \bibinfo {author} {\bibfnamefont
  {F.}~\bibnamefont {Krausz}},\ }\href@noop {} {\bibfield  {journal} {\bibinfo
  {journal} {Nature physics}\ }\textbf {\bibinfo {volume} {3}},\ \bibinfo
  {pages} {381} (\bibinfo {year} {2007})}\BibitemShut {NoStop}%
\bibitem [{\citenamefont {Hassan}\ \emph {et~al.}(2016)\citenamefont {Hassan},
  \citenamefont {Luu}, \citenamefont {Moulet}, \citenamefont {Raskazovskaya},
  \citenamefont {Zhokhov}, \citenamefont {Garg}, \citenamefont {Karpowicz},
  \citenamefont {Zheltikov}, \citenamefont {Pervak}, \citenamefont {Krausz}
  \emph {et~al.}}]{hassan2016optical}%
  \BibitemOpen
  \bibfield  {author} {\bibinfo {author} {\bibfnamefont {M.~T.}\ \bibnamefont
  {Hassan}}, \bibinfo {author} {\bibfnamefont {T.~T.}\ \bibnamefont {Luu}},
  \bibinfo {author} {\bibfnamefont {A.}~\bibnamefont {Moulet}}, \bibinfo
  {author} {\bibfnamefont {O.}~\bibnamefont {Raskazovskaya}}, \bibinfo {author}
  {\bibfnamefont {P.}~\bibnamefont {Zhokhov}}, \bibinfo {author} {\bibfnamefont
  {M.}~\bibnamefont {Garg}}, \bibinfo {author} {\bibfnamefont {N.}~\bibnamefont
  {Karpowicz}}, \bibinfo {author} {\bibfnamefont {A.}~\bibnamefont
  {Zheltikov}}, \bibinfo {author} {\bibfnamefont {V.}~\bibnamefont {Pervak}},
  \bibinfo {author} {\bibfnamefont {F.}~\bibnamefont {Krausz}},  \emph
  {et~al.},\ }\href@noop {} {\bibfield  {journal} {\bibinfo  {journal}
  {Nature}\ }\textbf {\bibinfo {volume} {530}},\ \bibinfo {pages} {66}
  (\bibinfo {year} {2016})}\BibitemShut {NoStop}%
\bibitem [{\citenamefont {W{\"o}rner}\ \emph {et~al.}(2011)\citenamefont
  {W{\"o}rner}, \citenamefont {Bertrand}, \citenamefont {Fabre}, \citenamefont
  {Higuet}, \citenamefont {Ruf}, \citenamefont {Dubrouil}, \citenamefont
  {Patchkovskii}, \citenamefont {Spanner}, \citenamefont {Mairesse},
  \citenamefont {Blanchet} \emph {et~al.}}]{worner2011conical}%
  \BibitemOpen
  \bibfield  {author} {\bibinfo {author} {\bibfnamefont {H.~J.}\ \bibnamefont
  {W{\"o}rner}}, \bibinfo {author} {\bibfnamefont {J.~B.}\ \bibnamefont
  {Bertrand}}, \bibinfo {author} {\bibfnamefont {B.}~\bibnamefont {Fabre}},
  \bibinfo {author} {\bibfnamefont {J.}~\bibnamefont {Higuet}}, \bibinfo
  {author} {\bibfnamefont {H.}~\bibnamefont {Ruf}}, \bibinfo {author}
  {\bibfnamefont {A.}~\bibnamefont {Dubrouil}}, \bibinfo {author}
  {\bibfnamefont {S.}~\bibnamefont {Patchkovskii}}, \bibinfo {author}
  {\bibfnamefont {M.}~\bibnamefont {Spanner}}, \bibinfo {author} {\bibfnamefont
  {Y.}~\bibnamefont {Mairesse}}, \bibinfo {author} {\bibfnamefont
  {V.}~\bibnamefont {Blanchet}},  \emph {et~al.},\ }\href@noop {} {\bibfield
  {journal} {\bibinfo  {journal} {Science}\ }\textbf {\bibinfo {volume}
  {334}},\ \bibinfo {pages} {208} (\bibinfo {year} {2011})}\BibitemShut
  {NoStop}%
\bibitem [{\citenamefont {Luzon}\ \emph {et~al.}(2017)\citenamefont {Luzon},
  \citenamefont {Jagtap}, \citenamefont {Livshits}, \citenamefont
  {Lioubashevski}, \citenamefont {Baer},\ and\ \citenamefont
  {Strasser}}]{luzon2017single}%
  \BibitemOpen
  \bibfield  {author} {\bibinfo {author} {\bibfnamefont {I.}~\bibnamefont
  {Luzon}}, \bibinfo {author} {\bibfnamefont {K.}~\bibnamefont {Jagtap}},
  \bibinfo {author} {\bibfnamefont {E.}~\bibnamefont {Livshits}}, \bibinfo
  {author} {\bibfnamefont {O.}~\bibnamefont {Lioubashevski}}, \bibinfo {author}
  {\bibfnamefont {R.}~\bibnamefont {Baer}}, \ and\ \bibinfo {author}
  {\bibfnamefont {D.}~\bibnamefont {Strasser}},\ }\href@noop {} {\bibfield
  {journal} {\bibinfo  {journal} {Physical Chemistry Chemical Physics}\
  }\textbf {\bibinfo {volume} {19}},\ \bibinfo {pages} {13488} (\bibinfo {year}
  {2017})}\BibitemShut {NoStop}%
\bibitem [{\citenamefont {Bhattacherjee}\ and\ \citenamefont
  {Leone}(2018)}]{bhattacherjee2018ultrafast}%
  \BibitemOpen
  \bibfield  {author} {\bibinfo {author} {\bibfnamefont {A.}~\bibnamefont
  {Bhattacherjee}}\ and\ \bibinfo {author} {\bibfnamefont {S.~R.}\ \bibnamefont
  {Leone}},\ }\href@noop {} {\bibfield  {journal} {\bibinfo  {journal}
  {Accounts of chemical research}\ }\textbf {\bibinfo {volume} {51}},\ \bibinfo
  {pages} {3203} (\bibinfo {year} {2018})}\BibitemShut {NoStop}%
\bibitem [{\citenamefont {Corkum}(1993)}]{corkum93}%
  \BibitemOpen
  \bibfield  {author} {\bibinfo {author} {\bibfnamefont {P.~B.}\ \bibnamefont
  {Corkum}},\ }\href@noop {} {\bibfield  {journal} {\bibinfo  {journal}
  {Physical review letters}\ }\textbf {\bibinfo {volume} {71}},\ \bibinfo
  {pages} {1994} (\bibinfo {year} {1993})}\BibitemShut {NoStop}%
\bibitem [{\citenamefont {{M. Lewenstein, P. Balcou, M. Y. Ivanov, A. Lhuillier
  and P. B. Corkum}}(1994)}]{lewenstein94}%
  \BibitemOpen
  \bibfield  {author} {\bibinfo {author} {\bibnamefont {{M. Lewenstein, P.
  Balcou, M. Y. Ivanov, A. Lhuillier and P. B. Corkum}}},\ }\href@noop {}
  {\bibfield  {journal} {\bibinfo  {journal} {Phys. Rev. A}\ }\textbf {\bibinfo
  {volume} {49}},\ \bibinfo {pages} {2117} (\bibinfo {year}
  {1994})}\BibitemShut {NoStop}%
\bibitem [{\citenamefont {Smirnova}\ and\ \citenamefont
  {Ivanov}(2014)}]{IvanovTutorial}%
  \BibitemOpen
  \bibfield  {author} {\bibinfo {author} {\bibfnamefont {O.}~\bibnamefont
  {Smirnova}}\ and\ \bibinfo {author} {\bibfnamefont {M.}~\bibnamefont
  {Ivanov}},\ }in\ \href@noop {} {\emph {\bibinfo {booktitle} {Attosecond and
  XUV Physics}}},\ \bibinfo {editor} {edited by\ \bibinfo {editor}
  {\bibfnamefont {T.}~\bibnamefont {Schultz}}\ and\ \bibinfo {editor}
  {\bibfnamefont {M.}~\bibnamefont {Vrakking}}}\ (\bibinfo  {publisher} {John
  Wiley and Sons},\ \bibinfo {year} {2014})\ Chap.~\bibinfo {chapter} {7}, pp.\
  \bibinfo {pages} {201--256}\BibitemShut {NoStop}%
\bibitem [{\citenamefont {Smirnova}\ \emph {et~al.}(2009)\citenamefont
  {Smirnova}, \citenamefont {Mairesse}, \citenamefont {Patchkovskii},
  \citenamefont {Dudovich}, \citenamefont {Villeneuve}, \citenamefont
  {Corkum},\ and\ \citenamefont {Ivanov}}]{smirnova2009high}%
  \BibitemOpen
  \bibfield  {author} {\bibinfo {author} {\bibfnamefont {O.}~\bibnamefont
  {Smirnova}}, \bibinfo {author} {\bibfnamefont {Y.}~\bibnamefont {Mairesse}},
  \bibinfo {author} {\bibfnamefont {S.}~\bibnamefont {Patchkovskii}}, \bibinfo
  {author} {\bibfnamefont {N.}~\bibnamefont {Dudovich}}, \bibinfo {author}
  {\bibfnamefont {D.}~\bibnamefont {Villeneuve}}, \bibinfo {author}
  {\bibfnamefont {P.}~\bibnamefont {Corkum}}, \ and\ \bibinfo {author}
  {\bibfnamefont {M.~Y.}\ \bibnamefont {Ivanov}},\ }\href@noop {} {\bibfield
  {journal} {\bibinfo  {journal} {Nature}\ }\textbf {\bibinfo {volume} {460}},\
  \bibinfo {pages} {972} (\bibinfo {year} {2009})}\BibitemShut {NoStop}%
\bibitem [{\citenamefont {Fleischer}\ \emph {et~al.}(2014)\citenamefont
  {Fleischer}, \citenamefont {Kfir}, \citenamefont {Diskin}, \citenamefont
  {Sidorenko},\ and\ \citenamefont {Cohen}}]{fleischer2014spin}%
  \BibitemOpen
  \bibfield  {author} {\bibinfo {author} {\bibfnamefont {A.}~\bibnamefont
  {Fleischer}}, \bibinfo {author} {\bibfnamefont {O.}~\bibnamefont {Kfir}},
  \bibinfo {author} {\bibfnamefont {T.}~\bibnamefont {Diskin}}, \bibinfo
  {author} {\bibfnamefont {P.}~\bibnamefont {Sidorenko}}, \ and\ \bibinfo
  {author} {\bibfnamefont {O.}~\bibnamefont {Cohen}},\ }\href@noop {}
  {\bibfield  {journal} {\bibinfo  {journal} {Nature Photonics}\ }\textbf
  {\bibinfo {volume} {8}},\ \bibinfo {pages} {543} (\bibinfo {year}
  {2014})}\BibitemShut {NoStop}%
\bibitem [{\citenamefont {Neufeld}\ \emph {et~al.}(2017)\citenamefont
  {Neufeld}, \citenamefont {Bordo}, \citenamefont {Fleischer},\ and\
  \citenamefont {Cohen}}]{neufeld2017high}%
  \BibitemOpen
  \bibfield  {author} {\bibinfo {author} {\bibfnamefont {O.}~\bibnamefont
  {Neufeld}}, \bibinfo {author} {\bibfnamefont {E.}~\bibnamefont {Bordo}},
  \bibinfo {author} {\bibfnamefont {A.}~\bibnamefont {Fleischer}}, \ and\
  \bibinfo {author} {\bibfnamefont {O.}~\bibnamefont {Cohen}},\ }\href@noop {}
  {\bibfield  {journal} {\bibinfo  {journal} {New Journal of Physics}\ }\textbf
  {\bibinfo {volume} {19}},\ \bibinfo {pages} {023051} (\bibinfo {year}
  {2017})}\BibitemShut {NoStop}%
\bibitem [{\citenamefont {Rice}(1992)}]{rice1992new}%
  \BibitemOpen
  \bibfield  {author} {\bibinfo {author} {\bibfnamefont {S.~A.}\ \bibnamefont
  {Rice}},\ }\href@noop {} {\bibfield  {journal} {\bibinfo  {journal}
  {Science}\ }\textbf {\bibinfo {volume} {258}},\ \bibinfo {pages} {412}
  (\bibinfo {year} {1992})}\BibitemShut {NoStop}%
\bibitem [{\citenamefont {Burnett}\ \emph {et~al.}(1993)\citenamefont
  {Burnett}, \citenamefont {Reed},\ and\ \citenamefont {Knight}}]{Knight93}%
  \BibitemOpen
  \bibfield  {author} {\bibinfo {author} {\bibfnamefont {K.}~\bibnamefont
  {Burnett}}, \bibinfo {author} {\bibfnamefont {V.~C.}\ \bibnamefont {Reed}}, \
  and\ \bibinfo {author} {\bibfnamefont {P.~L.}\ \bibnamefont {Knight}},\
  }\href {\doibase 10.1088/0953-4075/26/4/003} {\bibfield  {journal} {\bibinfo
  {journal} {Journal of Physics B: Atomic, Molecular and Optical Physics}\
  }\textbf {\bibinfo {volume} {26}},\ \bibinfo {pages} {561} (\bibinfo {year}
  {1993})}\BibitemShut {NoStop}%
\bibitem [{\citenamefont {Cerjan}\ and\ \citenamefont
  {Kosloff}(1993)}]{cerjan1993efficient}%
  \BibitemOpen
  \bibfield  {author} {\bibinfo {author} {\bibfnamefont {C.}~\bibnamefont
  {Cerjan}}\ and\ \bibinfo {author} {\bibfnamefont {R.}~\bibnamefont
  {Kosloff}},\ }\href@noop {} {\bibfield  {journal} {\bibinfo  {journal}
  {Physical Review A}\ }\textbf {\bibinfo {volume} {47}},\ \bibinfo {pages}
  {1852} (\bibinfo {year} {1993})}\BibitemShut {NoStop}%
\bibitem [{\citenamefont {Ben-Tal}\ \emph
  {et~al.}(1993{\natexlab{a}})\citenamefont {Ben-Tal}, \citenamefont
  {Moiseyev}, \citenamefont {Kosloff},\ and\ \citenamefont
  {Cerjan}}]{bental93}%
  \BibitemOpen
  \bibfield  {author} {\bibinfo {author} {\bibfnamefont {N.}~\bibnamefont
  {Ben-Tal}}, \bibinfo {author} {\bibfnamefont {N.}~\bibnamefont {Moiseyev}},
  \bibinfo {author} {\bibfnamefont {R.}~\bibnamefont {Kosloff}}, \ and\
  \bibinfo {author} {\bibfnamefont {C.}~\bibnamefont {Cerjan}},\ }\href@noop {}
  {\bibfield  {journal} {\bibinfo  {journal} {Journal of Physics B: Atomic,
  Molecular and Optical Physics}\ }\textbf {\bibinfo {volume} {26}},\ \bibinfo
  {pages} {1445} (\bibinfo {year} {1993}{\natexlab{a}})}\BibitemShut {NoStop}%
\bibitem [{\citenamefont {MacKay}(2003)}]{mackay2003information}%
  \BibitemOpen
  \bibfield  {author} {\bibinfo {author} {\bibfnamefont {D.~J.~C.}\
  \bibnamefont {MacKay}},\ }\href@noop {} {\emph {\bibinfo {title} {Information
  theory, inference and learning algorithms}}}\ (\bibinfo  {publisher}
  {Cambridge university press},\ \bibinfo {year} {2003})\BibitemShut {NoStop}%
\bibitem [{\citenamefont {Chu}\ and\ \citenamefont
  {Chu}(2001)}]{chu2001optimization}%
  \BibitemOpen
  \bibfield  {author} {\bibinfo {author} {\bibfnamefont {X.}~\bibnamefont
  {Chu}}\ and\ \bibinfo {author} {\bibfnamefont {S.-I.}\ \bibnamefont {Chu}},\
  }\href@noop {} {\bibfield  {journal} {\bibinfo  {journal} {Physical Review
  A}\ }\textbf {\bibinfo {volume} {64}},\ \bibinfo {pages} {021403} (\bibinfo
  {year} {2001})}\BibitemShut {NoStop}%
\bibitem [{\citenamefont {Villoresi}\ \emph {et~al.}(2004)\citenamefont
  {Villoresi}, \citenamefont {Bonora}, \citenamefont {Pascolini}, \citenamefont
  {Poletto}, \citenamefont {Tondello}, \citenamefont {Vozzi}, \citenamefont
  {Nisoli}, \citenamefont {Sansone}, \citenamefont {Stagira},\ and\
  \citenamefont {De~Silvestri}}]{villoresi2004optimization}%
  \BibitemOpen
  \bibfield  {author} {\bibinfo {author} {\bibfnamefont {P.}~\bibnamefont
  {Villoresi}}, \bibinfo {author} {\bibfnamefont {S.}~\bibnamefont {Bonora}},
  \bibinfo {author} {\bibfnamefont {M.}~\bibnamefont {Pascolini}}, \bibinfo
  {author} {\bibfnamefont {L.}~\bibnamefont {Poletto}}, \bibinfo {author}
  {\bibfnamefont {G.}~\bibnamefont {Tondello}}, \bibinfo {author}
  {\bibfnamefont {C.}~\bibnamefont {Vozzi}}, \bibinfo {author} {\bibfnamefont
  {M.}~\bibnamefont {Nisoli}}, \bibinfo {author} {\bibfnamefont
  {G.}~\bibnamefont {Sansone}}, \bibinfo {author} {\bibfnamefont
  {S.}~\bibnamefont {Stagira}}, \ and\ \bibinfo {author} {\bibfnamefont
  {S.}~\bibnamefont {De~Silvestri}},\ }\href@noop {} {\bibfield  {journal}
  {\bibinfo  {journal} {Optics letters}\ }\textbf {\bibinfo {volume} {29}},\
  \bibinfo {pages} {207} (\bibinfo {year} {2004})}\BibitemShut {NoStop}%
\bibitem [{\citenamefont {{R. A. Bartels, M. M. Murnane, H. C. Kapteyn, I
  Christov, and H. Rabitz }}(2004)}]{murnane04}%
  \BibitemOpen
  \bibfield  {author} {\bibinfo {author} {\bibnamefont {{R. A. Bartels, M. M.
  Murnane, H. C. Kapteyn, I Christov, and H. Rabitz }}},\ }\href@noop {}
  {\bibfield  {journal} {\bibinfo  {journal} {Phys. Rev. A}\ }\textbf {\bibinfo
  {volume} {{70}}},\ \bibinfo {pages} {{043404}} (\bibinfo {year}
  {{2004}})}\BibitemShut {NoStop}%
\bibitem [{\citenamefont {Winterfeldt}\ \emph {et~al.}(2008)\citenamefont
  {Winterfeldt}, \citenamefont {Spielmann},\ and\ \citenamefont
  {Gerber}}]{HHGcol}%
  \BibitemOpen
  \bibfield  {author} {\bibinfo {author} {\bibfnamefont {C.}~\bibnamefont
  {Winterfeldt}}, \bibinfo {author} {\bibfnamefont {C.}~\bibnamefont
  {Spielmann}}, \ and\ \bibinfo {author} {\bibfnamefont {G.}~\bibnamefont
  {Gerber}},\ }\href {\doibase 10.1103/RevModPhys.80.117} {\bibfield  {journal}
  {\bibinfo  {journal} {Rev. Mod. Phys.}\ }\textbf {\bibinfo {volume} {80}},\
  \bibinfo {pages} {117} (\bibinfo {year} {2008})}\BibitemShut {NoStop}%
\bibitem [{\citenamefont {Johnson}\ \emph {et~al.}(2018)\citenamefont
  {Johnson}, \citenamefont {Austin}, \citenamefont {Wood}, \citenamefont
  {Brahms}, \citenamefont {Gregory}, \citenamefont {Holzner}, \citenamefont
  {Jarosch}, \citenamefont {Larsen}, \citenamefont {Parker}, \citenamefont
  {Str{\"u}ber} \emph {et~al.}}]{johnson2018high}%
  \BibitemOpen
  \bibfield  {author} {\bibinfo {author} {\bibfnamefont {A.~S.}\ \bibnamefont
  {Johnson}}, \bibinfo {author} {\bibfnamefont {D.~R.}\ \bibnamefont {Austin}},
  \bibinfo {author} {\bibfnamefont {D.~A.}\ \bibnamefont {Wood}}, \bibinfo
  {author} {\bibfnamefont {C.}~\bibnamefont {Brahms}}, \bibinfo {author}
  {\bibfnamefont {A.}~\bibnamefont {Gregory}}, \bibinfo {author} {\bibfnamefont
  {K.~B.}\ \bibnamefont {Holzner}}, \bibinfo {author} {\bibfnamefont
  {S.}~\bibnamefont {Jarosch}}, \bibinfo {author} {\bibfnamefont {E.~W.}\
  \bibnamefont {Larsen}}, \bibinfo {author} {\bibfnamefont {S.}~\bibnamefont
  {Parker}}, \bibinfo {author} {\bibfnamefont {C.~S.}\ \bibnamefont
  {Str{\"u}ber}},  \emph {et~al.},\ }\href@noop {} {\bibfield  {journal}
  {\bibinfo  {journal} {Science advances}\ }\textbf {\bibinfo {volume} {4}},\
  \bibinfo {pages} {eaar3761} (\bibinfo {year} {2018})}\BibitemShut {NoStop}%
\bibitem [{\citenamefont {{A. P. Peirce, M. A. Dahleh, H.
  Rabitz}}(1988)}]{Rabitz}%
  \BibitemOpen
  \bibfield  {author} {\bibinfo {author} {\bibnamefont {{A. P. Peirce, M. A.
  Dahleh, H. Rabitz}}},\ }\href@noop {} {\bibfield  {journal} {\bibinfo
  {journal} {Phys. Rev. A}\ }\textbf {\bibinfo {volume} {{37}}},\ \bibinfo
  {pages} {{4950}} (\bibinfo {year} {{1988}})}\BibitemShut {NoStop}%
\bibitem [{\citenamefont {{Ronnie Kosloff, Stuart A. Rice, Pier Gaspard, Sam
  Tersigni and David Tannor}}(1989)}]{k67}%
  \BibitemOpen
  \bibfield  {author} {\bibinfo {author} {\bibnamefont {{Ronnie Kosloff, Stuart
  A. Rice, Pier Gaspard, Sam Tersigni and David Tannor}}},\ }\href@noop {}
  {\bibfield  {journal} {\bibinfo  {journal} {Chem. Phys.}\ }\textbf {\bibinfo
  {volume} {139}},\ \bibinfo {pages} {201} (\bibinfo {year}
  {1989})}\BibitemShut {NoStop}%
\bibitem [{\citenamefont {Glaser}\ \emph {et~al.}(2015)\citenamefont {Glaser},
  \citenamefont {Boscain}, \citenamefont {Calarco}, \citenamefont {Koch},
  \citenamefont {K{\"o}ckenberger}, \citenamefont {Kosloff}, \citenamefont
  {Kuprov}, \citenamefont {Luy}, \citenamefont {Schirmer}, \citenamefont
  {Schulte-Herbr{\"u}ggen} \emph {et~al.}}]{glaser2015training}%
  \BibitemOpen
  \bibfield  {author} {\bibinfo {author} {\bibfnamefont {S.~J.}\ \bibnamefont
  {Glaser}}, \bibinfo {author} {\bibfnamefont {U.}~\bibnamefont {Boscain}},
  \bibinfo {author} {\bibfnamefont {T.}~\bibnamefont {Calarco}}, \bibinfo
  {author} {\bibfnamefont {C.~P.}\ \bibnamefont {Koch}}, \bibinfo {author}
  {\bibfnamefont {W.}~\bibnamefont {K{\"o}ckenberger}}, \bibinfo {author}
  {\bibfnamefont {R.}~\bibnamefont {Kosloff}}, \bibinfo {author} {\bibfnamefont
  {I.}~\bibnamefont {Kuprov}}, \bibinfo {author} {\bibfnamefont
  {B.}~\bibnamefont {Luy}}, \bibinfo {author} {\bibfnamefont {S.}~\bibnamefont
  {Schirmer}}, \bibinfo {author} {\bibfnamefont {T.}~\bibnamefont
  {Schulte-Herbr{\"u}ggen}},  \emph {et~al.},\ }\href@noop {} {\bibfield
  {journal} {\bibinfo  {journal} {The European Physical Journal D}\ }\textbf
  {\bibinfo {volume} {69}},\ \bibinfo {pages} {279} (\bibinfo {year}
  {2015})}\BibitemShut {NoStop}%
\bibitem [{\citenamefont {Aroch}\ \emph {et~al.}(2018)\citenamefont {Aroch},
  \citenamefont {Kallush},\ and\ \citenamefont
  {Kosloff}}]{aroch2018optimizing}%
  \BibitemOpen
  \bibfield  {author} {\bibinfo {author} {\bibfnamefont {A.}~\bibnamefont
  {Aroch}}, \bibinfo {author} {\bibfnamefont {S.}~\bibnamefont {Kallush}}, \
  and\ \bibinfo {author} {\bibfnamefont {R.}~\bibnamefont {Kosloff}},\
  }\href@noop {} {\bibfield  {journal} {\bibinfo  {journal} {Physical Review
  A}\ }\textbf {\bibinfo {volume} {97}},\ \bibinfo {pages} {053405} (\bibinfo
  {year} {2018})}\BibitemShut {NoStop}%
\bibitem [{\citenamefont {{Jos\'e P. Palao and Ronnie Kosloff}}(2003)}]{k193}%
  \BibitemOpen
  \bibfield  {author} {\bibinfo {author} {\bibnamefont {{Jos\'e P. Palao and
  Ronnie Kosloff}}},\ }\href@noop {} {\bibfield  {journal} {\bibinfo  {journal}
  {Phys. Rev. A}\ }\textbf {\bibinfo {volume} {68}},\ \bibinfo {pages} {062308}
  (\bibinfo {year} {2003})}\BibitemShut {NoStop}%
\bibitem [{\citenamefont {Soml{\'o}i}\ \emph {et~al.}(1993)\citenamefont
  {Soml{\'o}i}, \citenamefont {Kazakov},\ and\ \citenamefont
  {Tannor}}]{somloi1993controlled}%
  \BibitemOpen
  \bibfield  {author} {\bibinfo {author} {\bibfnamefont {J.}~\bibnamefont
  {Soml{\'o}i}}, \bibinfo {author} {\bibfnamefont {V.~A.}\ \bibnamefont
  {Kazakov}}, \ and\ \bibinfo {author} {\bibfnamefont {D.~J.}\ \bibnamefont
  {Tannor}},\ }\href@noop {} {\bibfield  {journal} {\bibinfo  {journal}
  {Chemical physics}\ }\textbf {\bibinfo {volume} {172}},\ \bibinfo {pages}
  {85} (\bibinfo {year} {1993})}\BibitemShut {NoStop}%
\bibitem [{\citenamefont {Ohtsuki}\ \emph {et~al.}(2004)\citenamefont
  {Ohtsuki}, \citenamefont {Turinici},\ and\ \citenamefont
  {Rabitz}}]{ohtsuki2004generalized}%
  \BibitemOpen
  \bibfield  {author} {\bibinfo {author} {\bibfnamefont {Y.}~\bibnamefont
  {Ohtsuki}}, \bibinfo {author} {\bibfnamefont {G.}~\bibnamefont {Turinici}}, \
  and\ \bibinfo {author} {\bibfnamefont {H.}~\bibnamefont {Rabitz}},\
  }\href@noop {} {\bibfield  {journal} {\bibinfo  {journal} {The Journal of
  chemical physics}\ }\textbf {\bibinfo {volume} {120}},\ \bibinfo {pages}
  {5509} (\bibinfo {year} {2004})}\BibitemShut {NoStop}%
\bibitem [{\citenamefont {Eitan}\ \emph {et~al.}(2011)\citenamefont {Eitan},
  \citenamefont {Mundt},\ and\ \citenamefont {Tannor}}]{eitan2011optimal}%
  \BibitemOpen
  \bibfield  {author} {\bibinfo {author} {\bibfnamefont {R.}~\bibnamefont
  {Eitan}}, \bibinfo {author} {\bibfnamefont {M.}~\bibnamefont {Mundt}}, \ and\
  \bibinfo {author} {\bibfnamefont {D.~J.}\ \bibnamefont {Tannor}},\
  }\href@noop {} {\bibfield  {journal} {\bibinfo  {journal} {Physical Review
  A}\ }\textbf {\bibinfo {volume} {83}},\ \bibinfo {pages} {053426} (\bibinfo
  {year} {2011})}\BibitemShut {NoStop}%
\bibitem [{\citenamefont {{I. Degani, A. Zanna, L. S{\ae}len, and R.
  Nepstad}}(2009)}]{Degani}%
  \BibitemOpen
  \bibfield  {author} {\bibinfo {author} {\bibnamefont {{I. Degani, A. Zanna,
  L. S{\ae}len, and R. Nepstad}}},\ }\href@noop {} {\bibfield  {journal}
  {\bibinfo  {journal} {{SIAM J. Sci. Comput.}}\ }\textbf {\bibinfo {volume}
  {{31}}},\ \bibinfo {pages} {{3566}} (\bibinfo {year} {{2009}})}\BibitemShut
  {NoStop}%
\bibitem [{\citenamefont {Schirmer}\ and\ \citenamefont
  {de~Fouquieres}(2011)}]{schirmer2011efficient}%
  \BibitemOpen
  \bibfield  {author} {\bibinfo {author} {\bibfnamefont {S.~G.}\ \bibnamefont
  {Schirmer}}\ and\ \bibinfo {author} {\bibfnamefont {P.}~\bibnamefont
  {de~Fouquieres}},\ }\href@noop {} {\bibfield  {journal} {\bibinfo  {journal}
  {New Journal of Physics}\ }\textbf {\bibinfo {volume} {13}},\ \bibinfo
  {pages} {073029} (\bibinfo {year} {2011})}\BibitemShut {NoStop}%
\bibitem [{\citenamefont {Reich}\ \emph {et~al.}(2012)\citenamefont {Reich},
  \citenamefont {Ndong},\ and\ \citenamefont {Koch}}]{reich2012monotonically}%
  \BibitemOpen
  \bibfield  {author} {\bibinfo {author} {\bibfnamefont {D.~M.}\ \bibnamefont
  {Reich}}, \bibinfo {author} {\bibfnamefont {M.}~\bibnamefont {Ndong}}, \ and\
  \bibinfo {author} {\bibfnamefont {C.~P.}\ \bibnamefont {Koch}},\ }\href@noop
  {} {\bibfield  {journal} {\bibinfo  {journal} {The Journal of chemical
  physics}\ }\textbf {\bibinfo {volume} {136}},\ \bibinfo {pages} {104103}
  (\bibinfo {year} {2012})}\BibitemShut {NoStop}%
\bibitem [{\citenamefont {{I. Serban, J. Werschnik, and E. K. U.
  Gross}}(2005)}]{serban05}%
  \BibitemOpen
  \bibfield  {author} {\bibinfo {author} {\bibnamefont {{I. Serban, J.
  Werschnik, and E. K. U. Gross}}},\ }\href@noop {} {\bibfield  {journal}
  {\bibinfo  {journal} {Phys. Rev. A}\ }\textbf {\bibinfo {volume} {71}},\
  \bibinfo {pages} {053810} (\bibinfo {year} {2005})}\BibitemShut {NoStop}%
\bibitem [{\citenamefont {Werschnik}\ and\ \citenamefont
  {Gross}(2007)}]{gross07}%
  \BibitemOpen
  \bibfield  {author} {\bibinfo {author} {\bibfnamefont {J.}~\bibnamefont
  {Werschnik}}\ and\ \bibinfo {author} {\bibfnamefont {E.~K.~U.}\ \bibnamefont
  {Gross}},\ }\href {\doibase 10.1088/0953-4075/40/18/r01} {\bibfield
  {journal} {\bibinfo  {journal} {Journal of Physics B: Atomic, Molecular and
  Optical Physics}\ }\textbf {\bibinfo {volume} {40}},\ \bibinfo {pages} {R175}
  (\bibinfo {year} {2007})}\BibitemShut {NoStop}%
\bibitem [{\citenamefont {{Jos\'e P. Palao, Ronnie Kosloff, and Christiane P.
  Koch}}(2008)}]{k236}%
  \BibitemOpen
  \bibfield  {author} {\bibinfo {author} {\bibnamefont {{Jos\'e P. Palao,
  Ronnie Kosloff, and Christiane P. Koch}}},\ }\href@noop {} {\bibfield
  {journal} {\bibinfo  {journal} {Phys. Rev. A}\ }\textbf {\bibinfo {volume}
  {77}},\ \bibinfo {pages} {063412} (\bibinfo {year} {2008})}\BibitemShut
  {NoStop}%
\bibitem [{\citenamefont {Pezeshki}\ \emph {et~al.}(2008)\citenamefont
  {Pezeshki}, \citenamefont {Schreiber},\ and\ \citenamefont
  {Kleinekath{\"o}fer}}]{RamanASOCT}%
  \BibitemOpen
  \bibfield  {author} {\bibinfo {author} {\bibfnamefont {S.}~\bibnamefont
  {Pezeshki}}, \bibinfo {author} {\bibfnamefont {M.}~\bibnamefont {Schreiber}},
  \ and\ \bibinfo {author} {\bibfnamefont {U.}~\bibnamefont
  {Kleinekath{\"o}fer}},\ }\href {\doibase 10.1039/B714268D} {\bibfield
  {journal} {\bibinfo  {journal} {Phys. Chem. Chem. Phys.}\ }\textbf {\bibinfo
  {volume} {10}},\ \bibinfo {pages} {2058} (\bibinfo {year}
  {2008})}\BibitemShut {NoStop}%
\bibitem [{\citenamefont {Schaefer}(2012)}]{thesis}%
  \BibitemOpen
  \bibfield  {author} {\bibinfo {author} {\bibfnamefont {I.}~\bibnamefont
  {Schaefer}},\ }\href {https://arxiv.org/abs/1202.6520} {\bibfield  {journal}
  {\bibinfo  {journal} {{arXiv:1202.6520}}\ } (\bibinfo {year}
  {{2012}})}\BibitemShut {NoStop}%
\bibitem [{\citenamefont {Schaefer}\ and\ \citenamefont
  {Kosloff}(2012)}]{OCTHG}%
  \BibitemOpen
  \bibfield  {author} {\bibinfo {author} {\bibfnamefont {I.}~\bibnamefont
  {Schaefer}}\ and\ \bibinfo {author} {\bibfnamefont {R.}~\bibnamefont
  {Kosloff}},\ }\href {\doibase 10.1103/PhysRevA.86.063417} {\bibfield
  {journal} {\bibinfo  {journal} {Phys. Rev. A}\ }\textbf {\bibinfo {volume}
  {86}},\ \bibinfo {pages} {063417} (\bibinfo {year} {2012})}\BibitemShut
  {NoStop}%
\bibitem [{\citenamefont {Krause}\ \emph {et~al.}(1992)\citenamefont {Krause},
  \citenamefont {Schafer},\ and\ \citenamefont {Kulander}}]{krause92}%
  \BibitemOpen
  \bibfield  {author} {\bibinfo {author} {\bibfnamefont {J.~L.}\ \bibnamefont
  {Krause}}, \bibinfo {author} {\bibfnamefont {K.~J.}\ \bibnamefont {Schafer}},
  \ and\ \bibinfo {author} {\bibfnamefont {K.~C.}\ \bibnamefont {Kulander}},\
  }\href@noop {} {\bibfield  {journal} {\bibinfo  {journal} {Phys. Rev. A}\
  }\textbf {\bibinfo {volume} {45}},\ \bibinfo {pages} {4998} (\bibinfo {year}
  {1992})}\BibitemShut {NoStop}%
\bibitem [{\citenamefont {Yu}\ and\ \citenamefont {Esry}(2018)}]{EsryCAP}%
  \BibitemOpen
  \bibfield  {author} {\bibinfo {author} {\bibfnamefont {Y.}~\bibnamefont
  {Yu}}\ and\ \bibinfo {author} {\bibfnamefont {B.~D.}\ \bibnamefont {Esry}},\
  }\href {\doibase 10.1088/1361-6455/aab5d6} {\bibfield  {journal} {\bibinfo
  {journal} {Journal of Physics B: Atomic, Molecular and Optical Physics}\
  }\textbf {\bibinfo {volume} {51}},\ \bibinfo {pages} {095601} (\bibinfo
  {year} {2018})}\BibitemShut {NoStop}%
\bibitem [{\citenamefont {Solanp\"a\"a}\ \emph {et~al.}(2014)\citenamefont
  {Solanp\"a\"a}, \citenamefont {Budagosky}, \citenamefont
  {Shvetsov-Shilovski}, \citenamefont {Castro}, \citenamefont {Rubio},\ and\
  \citenamefont {R\"as\"anen}}]{RasanenHHG}%
  \BibitemOpen
  \bibfield  {author} {\bibinfo {author} {\bibfnamefont {J.}~\bibnamefont
  {Solanp\"a\"a}}, \bibinfo {author} {\bibfnamefont {J.~A.}\ \bibnamefont
  {Budagosky}}, \bibinfo {author} {\bibfnamefont {N.~I.}\ \bibnamefont
  {Shvetsov-Shilovski}}, \bibinfo {author} {\bibfnamefont {A.}~\bibnamefont
  {Castro}}, \bibinfo {author} {\bibfnamefont {A.}~\bibnamefont {Rubio}}, \
  and\ \bibinfo {author} {\bibfnamefont {E.}~\bibnamefont {R\"as\"anen}},\
  }\href {\doibase 10.1103/PhysRevA.90.053402} {\bibfield  {journal} {\bibinfo
  {journal} {Phys. Rev. A}\ }\textbf {\bibinfo {volume} {90}},\ \bibinfo
  {pages} {053402} (\bibinfo {year} {2014})}\BibitemShut {NoStop}%
\bibitem [{\citenamefont {Castro}\ \emph {et~al.}(2015)\citenamefont {Castro},
  \citenamefont {Rubio},\ and\ \citenamefont {Gross}}]{DFTHHG}%
  \BibitemOpen
  \bibfield  {author} {\bibinfo {author} {\bibfnamefont {A.}~\bibnamefont
  {Castro}}, \bibinfo {author} {\bibfnamefont {A.}~\bibnamefont {Rubio}}, \
  and\ \bibinfo {author} {\bibfnamefont {E.~K.~U.}\ \bibnamefont {Gross}},\
  }\href {\doibase 10.1140/epjb/e2015-50889-7} {\bibfield  {journal} {\bibinfo
  {journal} {The European Physical Journal B}\ }\textbf {\bibinfo {volume}
  {88}},\ \bibinfo {pages} {191} (\bibinfo {year} {2015})}\BibitemShut
  {NoStop}%
\bibitem [{\citenamefont {Balogh}\ \emph {et~al.}(2014)\citenamefont {Balogh},
  \citenamefont {B\'odi}, \citenamefont {Tosa}, \citenamefont {Goulielmakis},
  \citenamefont {Varj\'u},\ and\ \citenamefont {Dombi}}]{DombiGeneticHHG}%
  \BibitemOpen
  \bibfield  {author} {\bibinfo {author} {\bibfnamefont {E.}~\bibnamefont
  {Balogh}}, \bibinfo {author} {\bibfnamefont {B.}~\bibnamefont {B\'odi}},
  \bibinfo {author} {\bibfnamefont {V.}~\bibnamefont {Tosa}}, \bibinfo {author}
  {\bibfnamefont {E.}~\bibnamefont {Goulielmakis}}, \bibinfo {author}
  {\bibfnamefont {K.}~\bibnamefont {Varj\'u}}, \ and\ \bibinfo {author}
  {\bibfnamefont {P.}~\bibnamefont {Dombi}},\ }\href {\doibase
  10.1103/PhysRevA.90.023855} {\bibfield  {journal} {\bibinfo  {journal} {Phys.
  Rev. A}\ }\textbf {\bibinfo {volume} {90}},\ \bibinfo {pages} {023855}
  (\bibinfo {year} {2014})}\BibitemShut {NoStop}%
\bibitem [{\citenamefont {Jin}\ and\ \citenamefont {Lin}(2016)}]{Jin2016}%
  \BibitemOpen
  \bibfield  {author} {\bibinfo {author} {\bibfnamefont {C.}~\bibnamefont
  {Jin}}\ and\ \bibinfo {author} {\bibfnamefont {C.~D.}\ \bibnamefont {Lin}},\
  }\href {\doibase 10.1088/1674-1056/25/9/094213} {\bibfield  {journal}
  {\bibinfo  {journal} {Chinese Physics B}\ }\textbf {\bibinfo {volume} {25}},\
  \bibinfo {pages} {094213} (\bibinfo {year} {2016})}\BibitemShut {NoStop}%
\bibitem [{\citenamefont {Sch\"onborn}\ \emph {et~al.}(2016)\citenamefont
  {Sch\"onborn}, \citenamefont {Saalfrank},\ and\ \citenamefont
  {Klamroth}}]{SPO_HHG}%
  \BibitemOpen
  \bibfield  {author} {\bibinfo {author} {\bibfnamefont {J.~B.}\ \bibnamefont
  {Sch\"onborn}}, \bibinfo {author} {\bibfnamefont {P.}~\bibnamefont
  {Saalfrank}}, \ and\ \bibinfo {author} {\bibfnamefont {T.}~\bibnamefont
  {Klamroth}},\ }\href {\doibase 10.1063/1.4940316} {\bibfield  {journal}
  {\bibinfo  {journal} {The Journal of Chemical Physics}\ }\textbf {\bibinfo
  {volume} {144}},\ \bibinfo {pages} {044301} (\bibinfo {year} {2016})},\
  \Eprint {http://arxiv.org/abs/https://doi.org/10.1063/1.4940316}
  {https://doi.org/10.1063/1.4940316} \BibitemShut {NoStop}%
\bibitem [{\citenamefont {{T. E. Skinner, N. I. Gershenzon}}(2010)}]{Skinner}%
  \BibitemOpen
  \bibfield  {author} {\bibinfo {author} {\bibnamefont {{T. E. Skinner, N. I.
  Gershenzon}}},\ }\href@noop {} {\bibfield  {journal} {\bibinfo  {journal}
  {{J. Mag. Res.}}\ }\textbf {\bibinfo {volume} {{204}}},\ \bibinfo {pages}
  {{248}} (\bibinfo {year} {{2010}})}\BibitemShut {NoStop}%
\bibitem [{\citenamefont {Gaarde}\ \emph {et~al.}(2008)\citenamefont {Gaarde},
  \citenamefont {Tate},\ and\ \citenamefont {Schafer}}]{MacroscopicHHG}%
  \BibitemOpen
  \bibfield  {author} {\bibinfo {author} {\bibfnamefont {M.~B.}\ \bibnamefont
  {Gaarde}}, \bibinfo {author} {\bibfnamefont {J.~L.}\ \bibnamefont {Tate}}, \
  and\ \bibinfo {author} {\bibfnamefont {K.~J.}\ \bibnamefont {Schafer}},\
  }\href {\doibase 10.1088/0953-4075/41/13/132001} {\bibfield  {journal}
  {\bibinfo  {journal} {Journal of Physics B: Atomic, Molecular and Optical
  Physics}\ }\textbf {\bibinfo {volume} {41}},\ \bibinfo {pages} {132001}
  (\bibinfo {year} {2008})}\BibitemShut {NoStop}%
\bibitem [{\citenamefont {Fletcher}(1987)}]{Fletcher}%
  \BibitemOpen
  \bibfield  {author} {\bibinfo {author} {\bibfnamefont {R.}~\bibnamefont
  {Fletcher}},\ }\href@noop {} {\emph {\bibinfo {title} {Practical methods of
  optimization}}}\ (\bibinfo  {publisher} {John Wiley \& Sons},\ \bibinfo
  {year} {1987})\BibitemShut {NoStop}%
\bibitem [{\citenamefont {{Hillel Tal-Ezer, Ronnie Kosloff, Ido
  Schaefer}}(2012)}]{k273}%
  \BibitemOpen
  \bibfield  {author} {\bibinfo {author} {\bibnamefont {{Hillel Tal-Ezer,
  Ronnie Kosloff, Ido Schaefer}}},\ }\href@noop {} {\bibfield  {journal}
  {\bibinfo  {journal} {{J. Sci. Comput.}}\ }\textbf {\bibinfo {volume} {53}},\
  \bibinfo {pages} {211} (\bibinfo {year} {2012})}\BibitemShut {NoStop}%
\bibitem [{\citenamefont {Schaefer}\ \emph {et~al.}(2017)\citenamefont
  {Schaefer}, \citenamefont {Tal-Ezer},\ and\ \citenamefont
  {Kosloff}}]{SemiGlobal}%
  \BibitemOpen
  \bibfield  {author} {\bibinfo {author} {\bibfnamefont {I.}~\bibnamefont
  {Schaefer}}, \bibinfo {author} {\bibfnamefont {H.}~\bibnamefont {Tal-Ezer}},
  \ and\ \bibinfo {author} {\bibfnamefont {R.}~\bibnamefont {Kosloff}},\ }\href
  {\doibase https://doi.org/10.1016/j.jcp.2017.04.017} {\bibfield  {journal}
  {\bibinfo  {journal} {Journal of Computational Physics}\ }\textbf {\bibinfo
  {volume} {343}},\ \bibinfo {pages} {368 } (\bibinfo {year}
  {2017})}\BibitemShut {NoStop}%
\bibitem [{\citenamefont {Kosloff}\ and\ \citenamefont
  {Kosloff}(1983)}]{FourierGrid}%
  \BibitemOpen
  \bibfield  {author} {\bibinfo {author} {\bibfnamefont {D.}~\bibnamefont
  {Kosloff}}\ and\ \bibinfo {author} {\bibfnamefont {R.}~\bibnamefont
  {Kosloff}},\ }\href@noop {} {\bibfield  {journal} {\bibinfo  {journal}
  {Journal of Computational Physics}\ }\textbf {\bibinfo {volume} {52}},\
  \bibinfo {pages} {35} (\bibinfo {year} {1983})}\BibitemShut {NoStop}%
\bibitem [{\citenamefont {Kosloff}(1988)}]{k56}%
  \BibitemOpen
  \bibfield  {author} {\bibinfo {author} {\bibfnamefont {R.}~\bibnamefont
  {Kosloff}},\ }\href@noop {} {\bibfield  {journal} {\bibinfo  {journal} {The
  Journal of Physical Chemistry}\ }\textbf {\bibinfo {volume} {92}},\ \bibinfo
  {pages} {2087} (\bibinfo {year} {1988})}\BibitemShut {NoStop}%
\bibitem [{\citenamefont {Muga}\ \emph {et~al.}(2004)\citenamefont {Muga},
  \citenamefont {Palao}, \citenamefont {Navarro},\ and\ \citenamefont
  {Egusquiza}}]{CAP}%
  \BibitemOpen
  \bibfield  {author} {\bibinfo {author} {\bibfnamefont {J.}~\bibnamefont
  {Muga}}, \bibinfo {author} {\bibfnamefont {J.}~\bibnamefont {Palao}},
  \bibinfo {author} {\bibfnamefont {B.}~\bibnamefont {Navarro}}, \ and\
  \bibinfo {author} {\bibfnamefont {I.}~\bibnamefont {Egusquiza}},\ }\href
  {\doibase http://dx.doi.org/10.1016/j.physrep.2004.03.002} {\bibfield
  {journal} {\bibinfo  {journal} {Physics Reports}\ }\textbf {\bibinfo {volume}
  {395}},\ \bibinfo {pages} {357 } (\bibinfo {year} {2004})}\BibitemShut
  {NoStop}%
\bibitem [{\citenamefont {Palao}\ and\ \citenamefont
  {Muga}(1998)}]{squareBarriers}%
  \BibitemOpen
  \bibfield  {author} {\bibinfo {author} {\bibfnamefont {J.}~\bibnamefont
  {Palao}}\ and\ \bibinfo {author} {\bibfnamefont {J.}~\bibnamefont {Muga}},\
  }\href {\doibase http://dx.doi.org/10.1016/S0009-2614(98)00635-6} {\bibfield
  {journal} {\bibinfo  {journal} {Chemical Physics Letters}\ }\textbf {\bibinfo
  {volume} {292}},\ \bibinfo {pages} {1 } (\bibinfo {year} {1998})}\BibitemShut
  {NoStop}%
\bibitem [{\citenamefont {Kalotas}\ and\ \citenamefont
  {Lee}(1991)}]{staticScattering}%
  \BibitemOpen
  \bibfield  {author} {\bibinfo {author} {\bibfnamefont {T.}~\bibnamefont
  {Kalotas}}\ and\ \bibinfo {author} {\bibfnamefont {A.}~\bibnamefont {Lee}},\
  }\href@noop {} {\bibfield  {journal} {\bibinfo  {journal} {American Journal
  of Physics}\ }\textbf {\bibinfo {volume} {59}},\ \bibinfo {pages} {48}
  (\bibinfo {year} {1991})}\BibitemShut {NoStop}%
\bibitem [{\citenamefont {Ben-Tal}\ \emph
  {et~al.}(1993{\natexlab{b}})\citenamefont {Ben-Tal}, \citenamefont
  {Moiseyev},\ and\ \citenamefont {Beswick}}]{BenTal_symmetry}%
  \BibitemOpen
  \bibfield  {author} {\bibinfo {author} {\bibfnamefont {N.}~\bibnamefont
  {Ben-Tal}}, \bibinfo {author} {\bibfnamefont {N.}~\bibnamefont {Moiseyev}}, \
  and\ \bibinfo {author} {\bibfnamefont {A.}~\bibnamefont {Beswick}},\ }\href
  {\doibase 10.1088/0953-4075/26/18/012} {\bibfield  {journal} {\bibinfo
  {journal} {Journal of Physics B: Atomic, Molecular and Optical Physics}\
  }\textbf {\bibinfo {volume} {26}},\ \bibinfo {pages} {3017} (\bibinfo {year}
  {1993}{\natexlab{b}})}\BibitemShut {NoStop}%
\bibitem [{\citenamefont {Alon}\ \emph {et~al.}(1998)\citenamefont {Alon},
  \citenamefont {Averbukh},\ and\ \citenamefont {Moiseyev}}]{ofir98}%
  \BibitemOpen
  \bibfield  {author} {\bibinfo {author} {\bibfnamefont {O.~E.}\ \bibnamefont
  {Alon}}, \bibinfo {author} {\bibfnamefont {V.}~\bibnamefont {Averbukh}}, \
  and\ \bibinfo {author} {\bibfnamefont {N.}~\bibnamefont {Moiseyev}},\ }\href
  {\doibase 10.1103/PhysRevLett.80.3743} {\bibfield  {journal} {\bibinfo
  {journal} {Phys. Rev. Lett.}\ }\textbf {\bibinfo {volume} {80}},\ \bibinfo
  {pages} {3743} (\bibinfo {year} {1998})}\BibitemShut {NoStop}%
\bibitem [{\citenamefont {Diestler}(2008)}]{Diestler08}%
  \BibitemOpen
  \bibfield  {author} {\bibinfo {author} {\bibfnamefont {D.~J.}\ \bibnamefont
  {Diestler}},\ }\href {\doibase 10.1103/PhysRevA.78.033814} {\bibfield
  {journal} {\bibinfo  {journal} {Phys. Rev. A}\ }\textbf {\bibinfo {volume}
  {78}},\ \bibinfo {pages} {033814} (\bibinfo {year} {2008})}\BibitemShut
  {NoStop}%
\bibitem [{\citenamefont {Baggesen}\ and\ \citenamefont
  {Madsen}(2011{\natexlab{a}})}]{xvaHHG}%
  \BibitemOpen
  \bibfield  {author} {\bibinfo {author} {\bibfnamefont {J.~C.}\ \bibnamefont
  {Baggesen}}\ and\ \bibinfo {author} {\bibfnamefont {L.~B.}\ \bibnamefont
  {Madsen}},\ }\href {\doibase 10.1088/0953-4075/44/11/115601} {\bibfield
  {journal} {\bibinfo  {journal} {Journal of Physics B: Atomic, Molecular and
  Optical Physics}\ }\textbf {\bibinfo {volume} {44}},\ \bibinfo {pages}
  {115601} (\bibinfo {year} {2011}{\natexlab{a}})}\BibitemShut {NoStop}%
\bibitem [{\citenamefont {P{\'{e}}rez-Hern{\'{a}}ndez}\ and\ \citenamefont
  {Plaja}(2011)}]{xvaHHGcomment}%
  \BibitemOpen
  \bibfield  {author} {\bibinfo {author} {\bibfnamefont {J.~A.}\ \bibnamefont
  {P{\'{e}}rez-Hern{\'{a}}ndez}}\ and\ \bibinfo {author} {\bibfnamefont
  {L.}~\bibnamefont {Plaja}},\ }\href {\doibase 10.1088/0953-4075/45/2/028001}
  {\bibfield  {journal} {\bibinfo  {journal} {Journal of Physics B: Atomic,
  Molecular and Optical Physics}\ }\textbf {\bibinfo {volume} {45}},\ \bibinfo
  {pages} {028001} (\bibinfo {year} {2011})}\BibitemShut {NoStop}%
\bibitem [{\citenamefont {Baggesen}\ and\ \citenamefont
  {Madsen}(2011{\natexlab{b}})}]{xvaHHGreply}%
  \BibitemOpen
  \bibfield  {author} {\bibinfo {author} {\bibfnamefont {J.~C.}\ \bibnamefont
  {Baggesen}}\ and\ \bibinfo {author} {\bibfnamefont {L.~B.}\ \bibnamefont
  {Madsen}},\ }\href {\doibase 10.1088/0953-4075/45/2/028002} {\bibfield
  {journal} {\bibinfo  {journal} {Journal of Physics B: Atomic, Molecular and
  Optical Physics}\ }\textbf {\bibinfo {volume} {45}},\ \bibinfo {pages}
  {028002} (\bibinfo {year} {2011}{\natexlab{b}})}\BibitemShut {NoStop}%
\end{thebibliography}%

\end{document}